\documentclass[preprint, 12pt, authoryear]{elsarticle}

\usepackage{libertinus}
\usepackage[T1]{fontenc}
\usepackage[utf8]{inputenc}
\usepackage[british]{babel}
\usepackage{amsmath,amssymb,amsfonts}
\usepackage{bm}
\usepackage{graphicx}
\graphicspath{{plotting_files/}}
\usepackage{booktabs}
\usepackage{float}
\usepackage{longtable}
\usepackage{siunitx}
\usepackage{multirow}
\usepackage{array}
\usepackage{makecell}
\usepackage{pdflscape}

\usepackage{subcaption}
\usepackage{xcolor}
\definecolor{colourone}{HTML}{E69F00}
\definecolor{colourtwo}{HTML}{56B4E9}
\definecolor{colourthree}{HTML}{009E73}
\definecolor{colourfour}{HTML}{CC79A7}
\definecolor{colourfive}{HTML}{0072B2}
\definecolor{coloursix}{HTML}{D55E00}
\usepackage{annotate-equations}
\usetikzlibrary{positioning,arrows.meta,fit,calc}
\usepackage{needspace}

\usepackage[htt]{hyphenat}
\usepackage{microtype}
\usepackage{orcidlink}

\usepackage{hyperref}
\hypersetup{colorlinks=true, allcolors={[rgb]{0.11,0.24,0.53}}}

\usepackage{cleveref}
\crefname{section}{Section}{Sections}
\Crefname{section}{Section}{Sections}
\crefname{subsection}{Section}{Sections}
\Crefname{subsection}{Section}{Sections}
\crefname{equation}{Equation}{Equations}
\Crefname{equation}{Equation}{Equations}
\crefname{table}{Table}{Tables}
\Crefname{table}{Table}{Tables}
\crefname{figure}{Figure}{Figures}
\Crefname{figure}{Figure}{Figures}

\newcommand{\ST}{S_{T}}
\newcommand{\Sone}{S_{1}}

\newcommand{\Si}{S_{i}}
\newcommand{\STi}{S_{T_{i}}}

\newcommand{\mustar}{\mu^{*}}

\newcommand{\E}{\mathbb{E}}
\newcommand{\Var}{\mathrm{Var}}

\hypersetup{pdfauthor={Daniele Pessina, Roberto A. Abbiati, Maria M. Papathanasiou}}

\journal{a peer-reviewed journal}

\begin{document}

\begin{frontmatter}

  \title{Amortised global sensitivity analysis of pharmacokinetic models for Simulated Clinical Trials}

  \author[chemeng,sargent]{Daniele Pessina\corref{cor1}\,\orcidlink{0009-0004-3107-6857}}
  \ead{daniele.pessina18@imperial.ac.uk}
  \author[roche]{Roberto A. Abbiati}
  \author[chemeng,sargent]{Maria M. Papathanasiou\corref{cor1}\,\orcidlink{0000-0002-8886-0624}}
  \ead{maria.papathanasiou11@imperial.ac.uk}
  \cortext[cor1]{Corresponding authors}
  \affiliation[chemeng]{organization={Department of Chemical Engineering, Imperial College London},
  city={London}, postcode={SW7 2AZ}, country={U.K.}}
  \affiliation[sargent]{organization={Sargent Centre for Process Systems Engineering, Imperial College London},
  city={London}, postcode={SW7 2AZ}, country={U.K.}}
  \affiliation[roche]{organization={Roche Pharma Research and Early Development, Quantitative Clinical Pharmacology},
    addressline={F. Hoffmann-La Roche}}

  \begin{abstract}

    Pharmacokinetic and pharmacokinetic--pharmacodynamic models are used in simulated clinical trials to compare dosing protocols and patient groups before a trial is run.
    Distinct virtual patient cohorts, representing renal impairment, obesity and overdosage, can impose different input distributions on the same model.
    Global sensitivity analysis (GSA) identifies which uncertain inputs drive the cohort's predictions, but a direct analysis incurs its full computational cost for every cohort.
    An alternative accounting is proposed here: a single surrogate is trained on an input envelope wide enough to contain every cohort, each cohort is subsequently analysed using that surrogate, and the reported quantity is the number of cohorts at which the surrogate route has cost fewer mechanistic evaluations than the direct route.
    The study tests three GSA methods: variance-based GSA, Morris screening and optimal transport. It covers four PK and PK/PD models with seven to twenty-three free inputs, 21 cohorts and five surrogates.
    Every estimate is scored against a converged mechanistic solution obtained with its own method, and all costs are expressed in mechanistic model evaluations.
    The parameter ranking is recovered at low cost: gradient-boosted trees reconstruct it on 20 of the 21 cohorts and cost fewer evaluations to train than one direct cohort analysis, on all four models, at break-even counts between 0.02 and 0.14 cohorts.
    Amortisation is fastest at the highest dimensions, where each direct variance-based sample of a $D$-input model requires $D+2$ evaluations while the training budget does not scale accordingly.
    Reconstruction of the absolute index values is a separate and more demanding problem that narrows with dimension: two of the five surrogates fail at fifteen inputs, where the gradient-boosted trees and the Gaussian process still succeed, and at twenty-three inputs no surrogate succeeds on a majority of cohorts.
    The break-even cohort count makes the repeated-use advantage explicit: for parameter rankings it is below one on all four models.
  \end{abstract}


\end{frontmatter}

\section{Introduction}
\label{sec:introduction}

Mathematical models of drug disposition and effect are used to make decisions under uncertainty: which parameters must be measured precisely, which can be fixed to a nominal value, and how confident a prediction is.
Global sensitivity analysis (GSA) can help with this by apportioning the uncertainty in a model output among its uncertain inputs, over the whole input space rather than at a single nominal point \citep{saltelli2008primer,iooss2015}.
In pharmacokinetic (PK), physiologically based PK (PBPK) and quantitative systems pharmacology (QSP) modelling, where models are high-dimensional, governed by stiff ordinary differential equations and driven by mixed input distributions, GSA has become a standard tool for model interrogation, reduction and reporting \citep{zhang2015sobol,mcnally2011workflow,marino2008methodology}.
Yet local one-at-a-time sensitivity, which perturbs one input around a nominal value, cannot capture interactions or the effect of the input distribution, and misleads on the non-linear, interaction-rich models typical of systems pharmacology; because it is cheaper, it remains common \citep{saltelli2008primer,nestorov1999sensitivity}.

Increasingly, the decisions these models support are taken before any clinical trial is run.
Pharmaceutical research and development has seen a sustained decline in new drugs approved per unit of spending \citep{scannellDiagnosingDeclinePharmaceutical2012}.
Simulated clinical trials generate and analyse virtual patient responses under a proposed protocol, yielding estimates of outcomes that can inform the design of a real trial \citep{arseneSilicoClinicalTrials2024,savySimulatedClinicalTrials2018,meyerDesignSimulateRefine2026}.
A survey of quantitative systems pharmacology practitioners identifies clinical trial design and patient stratification among the applications the approach has most helped with, alongside dosing and schedule selection and the evaluation of combination therapy \citep{lemaireColdHotChanging2023}.

In these trials, ensembles of model parametrisations represent patient-to-patient variability. Each sample corresponds to a virtual patient whose predicted response can be checked against clinical observations or predefined plausibility criteria; accepted patients form a virtual cohort \citep{chengVirtualPopulationsQuantitative2022a}.
Input distributions and correlations may be estimated from a non-linear mixed-effects population fit \citep{cortes-riosStepbyStepWorkflowPerforming2025}, while the construction and acceptance of virtual patients remain methodological choices, with approaches ranging from Markov chain Monte Carlo to stochastically bounded selection \citep{kolesovaApplicationDifferentApproaches2022}.

In this workflow, GSA has two potential roles: it explains which inputs drive variability within a generated cohort, and it can inform the plausibility assessment used when selecting virtual patients \citep{braniffIntegratedQuantitativeSystems2025}.
For example, \citet{hilhorstSensitivityAnalysisExploring} identified stenosis severity as the dominant driver of variability in a virtual cohort of the coronary circulation, using a variance-based GSA adapted for the correlations that acceptance criteria induce, and \citet{kolesovaApplicationDifferentApproaches2022} separated competing virtual cohort generation schemes with a sensitivity analysis across thirty-nine parameters.

GSA methods answer different pharmacological questions and require different simulation designs (\cref{tab:gsa-landscape}).

\begin{table}[tbp]
\centering
\caption{Main families of global sensitivity methods and the question each addresses. The three methods compared in this study are variance-based GSA, Morris screening and optimal transport.}
\label{tab:gsa-landscape}
\footnotesize
\setlength{\tabcolsep}{4pt}
\renewcommand{\arraystretch}{1.12}
\begin{tabular}{@{}>{\raggedright\arraybackslash}p{0.20\textwidth}>{\raggedright\arraybackslash}p{0.36\textwidth}>{\raggedright\arraybackslash}p{0.36\textwidth}@{}}
\toprule
Family & What it measures & Representative methods \\
\midrule
Variance-based & Shares of output variance attributable to inputs and their interactions & Sobol' indices \citep{sobol2001}; extended FAST \citep{saltelli1999fast} \\
Morris screening & Average input effects and their direction across the input space & Elementary effects \citep{morris1991factorial} \\
Derivative-based & Local gradients aggregated across the input distribution; bounds on total effects & Derivative-based global measures \citep{sobol2009derivative} \\
Surrogate decomposition & Sensitivity indices read from a fitted model expansion & Polynomial chaos \citep{sudret2008global}; high-dimensional model representation \citep{li2002practical} \\
Distributional and dependence & Changes in the output distribution, including effects beyond variance shares & $\delta$ importance \citep{borgonovo2007delta}; PAWN \citep{pianosi2015pawn}; Hilbert--Schmidt dependence \citep{daveiga2015dependence}; optimal transport \citep{borgonovo2025ot} \\
\bottomrule
\end{tabular}
\end{table}

Distributional and dependence measures are established in environmental and reliability analysis \citep{khorashadizadeh2017comparison,elamri2022optimized} but remain uncommon in pharmacokinetic modelling, where variance-based GSA and Morris screening dominate.
We therefore compare variance shares, screening effects and distributional changes on the same models and cohorts, and price each by the mechanistic simulations required for a stable answer.

A common challenge is that variance-based GSA is expensive and slow to converge, and is therefore impractical for the very models that most need it.
A variance-based GSA of $D$ inputs costs $N(D+2)$ model evaluations for the first- and total-order indices, with $N$ often in the thousands before the estimates stabilise \citep{sobol2001,saltelli2010}. %
Morris screening was developed for precisely this constraint, replacing the variance decomposition with an average of local slopes at $r(D+1)$ evaluations, with $r$ in the hundreds where $N$ is in the thousands, and returning a ranking rather than an apportioned share of variance \citep{morris1991factorial,campolongo2007effective}.
Optimal transport indices are nominally cheaper still, since they are estimated from any given sample of $N$ points with no factor in $D$ at all \citep{borgonovo2025ot,plischke2013givendata}, and they measure a shift in the whole output distribution rather than in its variance.
Two established research avenues address this cost: replacing the mechanistic model with a cheap surrogate and reading or estimating the indices from it \citep{sudret2008global,legratiet2017,schafer2024multifidelity}, and using GSA itself to fix non-influential parameters and so shrink the problem before an expensive downstream task \citep{hsieh2018applying,hsieh2020pksensi}.
Both are now established in the PBPK and QSP literature \citep{mcnally2011workflow,melillo2019variance,loizou2015application,sanchezrestrepo2022global}.
Convergence dynamics are strongly case-dependent; a stable ranking or screening can be reached well before the absolute index values settle, and no single sample size serves every problem \citep{sarrazin2016}.
Existing comparisons often cover a small number of methods on a single model or a static analytical benchmark, and rarely quantify convergence as a function of the number of expensive mechanistic evaluations.

An open challenge is to compare cost across methods and models while accounting for repeated analyses of one model under different input distributions.
A surrogate amortises that cost as it is trained once on an envelope wide enough to contain every cohort, and each further cohort is then read from it at no additional cost in mechanistic evaluations.

One competing route which does not require a surrogate is to reweight a single Monte Carlo sample to yield variance-based indices under a different input distribution.
\citet{boucharif2025} applied importance sampling to variance-based indices specifically, weighting each stored point by the ratio of the cohort density to the envelope density, so that one mechanistic campaign would yield every cohort analysis with no surrogate approximation error at all.
Its efficiency falls geometrically as a cohort narrows, and fails entirely if a cohort pins an axis to a single value, because the envelope samples that value with probability zero and so no density ratio exists.

Responding to these challenges, our contributions are threefold.
First, we introduce the envelope-and-cohort design, one wide uniform envelope per model with clinically motivated cohorts nested inside it and one surrogate trained on the envelope serving every cohort, and place every method and every surrogate on a single cost axis measured in mechanistic model evaluations.
Second, we report the break-even cohort count per model and per surrogate, separated by a two-tier accuracy gate into the cost of a correct ranking and the cost of a correct index value, which differ by an order of magnitude and diverge with input dimension.
Third, we compare variance-based, Morris screening and optimal transport as peers on identical models, cohorts, cost axis and designs, and ask what a second method adds for its computational cost.
Reconstructing absolute index values becomes harder with dimension until it is reachable only for the best surrogate, and the distributional lens adds distributional information without reordering the parameters that matter on these models.

\section{Methodology}
\label{sec:materials}

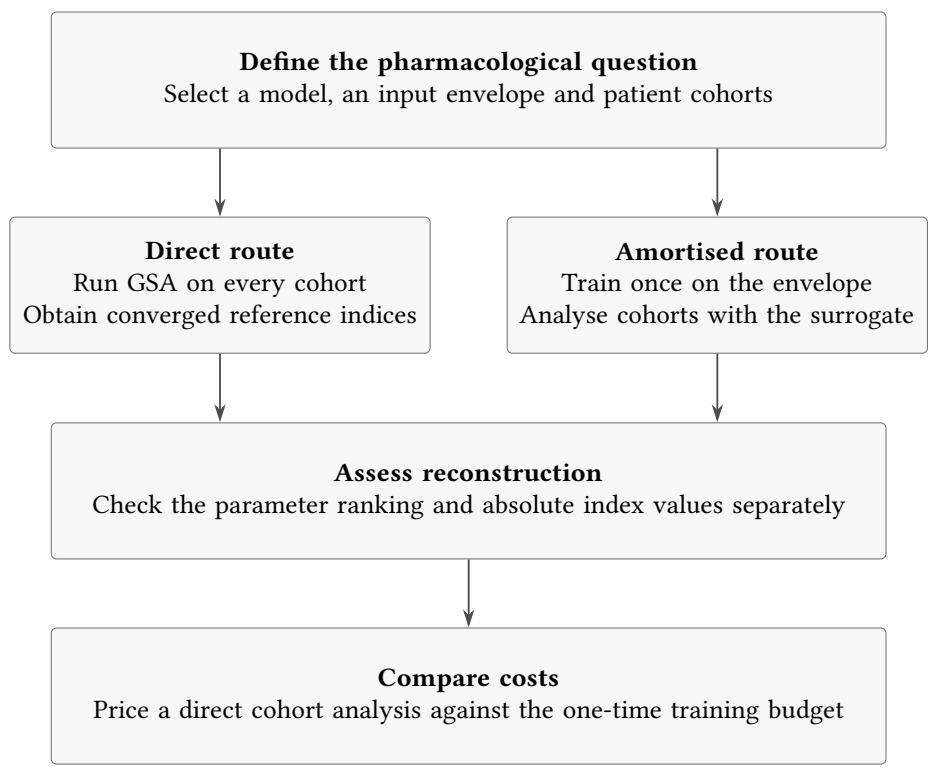
\begin{figure}[H]
\centering
\begin{tikzpicture}[
  font=\footnotesize,
  wfbox/.style={draw=black!55, rounded corners=2pt, fill=black!3,
    align=center, text width=0.38\textwidth, minimum height=18mm, inner sep=5pt},
  route/.style={draw=black!70, -{Stealth[length=2.4mm,width=1.5mm]},
    line width=0.7pt}
]
\node[wfbox, text width=0.78\textwidth, anchor=north] (define) at (0,0)
  {\textbf{Define the pharmacological question}\\Select a model, an input envelope and patient cohorts};
\node[wfbox, anchor=north] (direct)
  at ([xshift=-0.24\textwidth,yshift=-9mm]define.south)
  {\textbf{Direct route}\\Run GSA on every cohort\\Obtain converged reference indices};
\node[wfbox, anchor=north] (surrogate)
  at ([xshift=0.24\textwidth,yshift=-9mm]define.south)
  {\textbf{Amortised route}\\Train once on the envelope\\Analyse cohorts with the surrogate};
\node[fit=(direct)(surrogate), inner sep=0pt] (routes) {};
\node[wfbox, text width=0.78\textwidth, anchor=north] (assess)
  at ([yshift=-9mm]routes.south)
  {\textbf{Assess reconstruction}\\Check the parameter ranking and absolute index values separately};
\node[wfbox, text width=0.78\textwidth, anchor=north] (compare)
  at ([yshift=-9mm]assess.south)
  {\textbf{Compare costs}\\Price a direct cohort analysis against the one-time training budget};

\draw[route] ([xshift=-0.24\textwidth]define.south) -- (direct.north);
\draw[route] ([xshift=0.24\textwidth]define.south) -- (surrogate.north);
\draw[route] (direct.south) -- ([xshift=-0.24\textwidth]assess.north);
\draw[route] (surrogate.south) -- ([xshift=0.24\textwidth]assess.north);
\draw[route] (assess.south) -- (compare.north);
\end{tikzpicture}
\caption{Study workflow. The direct and amortised routes are applied to the same cohorts, scored against converged mechanistic indices at the ranking and absolute tiers, and priced in mechanistic model evaluations.}
\label{fig:study-workflow}
\end{figure}

The comparison has two routes over the same cohorts (\cref{fig:study-workflow}).
The direct route runs each sensitivity design on the mechanistic model for every cohort; the amortised route trains one surrogate on a wide input envelope and runs those cohort designs through it.
Each estimate is judged against its own method's converged mechanistic solution, first on the parameter ranking and then on the absolute index values.
The cost on both routes is the number of mechanistic evaluations needed to reach the same tier, so the one-time training expense can be compared with the repeated cost of a direct cohort analysis.

\subsection{The envelope-and-cohort design}
\label{sec:envelope}

\begin{figure}[htbp]
\centering
\makebox[\textwidth][c]{\includegraphics[width=0.8\textwidth]{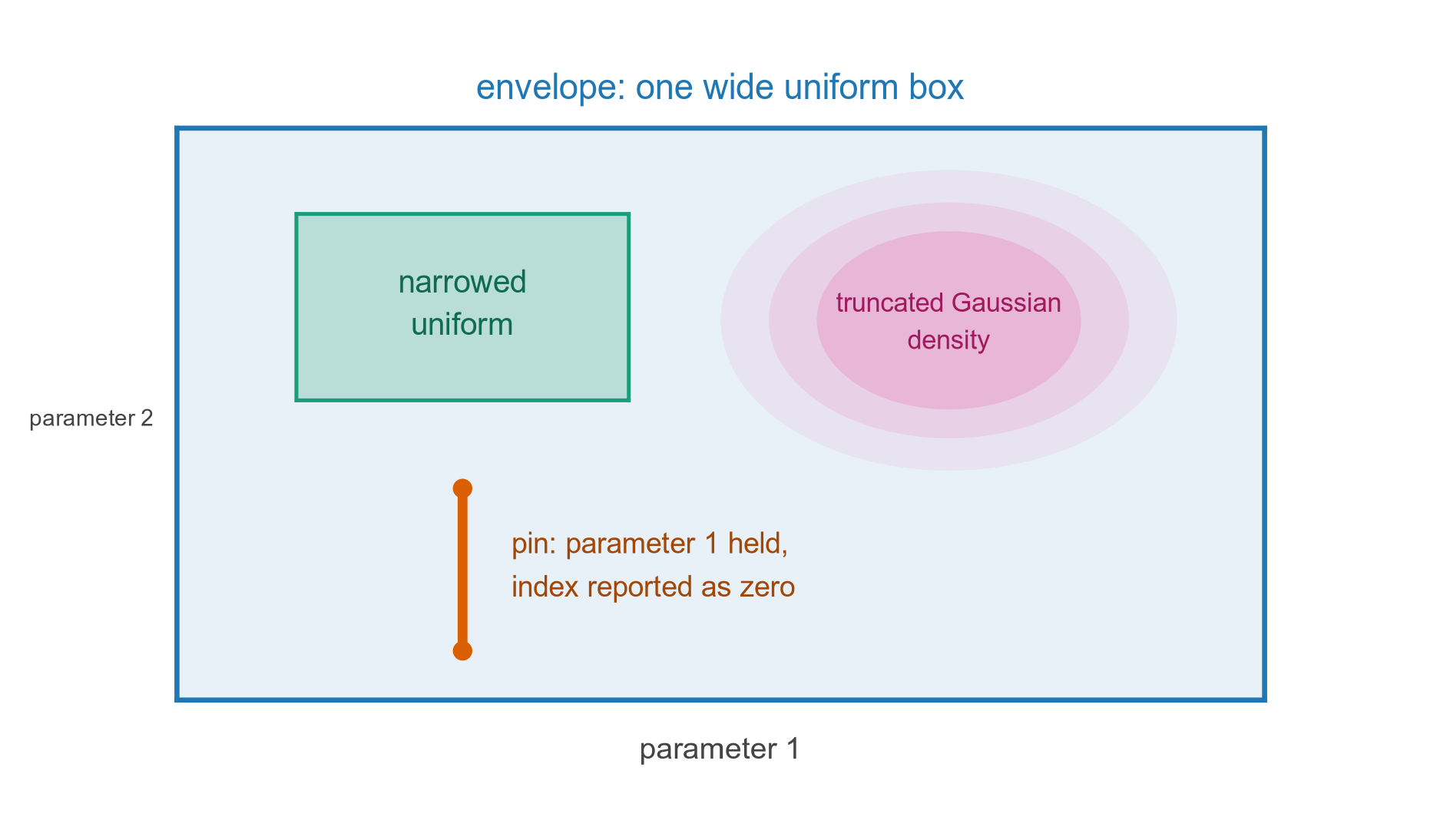}}
\caption{The envelope-and-cohort design. One wide uniform envelope contains every cohort distribution.
A cohort may narrow an input to a uniform sub-band, concentrate it under a truncated Gaussian density, or pin it to one value, removing that axis from the sensitivity design and reporting its index as zero.
The Gaussian is truncated at the envelope bounds although only its density contours are drawn; this containment lets one surrogate trained on the envelope serve every cohort.
}
\label{fig:schematic}
\end{figure}

Each model has kinetic parameters, patient covariates and dose as free inputs that have a range of plausible values.
A sensitivity analysis can be run over the whole range, but a scientist may also want to know how the sensitivity structure changes when a patient group is restricted to a simulated clinical trial, represented by a specific range of parameter bounds.
\cref{fig:schematic} shows the arrangement.
An envelope is a single wide uniform box over the whole range of each axis, while a simulated-trial cohort is a narrower distribution inside that box, representing a patient group of interest.
It might have a truncated Gaussian prior over one axis, a narrowed uniform sub-band over another, and a pinned value on a third.

A surrogate is trained once on a space-filling design over the uniform envelope, and every cohort analysis is done via the surrogate instead of the mechanistic model.
Every override range lies inside the envelope box to avoid surrogate extrapolation.
For a cohort with no pinned axis, the geometric support volume relative to the envelope is the product of its per-axis width ratios; a one-third restriction therefore retains one third of the envelope, while two independent one-third restrictions retain one ninth.
A pin has zero full-dimensional volume, while a Gaussian cohort has full support because it is truncated to the envelope; its concentration is recorded by the effective sample size in Appendix~\ref{app:reweighting}.

\subsection{Sensitivity analysis methods}
\label{sec:methods-gsa}

Three methods are used throughout this work, each giving different information about the same model and each with a different cost.

\subsubsection{Variance-based GSA}
\label{sec:sobol}

Variance-based sensitivity analysis apportions the variance of a square-integrable output $Y = f(X_1,\dots,X_D)$ among $D$ mutually independent inputs \citep{sobol1993,sobol2001}.
\cref{eq:s1} defines the first-order index of input $i$, the share of output variance explained by variation in $X_i$ alone under the specified input distribution.
This is an expected variance contribution, not the effect of fixing $X_i$ at a particular value.
\begin{equation}
\Si = \frac{\Var\!\left(\E\!\left[Y \mid X_i\right]\right)}{\Var(Y)},
\label{eq:s1}
\end{equation}
The total-effect index of \cref{eq:st} is the share that remains when every input except $X_i$ is fixed, and therefore includes every interaction in which $X_i$ participates \citep{homma1996}.
\begin{equation}
\STi = \frac{\E\!\left[\Var\!\left(Y \mid \bm{X}_{\sim i}\right)\right]}{\Var(Y)},
\label{eq:st}
\end{equation}
Their difference $\STi - \Si$ is the part of an input's influence that acts only through interaction, and is zero when the model is additive in that input.
Both are estimated on a hybrid-matrix design in which one column of a re-sample matrix is substituted into the base sample \citep{saltelli2010,owen2013}.
The Saltelli and Jansen estimators \citep{saltelli2010,jansen1999} are used with Owen-scrambled Sobol sequences \citep{owen1997,owen1998,sobol1967}, which Appendix~\ref{app:gsa-methods} constructs.
The combined design scales by the number of base samples $N$ and the number of inputs $D$, following \cref{eq:cost-sobol}.
\begin{equation}
N_{\text{Sobol}} = N\,(D+2)
\label{eq:cost-sobol}
\end{equation}
One property of that sampler is that the first $N' < N$ points of a scrambled Sobol sequence are themselves a valid design, so a single run at high $N$ contains every smaller design as a prefix, and one expensive analysis yields a whole convergence sweep at no additional cost.
A limitation of variance-based indices is that they are blind to distributional change that does not affect variance.
Since the indices are shares of variance, an input whose value changes the shape of the output distribution without changing its variance carries no first-order or total-effect signal at all.

\subsubsection{Morris screening by elementary effects}
\label{sec:morris}

The Morris method treats sensitivity as an average of local slopes taken across the whole input space \citep{morris1991factorial}.
For a point $\bm{x}$ and a step $\Delta_i$ in coordinate $i$, the elementary effect is shown in \cref{eq:ee}, where $\bm{e}_i$ is the $i$th unit vector. The distribution of $\mathrm{EE}_i$ over many starting points is summarised by the mean absolute effect $\mustar_i$, which is the importance measure, the signed mean $\mu_i$, which gives  the direction of the effect, and the standard deviation $\sigma_i$, which flags non-linearity or interaction without distinguishing between them \citep{campolongo2007effective,campolongo2011unified}.
Effects are computed on radial star trajectories following \cref{eq:cost-morris} \citep{ruano2012improved}.
\begin{equation}
\mathrm{EE}_i(\bm{x}) = \frac{f(\bm{x} + \Delta_i \bm{e}_i) - f(\bm{x})}{\Delta_i},
\label{eq:ee}
\end{equation}
\begin{equation}
N_{\text{Morris}} = r\,(D+1)
\label{eq:cost-morris}
\end{equation}

The mean absolute effect $\mustar$ carries the units of the output, so raw values are not comparable between cohorts unless normalised.
Additionally, the interpretation of $\mustar$ as a proxy for the total-effect Sobol index is established for uniform inputs on the unit interval and does not hold in general for other input distributions \citep{feng2019enhanced}.
Any agreement observed here between $\mustar$ and $\ST$ under non-uniform cohort priors is therefore an empirical result that runs ahead of the supporting theory, and is reported as such.

\subsubsection{Optimal transport}
\label{sec:ot}

Optimal transport is a third method that measures the influence of an input on the output distribution rather than on its variance.
Fixing an input can leave the output variance almost unchanged and still move the output distribution, shifting its mean, altering its spread, or making it skewed or bimodal.
Following \citet{borgonovo2025ot}, the influence of an input is measured by how far the output distribution moves when that input is fixed.
For scalar outputs the transport problem reduces to a quantile coupling and has a closed form, so no solver is required \citep{peyre2019computational}.
\cref{eq:w2} gives the squared $2$-Wasserstein distance between two distributions with quantile functions $F^{-1}$ and $G^{-1}$, the mean squared gap between the two sets of quantiles.
\begin{equation}
W_2^{2} = \int_{0}^{1} \left( F^{-1}(u) - G^{-1}(u) \right)^{2} \, \mathrm{d}u,
\label{eq:w2}
\end{equation}
The index of input $i$, \cref{eq:ot}, averages this distance over the value being conditioned on and divides by twice the output variance.
\begin{equation}
\mathrm{OT}_i = \frac{\E_{X_i}\!\left[
W_2^{2}\!\left(\, \mathcal{L}(Y \mid X_i),\; \mathcal{L}(Y) \,\right)\right]}
{2\,\Var(Y)}.
\label{eq:ot}
\end{equation}
The denominator is an upper bound on the numerator, so the index lies in $[0,1]$ and is zero exactly when fixing $X_i$ leaves the output distribution unchanged.
Each trajectory is therefore treated as a set of independent scalar problems, one per timepoint, and the index is reported on the same time grid as the variance-based and Morris screening indices.
The multivariate variant, which transports the output point clouds themselves and requires an entropic solver, is not used here.
The estimator supports any dataset of size $N$ \citep{plischke2013givendata,chiani2025gsaot}.
It is run here on the $A$ matrix of the variance-based design, so all three methods are read from one set of draws.
Cost is given by \cref{eq:cost-ot}, and carries no factor in $D$ at all.
\begin{equation}
N_{\text{OT}} = N
\label{eq:cost-ot}
\end{equation}

The optimal transport index decomposes into an \emph{advective} term measuring displacement of the conditional mean, a \emph{diffusive} term measuring change in spread, and a non-negative remainder carrying the higher moments.
This paper reports the split in two parts, advective against everything else, pooling the diffusive term with the remainder.

\subsection{The four models}
\label{sec:models}

To examine the effect of input dimension on the amortisation, four pharmacokinetic and pharmacokinetic--pharmacodynamic models are considered, with seven to twenty-three dimensional input vectors.
They cover the plasma concentrations of remifentanil, ibuprofen and paracetamol, and the body-weight response to a GLP-1 receptor agonist, so the same cost comparison is made across distinct drugs, outputs and sources of input uncertainty.
Details are collected in \cref{tab:models}, and the derivation of every bound is given in Appendix~\ref{app:provenance}.

\begin{table}[H]
\centering
\caption{The four models, their free input dimension $D$, the primary output analysed, the number of
cohorts, and the origin of the envelope bounds, which governs what a cohort can mean on each.
}
\label{tab:models}
\footnotesize
\setlength{\tabcolsep}{3pt}
\begin{tabular}{@{}p{0.17\textwidth}p{0.16\textwidth}ccp{0.17\textwidth}p{0.30\textwidth}@{}}
\toprule
Model             & Primary output                             & $D$ & Cohorts & Bound source           & Model and bound references \\
\midrule
Remifentanil PBPK & plasma concentration                       & 7   & 5       & original fit           & \citep{abbiati2016definition,minto1997pharmacokinetics} \\
GLP-1RA PK/PD     & body-weight change                         & 9   & 7       & published uncertainty  & \citep{bosch2024quantification} \\
Ibuprofen PBPK    & plasma concentration                       & 15  & 5       & population percentiles & \citep{pearce2017httk,ring2017identifying} \\
Paracetamol PBPK  & parent plasma concentration                & 23  & 4       & sampling convention    & \citep{zurlinden2016physiologically,hsieh2018applying} \\
\bottomrule
\end{tabular}
\end{table}

\subsubsection{Remifentanil PBPK}

Remifentanil is a short-acting opioid, and the model is the physiologically based structure of \citet{abbiati2016definition}.
Its envelope bounds come from a nonlinear mixed-effects fit performed for this work on the 65-patient, 1992-observation dataset distributed with the \texttt{nlme} package \citep{pinheirobates2000}, which records the population of \citet{minto1997influence,minto1997pharmacokinetics}.
The fit carries eight fixed effects on the log kinetic parameters, four of them with diagonal lognormal random effects, and found five of the eight kinetic parameters to be non-identifiable, so those are pinned.

\subsubsection{GLP-1RA PK/PD}

The GLP-1RA model is a semaglutide pharmacokinetic model driving an energy-intake and body-weight turnover response \citep{bosch2024quantification}, whose nine free axes are its uncertain pharmacokinetic and pharmacodynamic parameters together with the patient covariates and dose.
Its bounds are the parameter-uncertainty table of that same fit.

\subsubsection{Ibuprofen PBPK}

The ibuprofen model is the generic high-throughput toxicokinetic structure of the \texttt{httk} package \citep{wambaugh2015toxicokinetic,pearce2017httk} instantiated for that chemical.
Its bounds are not a fit but the 2.5 to 97.5 percentile spread of a 3000-individual virtual population generated with \texttt{httk-pop} \citep{pearce2017httk,ring2017identifying}, sampled independently over those distributions.

\subsubsection{Paracetamol PBPK}

A paracetamol model developed by \citet{zurlinden2016physiologically} is included, imported from the \texttt{pksensi} package \citep{hsieh2020pksensi}.
The sampling is done by sampling from synthetic bounds rather than an uncertainty estimate \citep{hsieh2018applying}.

\subsubsection{Cohort distributions}

\begin{figure}[p]
\centering
\makebox[\textwidth][c]{\includegraphics[width=1.35\textwidth]{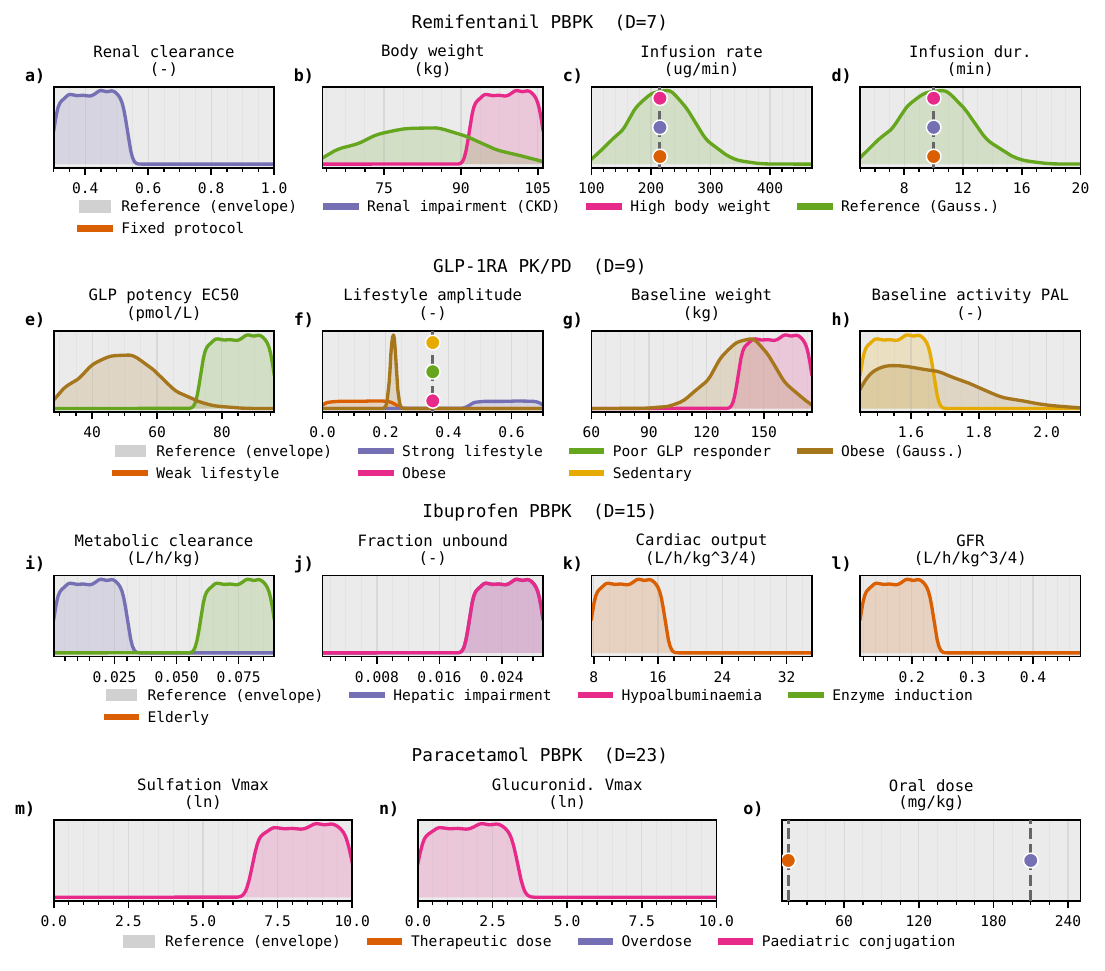}}
\caption{Cohort marginals for the four models, from four thousand draws of each cohort's own marginal.
Panels (a) to (d) are the remifentanil model, (e) to (h) the GLP-1RA model, (i) to (l) the ibuprofen model and (m) to (o) the paracetamol model.
}
\label{fig:marginals}
\end{figure}

\cref{fig:marginals} shows the different distributions of the cohorts on each model, with the envelope in grey and the cohorts in colour.
The grey band on a panel is the envelope and is the training data for the surrogates.
A filled curve is a cohort that narrows the axis to a sub-range, and a dashed line carrying one coloured dot per cohort is an axis pinned to a fixed value.
Every axis of every model is tabulated with its distribution in Appendix~\ref{app:provenance}.

For the two Gaussian cohorts, every axis takes a truncated Gaussian distribution.
The sampling is done by inverse transform sampling, with the cumulative distribution function of the truncated Gaussian used to map a uniform draw.
Numerical clipping of the sampling is avoided, so that the distribution density remains continuous.

\subsection{Sensitivity designs, converged solutions and cost accounting}
\label{sec:designs}

The variance-based GSA, Morris screening and OT are each scored against their own converged solution.
A converged solution is computed by running each method at a large base sample size $N$ and taking the element-wise median over twelve replicate seeds.
Direct convergence sweeps are obtained by prefix slicing, with $N$ doubling from one design to the next.
Each is a prefix of the scrambled sequence the design is drawn from, optimal transport included, since it is read from the $A$ matrix of the variance-based design, so no value of $N$ below the converged solution costs any additional simulation.

\subsection{Reduction over time}
\label{sec:reduction}

We use the generalised sensitivity index of \citet{lamboni2011multivariate} and \citet{gamboa2014multivariate}. Every global sensitivity index in this study is computed at each timepoint, and reduced to one number per parameter by a mean over the horizon weighted by the output variance at that timepoint and excluding the initial condition \cref{eq:reduction}. $\Var(Y(t))$ is taken over the cohort's own input distribution, from a held-out sample of 1024 draws.
The same weights are applied to all three methods, so that the panels stay comparable under one reduction.
\begin{equation}
\bar{s}_i = \frac{\sum_{t>0} \Var\!\left(Y(t)\right)\, s_i(t)}{\sum_{t>0} \Var\!\left(Y(t)\right)},
\label{eq:reduction}
\end{equation}

\subsection{Convergence criterion for GSA estimates}
\label{sec:gate}

An importance threshold defines the set of parameters whose indices are evaluated for ranking and absolute accuracy.
We use five per cent throughout as a user-chosen analysis cutoff and report the sensitivity of important-set membership to a two per cent cutoff in \cref{sec:results-structure}.
For total-effect Sobol' indices, five per cent means an input accounts for at least $0.05$ of the output variance, including its interactions.
For Morris screening, whose elementary effects carry output units, and optimal transport, whose normalised index measures distributional change, we use five per cent of the leading index within each cohort as a relative cutoff.
Neither cutoff is an observational-error or parameter-identifiability criterion.

Let $\hat{\bm{s}}$ be an estimated index vector and $\bm{s}$ the converged solution.
Convergence is assessed against a single criterion with two nested tiers, named the same way throughout.
The \emph{ranking tier} requires that the estimate reproduce the ordering of the converged solution over the \emph{important set} of axes, and that it recalls the unimportant set, so that no inert axis is promoted.
The \emph{absolute tier} requires that the estimate meet the ranking tier and in addition that every important axis lie within a mixed absolute and relative tolerance of the converged solution.
The word \emph{threshold} is reserved throughout for the five per cent importance threshold that defines the important set, and is never used for either tier.

\begin{description}
\item[Parameter ranking reconstruction.] The Spearman rank correlation between $\hat{\bm{s}}$ and
      $\bm{s}$, computed \emph{over the important set only}, is at least $0.90$; and the recall of the
      unimportant set is at least $0.95$, so that no inert axis is promoted.
      Where fewer than three axes are important, the rank correlation is replaced by exact
      agreement on important-set membership and on the leading axis, while the same
      unimportant-set recall requirement is retained.
\item[Absolute index reconstruction.] The ranking tier is met, and in addition every important axis
      lies inside the mixed tolerance of \cref{eq:tolerance}, with $\varepsilon_\text{abs}$ an absolute floor and $\varepsilon_\text{rel}$ a relative term.
      \begin{equation}
      \left|\hat{s}_i - s_i\right| \;\le\;
      \varepsilon_\text{abs} + \varepsilon_\text{rel} \left|s_i\right|,
      \qquad \varepsilon_\text{abs} = 0.01, \quad \varepsilon_\text{rel} = 0.10 ,
      \label{eq:tolerance}
      \end{equation}
      The absolute term governs the axes that are important but small.

\end{description}

\subsection{Surrogate methodology}
\label{sec:surrogates}

A surrogate here is a statistical model of the mechanistic model's output, trained once on a space-filling Latin Hypercube design over the envelope, and then used to evaluate the sensitivity design of any cohort.
The following surrogates are tested:

\begin{description}
\item[Gradient-boosted trees (\texttt{xgboost}, \texttt{catboost}).] An additive ensemble of shallow
      regression trees, each fitted to the residual of the ensemble so far
      \citep{friedman2001,chen2016xgboost,prokhorenkova2018catboost}.
      Trees are piecewise constant, so they represent thresholds and sharp interactions well and smooth responses less well.
      Time is handled as a feature: the trajectory is unpivoted so that normalised time becomes an extra input and one model covers the whole time course.
      Inputs are standardised, with a guard for pinned zero-variance axes.
      Neither carries a training-size limitation.
\item[Gaussian process (\texttt{gp}).] A probabilistic model placing a prior over functions through a
      covariance kernel \citep{rasmussen2006}.
      The kernel is a separable product of an automatic-relevance-determination Mat\'ern-5/2 over the standardised inputs and a Mat\'ern-3/2 over normalised time.
      Automatic relevance determination gives one learned lengthscale per input, so an unimportant input acquires a long lengthscale and is effectively switched off.
      The separable form admits a Kronecker factorisation, so that for $n$ training points over $T$ timepoints the cost is $O(n^3 + T^3)$ rather than $O((nT)^3)$ \citep{gilboa2015scaling,saatci2011structured,rougier2008outerproduct}.
      Hyperparameters are fitted by maximum marginal likelihood.
      Because cost grows with the cube of the training size, the training sweep is capped at 2048 points and at 1024 from fifteen inputs upward.
\item[Polynomial chaos expansion (\texttt{pce}).] A weighted sum of orthogonal polynomials in the
      input variables, with the basis matched to the input distribution
      \citep{xiu2002,sudret2008global}.
      Fitted per timepoint and per output channel.
      The number of basis terms grows combinatorially with dimension and order, so the default order is 2 at every dimension and a further guard lowers the order until the number of basis terms fits the number of training rows.
      That guard is driven by training size, not by a dimension threshold.
\item[RS-HDMR (\texttt{hdmr}).] A direct fit of the analysis-of-variance decomposition, with each
      low-order term expanded in cubic B-splines \citep{li2002practical}.
      The fit needs at least 300 points, so its training sweep starts at 512, and it is capped at 1024 above fourteen inputs.
\end{description}

The five that were kept are all tabular, consuming the sampled inputs and time directly as features, which is also what makes a pinned axis simple to express as it is simply held constant in the feature frame.

\subsubsection{Hyperparameter selection}
\label{sec:hyperparameters}

Hyperparameters were tuned separately for each surrogate on each setup, so a model contributing several outputs is tuned once per output.
The search minimised the envelope-normalised error of \cref{eq:nrmse} under five-fold cross-validation on the envelope corpus, using a single corpus of 1024 draws at one seed; the tuner is Optuna driving a tree-structured Parzen estimator sampler.
The selected set is then applied unchanged across the whole training sweep and all seeds, so no training size and no seed sees a hyperparameter fitted to itself.
The search covers tree depth, learning rate, tree count, subsampling and regularisation for the gradient-boosted ensembles, expansion order, ridge weight and fit fraction for the polynomial chaos expansion, and component order, spline count and penalty for RS-HDMR, at budgets of 30 trials for XGBoost, 25 for CatBoost, and 20 each for the polynomial chaos expansion and RS-HDMR.
The Gaussian process is outside this layer, fitting its own hyperparameters by maximum marginal likelihood.

\subsubsection{Fit assessment}
\label{sec:fit-assessment}

Fit quality is scored on a freshly simulated held-out set of 1024 draws from each cohort's own distribution, as the normalised root-mean-square error of \cref{eq:nrmse}.
The residual and total sums of squares are pooled over the horizon rather than divided at each timepoint, which both keeps the score finite where the output barely varies and applies the same output-variance weight as the index reduction of \cref{eq:reduction}.

\subsection{Assessing the break-even cohort count}
\label{sec:breakeven-method}

Two costs are compared, both in mechanistic model evaluations.
$N_\text{direct}$ is the number required by one direct analysis of one cohort to reach the \textit{ranking} or \textit{absolute} tiers following \cref{eq:cost-sobol,eq:cost-morris,eq:cost-ot}.
$N_\text{env}$ is the number required to train a surrogate that reaches the same tier, incurred once on the envelope and shared by every cohort.

An analysis of one cohort is called \emph{direct} when it is run on the mechanistic model and \emph{amortised} when it is read from a surrogate, since the surrogate's construction cost is shared across every cohort it serves.

Over $K$ cohorts the direct route needs $K N_\text{direct}$ mechanistic model simulations while the surrogate route requires $N_\text{env}$ once for training.
The break-even cohort count $K^{*}$ of \cref{eq:kstar} expresses that one-time training budget in direct-cohort equivalents, so it can take a fractional value.
\begin{equation}
K^{*} = \frac{N_\text{env}}{N_\text{direct}}
\label{eq:kstar}
\end{equation}
A value below one means that training already costs fewer evaluations than one direct cohort analysis; for an integer number of cohorts, the surrogate route is cheaper whenever $K > K^{*}$.

Training corpora exist only at the six sizes of the sweep, $512$ to $16{,}384$ in powers of two, so $N_\text{env}$ is always one of those six and any intermediate figure is rounded up to the smallest that exists.
Each cell is run at twelve seeds and reported at their median: $N_\text{env}$ is the smallest corpus at which the median estimate clears the tier and stays clear at every larger corpus.
A cell is summarised by the median over its cohorts, alongside the number of cohorts that never reach the tier at any training size.

\subsection{Implementation and reproducibility}
\label{sec:implementation}

Models are solved in JAX with Diffrax for the differentiable engine \citep{bradbury2018jax,kidger2022}, and in R with \texttt{deSolve} and compiled C for the remainder.
The variance-based, Morris screening and optimal transport estimators come from \texttt{jaxgsa} \citep{pessina2026jaxgsa}.
Twelve replicate seeds are used throughout, each a distinct envelope scramble driving both the training corpus and every cohort design draw.
The variance-based and optimal transport estimators are asserted against problems with known analytical indices and against an independent implementation; the Morris screening estimator is used as supplied and was not independently validated here, so a systematic error in it would appear in \cref{sec:results-agreement} as a disagreement between methods.

\section{Results and discussion}
\label{sec:results}

\subsection{The sensitivity structure of the four models}
\label{sec:results-structure}

\begin{figure}[p]
\centering
\makebox[\textwidth][c]{\includegraphics[width=1.35\textwidth]{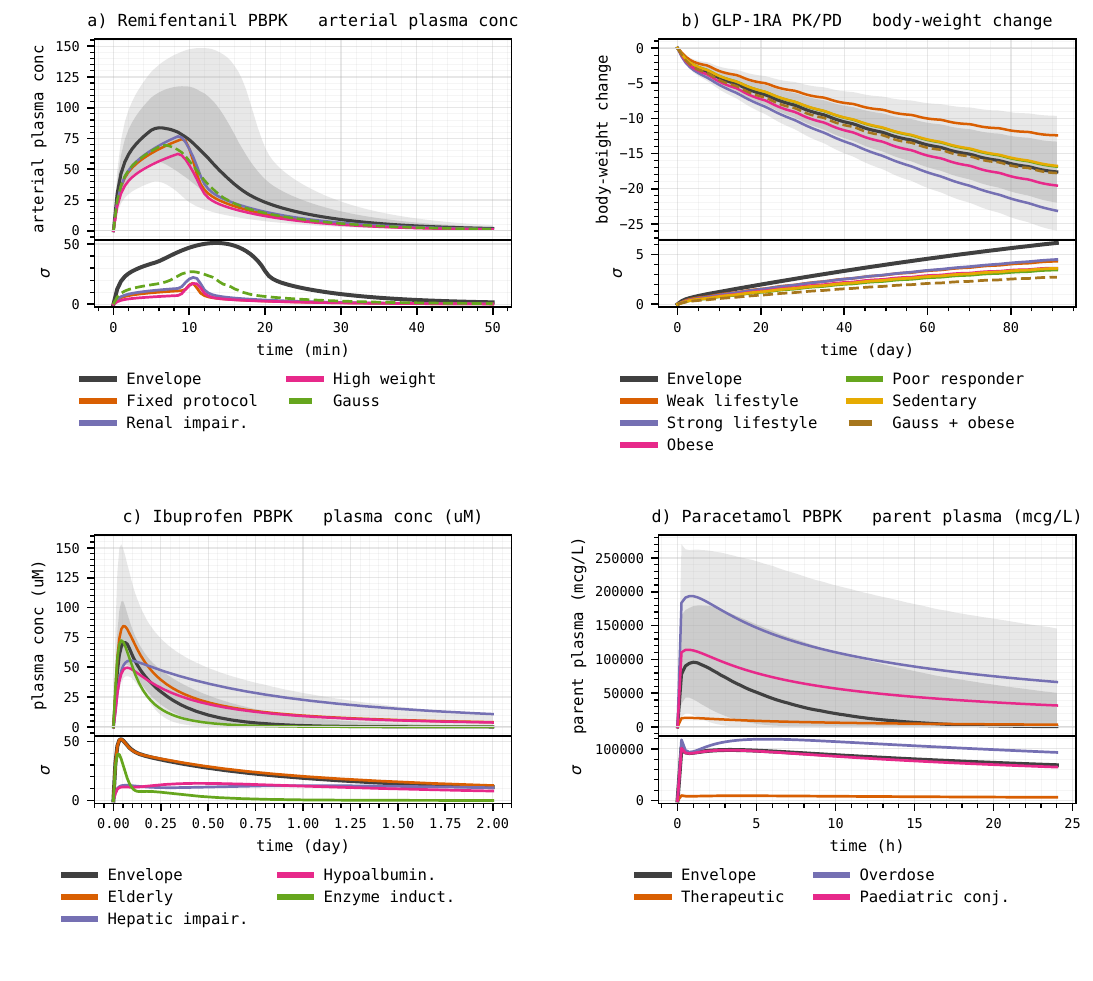}}
\caption{Output trajectory spread over the envelope for the primary output of each model, in dimension order: the envelope median with its 25--75 and 10--90 bands, each cohort's mean overlaid, and the output standard deviation $\sigma$ against time in the narrow panel beneath.
}
\label{fig:outputs}
\end{figure}

\cref{fig:outputs} shows the primary output of each model as a trajectory over the envelope, with each cohort's mean overlaid and the output standard deviation as a shaded region.
Each model has distinct characteristics: the predicted remifentanil concentration spreads in the infusion window and decays to near zero well before the end of the horizon, the paracetamol plasma concentration carries a standard deviation of the order of its own median across the whole twenty-four hours, the ibuprofen plasma concentration is a narrow peak that spreads and decays, and the GLP-1RA body-weight change starts at zero by construction and widens monotonically to the end of the three months.

A user-defined importance threshold divides the parameters into an \emph{important} set and an \emph{unimportant} set, as specified in \cref{sec:gate}.

For variance-based GSA the cutoff is an absolute share of output variance; for Morris screening and optimal transport it is relative to the cohort's leading index.
Changing that choice can change which parameters a modeller carries forward.

Applied to the whole-envelope cohort, lowering the threshold from five per cent to two adds between zero and four parameters to the variance-based important set, and how much it adds is unrelated to how many inputs the model has: the twenty-three-input paracetamol model gains one, and the nine-input GLP-1RA model gains four (\cref{tab:structure}).
The other two GSA methods respond to the same movement very differently.
The Morris set grows on all four models, and unlike the variance-based set its growth does track dimension, adding one axis on the two smallest models and five on the two largest, whereas the optimal-transport set is unchanged on two of the four and never grows by more than three axes.

Morris screening is the method whose set is least stable to a small movement of the threshold.
On the paracetamol model its important set is unchanged only between $0.0474$ and $0.0510$, so a threshold a little either side of five per cent selects different parameters, and of the twelve whole-envelope combinations of model and method the two narrowest such intervals are both Morris ones, at $0.0036$ on the paracetamol model and $0.0097$ on the ibuprofen model.
Fragility has a second sense, and over the wider move from five per cent to two it reverses the ranking: the identity of the important set changes in 45 of those 63 combinations, and it is the variance-based analysis that changes most often, in 17 of its 21 cohorts, against 13 for Morris screening.

\begin{figure}[htbp]
\centering
\makebox[\textwidth][c]{\includegraphics[width=1.35\textwidth]{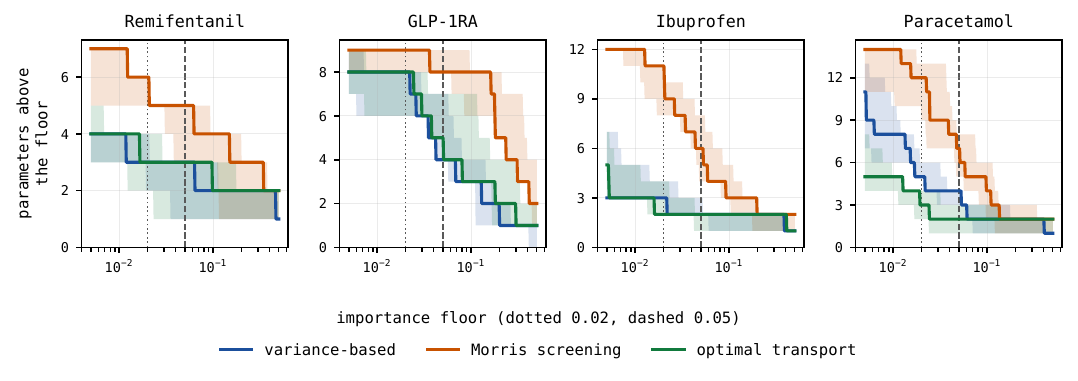}}
\caption{The size of the important set against the threshold for each GSA method.
}
\label{fig:floor}
\end{figure}

\subsection{Variance-based GSA}
\label{sec:results-sobol}

\begin{figure}[p]
\centering
\makebox[\textwidth][c]{\includegraphics[width=1.35\textwidth]{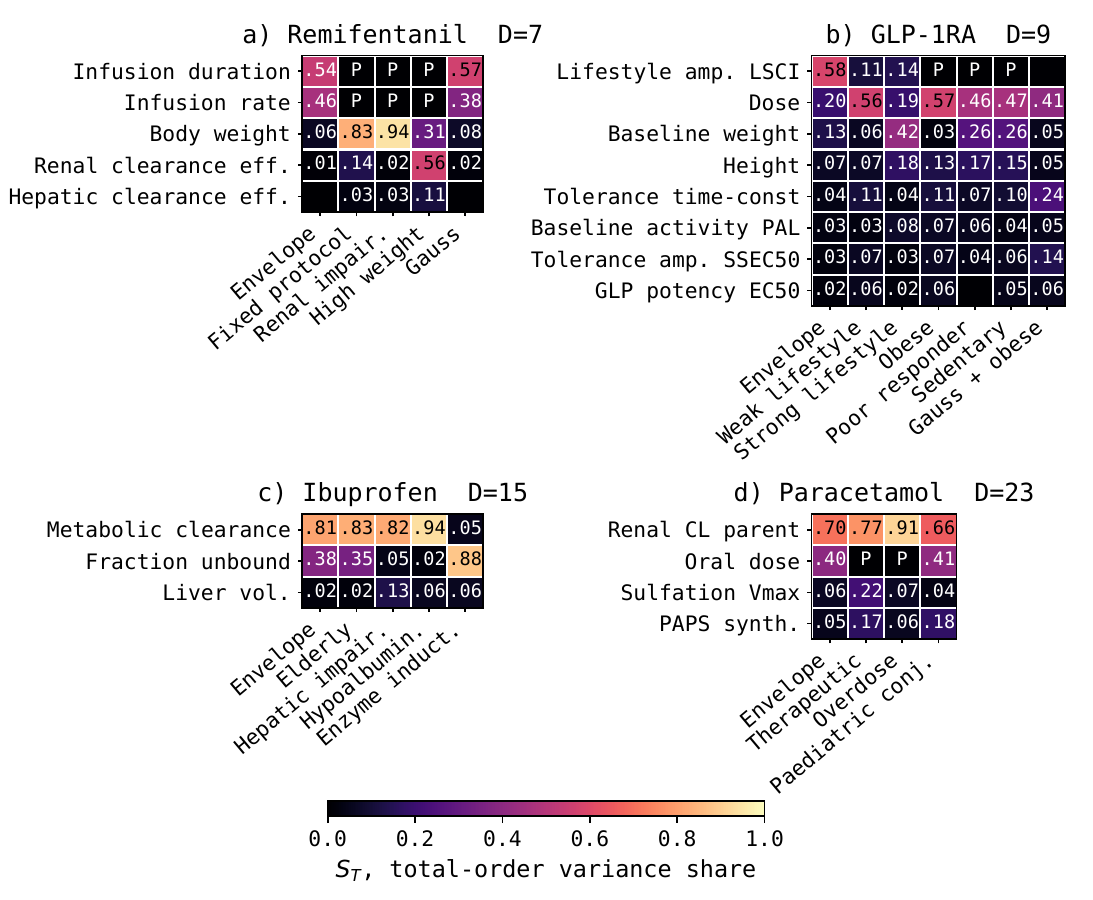}}
\caption{Converged variance-weighted total-effect Sobol index $\ST$ by parameter and cohort.
A cell marked \emph{P} is an axis the cohort pins, which is therefore held rather than estimated; a cell with no number is below $0.01$.
}
\label{fig:heatmaps}
\end{figure}

\cref{fig:heatmaps} gives the converged variance-based solution of the total-effect Sobol' index for every parameter and cohort of the four models.

\cref{tab:structure} reports the size of the important parameter set for each of the studied models and cohorts. For most models, the ratio of important axes over model dimensions does not fall smoothly with $D$. The absolute size of the important set does not fall with dimension either.
The GLP-1RA model carries more important axes than remifentanil at $D=7$ in every cohort and about twice as many as the paracetamol model at $D=23$.

\begin{table}[htbp]
\centering
\caption{The sensitivity structure of the four models, one column per model in dimension order.
Important-set sizes are counted at the five per cent importance threshold and given as the range over that model's cohorts.
The threshold rows give the size of the whole-envelope important set at the five per cent threshold and at two per cent, in that order, for each of the three methods.
The Morris non-linearity row counts the important axis-cohort pairs at which $\sigma$ exceeds $\mustar$.
The diffusive share is the ratio of summed diffusive to summed total transport over a cohort's important set, quoted for the whole-envelope cohort with the range over cohorts beneath it.
}
\label{tab:structure}
\footnotesize
\setlength{\tabcolsep}{5pt}
\begin{tabular}{@{}lrrrr@{}}
\toprule
                                     & Remifentanil & GLP-1RA      & Ibuprofen    & Paracetamol  \\
                                     & $D=7$        & $D=9$        & $D=15$       & $D=23$       \\
\midrule
Cohorts                              & 5            & 7            & 5            & 4            \\
Important axes, range over cohorts   & 1--3         & 4--7         & 2--3         & 3--4         \\
\addlinespace
\multicolumn{5}{@{}l}{\emph{Whole-envelope important set at the 5\% and 2\% thresholds}}         \\
\quad Variance-based                 & 3, 3         & 4, 8         & 2, 3         & 4, 5         \\
\quad Morris screening               & 5, 6         & 8, 9         & 6, 11        & 7, 12        \\
\quad Optimal transport              & 3, 3         & 5, 8         & 2, 2         & 2, 3         \\
\addlinespace
Morris pairs with $\sigma > \mustar$ & 2/21         & 33/57        & 12/31        & 21/26        \\
\addlinespace
Diffusive share, whole envelope      & 0.367        & 0.149        & 0.585        & 0.547        \\
\quad range over cohorts             & 0.167--0.414 & 0.092--0.153 & 0.342--0.585 & 0.480--0.581 \\
\bottomrule
\end{tabular}
\end{table}

The variance-based analysis also reports the first-order Sobol' index $\Sone$, which is the share of the output variance that a parameter explains on its own, and the gap $\ST - \Sone$, the cross-parameter contributions.

Changes in the cohort distribution alter which parameters cross the chosen importance threshold.
For example, renal clearance efficiency in the remifentanil model is the clearest case: over the whole envelope it carries $\ST = 0.012$, a fortieth of the importance threshold, so a study designed on the envelope alone would discard it.
The protocol cohort fixes the dose, and there it reaches $0.140$.
The heavy cohort fixes the dose and also restricts body weight to a high band.
That one further restriction raises renal clearance efficiency to $0.563$ and makes it the leading parameter, ahead of body weight at $0.311$.
This is a case where the variance-based analysis places a parameter below the chosen threshold over the envelope but above it in a patient group, a difference that could inform clinical trial design.

\subsection{Morris screening}
\label{sec:results-morris}
\begin{figure}[p]
\centering
\makebox[\textwidth][c]{\includegraphics[width=1.35\textwidth]{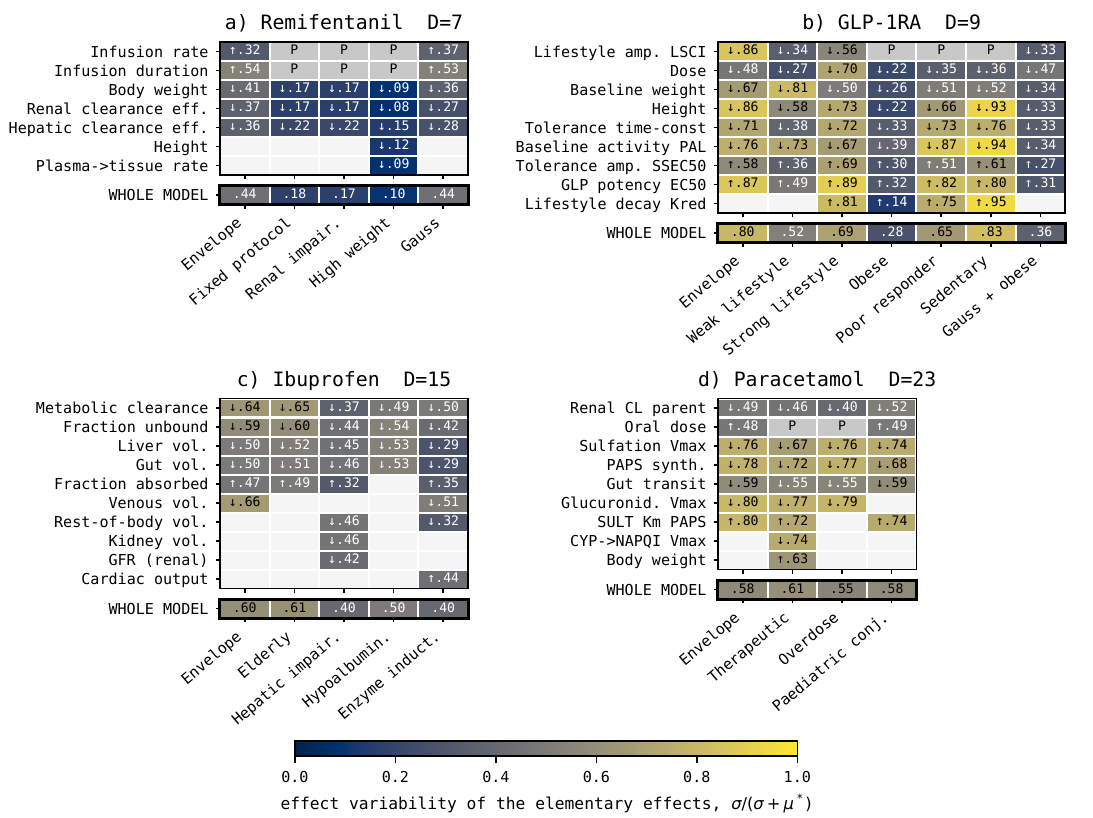}}
\caption{Morris screening by parameter and cohort, in dimension order, where the non-linearity flag is $\sigma/(\sigma+\mustar)$ and the arrow is the sign of the mean elementary effect $\mu$, pointing up where raising the input raises the output.
A cell marked \emph{P} is an axis the cohort pins, and an empty cell is one falling below the five per cent threshold.
The separated bottom row is the same ratio of the sums over that cohort's important set.
}
\label{fig:morris}
\end{figure}

Morris screening measures direction, returning the sign of an effect and a non-linearity flag.
\cref{fig:morris} reports both as $\sigma/(\sigma+\mustar)$ with the sign of $\mu$ as the arrow.
Within the 21 cohorts studied, none of the 35 sampled parameters reverses the sign of its mean elementary effect in a cohort where it meets the importance threshold.

Counted by axis, the fraction of important axis-cohort pairs at which $\sigma$ exceeds $\mustar$ runs from a tenth on remifentanil to four fifths on the paracetamol model, with no ordering by dimension (\cref{tab:structure}).
Only on remifentanil is a mean elementary effect a fair summary of the parameters; on the other three it hides slopes that change substantially across the input space.

Across the 21 cohorts Morris screening places 59 axis-cohort pairs in the important set that the variance-based GSA does not.
None have a total-effect index at or above the five per cent threshold, the largest reaching $0.0475$, with 28 of the 59 sitting between $0.02$ and the threshold.

\subsection{Optimal transport}
\label{sec:results-ot}

Optimal transport measures distributional shape.
The index splits into an advective component, the squared displacement of the conditional mean, and a diffusive remainder carrying the change in spread and shape.

Restricted to the important set, the diffusive share splits the four models into two groups: about half of the transport is a change in spread and shape on the ibuprofen and paracetamol models, and a seventh of it on the GLP-1RA model, with remifentanil between them (\cref{tab:structure}).

\begin{figure}[p]
\centering
\makebox[\textwidth][c]{\includegraphics[width=1.35\textwidth]{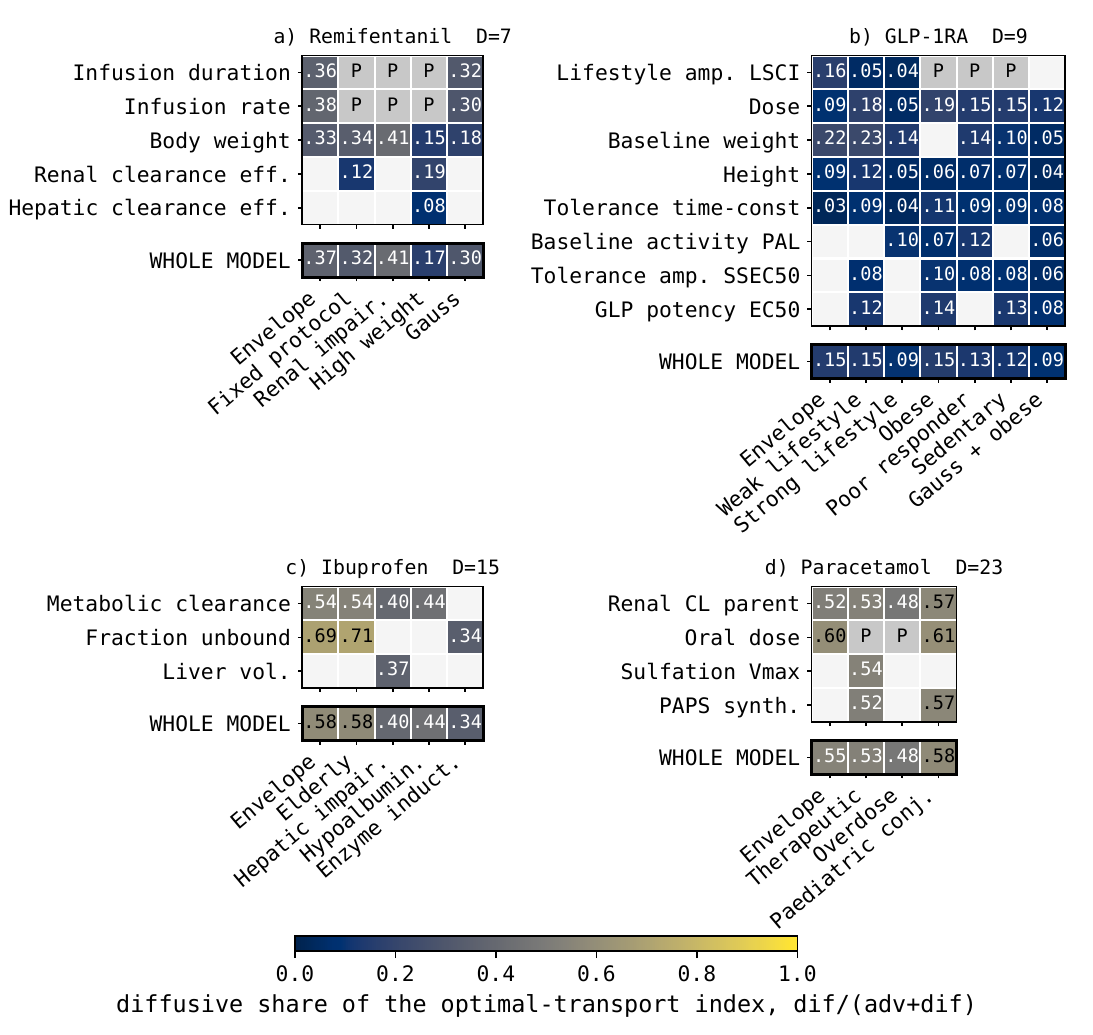}}
\caption{The diffusive share of the optimal-transport index by parameter and cohort, with the two components as defined in \cref{sec:results-ot}.
A cell marked \emph{P} is an axis the cohort pins, and an empty cell is one falling below the five per cent threshold.
}
\label{fig:ot}
\end{figure}

\cref{fig:ot} shows the diffusive share for every parameter and cohort of the four models. The share is a ratio of summed diffusive transport to summed total transport over the axes clearing the threshold, so an axis contributes in proportion to the effect.

The distributional lens adds substantial shape information on two of the four models and almost none on the GLP-1RA model.
On the ibuprofen and paracetamol models about half of what conditioning does to the output is to change its spread and shape, so a variance-based analysis, which registers only the displacement of the centre, describes half of what those inputs do.
On the GLP-1RA model nine tenths of the transport is displacement, and the distributional lens returns the same information as the first-order variance index.

An optimal-transport index is a squared Wasserstein distance and therefore cannot be negative regardless of sample size, whereas a variance-based estimator is a difference of sampled means and can return a negative index, which is a recognised and visible symptom of an unconverged analysis.
Optimal transport has no such warning: its finite-sample error can only appear as an apparent optimal-transport index on a parameter that has none, so an inert axis has indices which tend towards zero as rows are added.

\subsection{Comparing GSA methods}
\label{sec:results-lenses}
\label{sec:results-agreement}
\label{sec:results-cost}
\label{sec:results-convergence}

The three methods order the important parameters almost identically, though they disagree on which axes are important.
Restricted to the variance-based important set, which is the only scope carrying an absolute threshold, optimal transport reproduces the ordering exactly in 18 of the 21 cohorts and Morris screening in 16. All three methods agree on the leading parameter for 20 of the 21 cohorts.

\begin{figure}[htbp]
\centering
\makebox[\textwidth][c]{\includegraphics[width=1.35\textwidth]{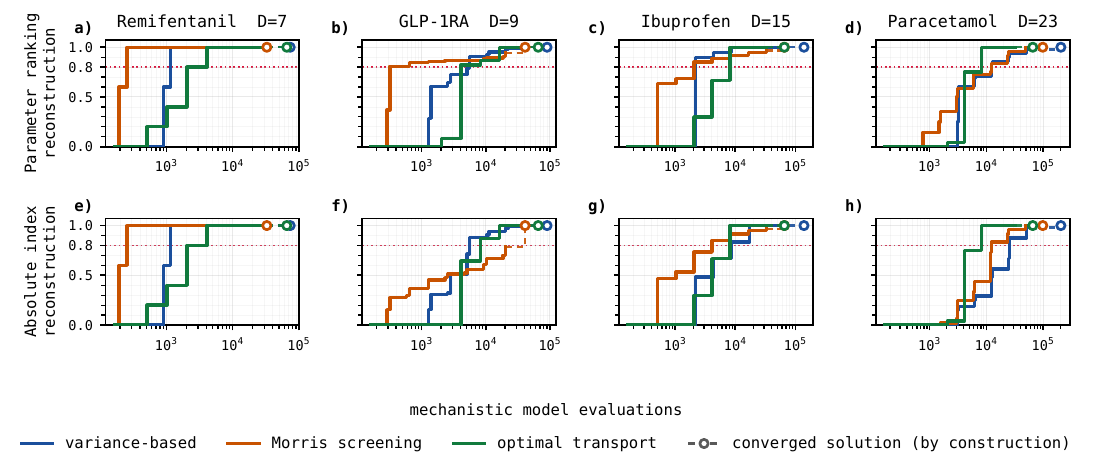}}
\caption{The fraction of a model's cohort and replicate runs whose direct analysis has reconstructed at or below a given budget.
Each method is scored against its own converged solution.}
\label{fig:direct-cdf}
\end{figure}

\cref{fig:direct-cdf} gives the cost of a converged direct analysis for each method on each model, at the ranking tier and at the absolute tier of \cref{sec:gate}.
Two of the three methods scale with the number of model dimensions, so every entry is quoted in mechanistic evaluations, with that factor already applied.
Cohorts with pinned axes cost less, since $D$ is the free dimension, so each curve pools every cohort and every replicate seed of a model: 12 seeds on each of its cohorts, corresponding to 48 to 84 runs per model.

Morris screening is the cheapest route to its own ranking on remifentanil and the GLP-1RA model when 80\% of runs must succeed, 4x and 18x cheaper than the variance-based GSA, though its lead has almost gone by fifteen inputs, where it needs 6\% fewer evaluations, and at twenty-three inputs the two cost the same and both are beaten by optimal transport.

Increasing the passing rate from 80\% to 90\% leaves the variance-based budgets untouched on the three lower-dimensional models and multiplies the Morris budget by up to fifty-seven.
Each method therefore has different robustness to the same convergence criterion across different models and cohorts, with Morris screening the most sensitive of the three.

At the 80\% level the optimal-transport budget spans a factor of four across the four models, from 2048 rows on remifentanil to 8192 on the ibuprofen and paracetamol models; the variance-based budget spans a factor of eleven and Morris screening a factor of forty-eight.
Against the variance-based analysis it runs from a quarter fewer evaluations on the GLP-1RA model to nearly four times as many on the ibuprofen model, and at twenty-three inputs it is the cheapest of the three, by a third at the 80\% level and by a factor of three at 90\%.

Reconstructing the absolute values on top of the ranking requires no additional evaluations at seven inputs for any method, and above that the additional cost is at most fourfold.
Morris screening on the GLP-1RA model is the exception: it is the only one of the twenty-four cells that reaches neither share within the measured sweep, so screening there recovers the ordering but never the magnitudes until the final sample size.

\subsection{Surrogate fit quality}
\label{sec:results-fit}
\begin{figure}[htbp]
\centering
\makebox[\textwidth][c]{\includegraphics[width=1.30\textwidth]{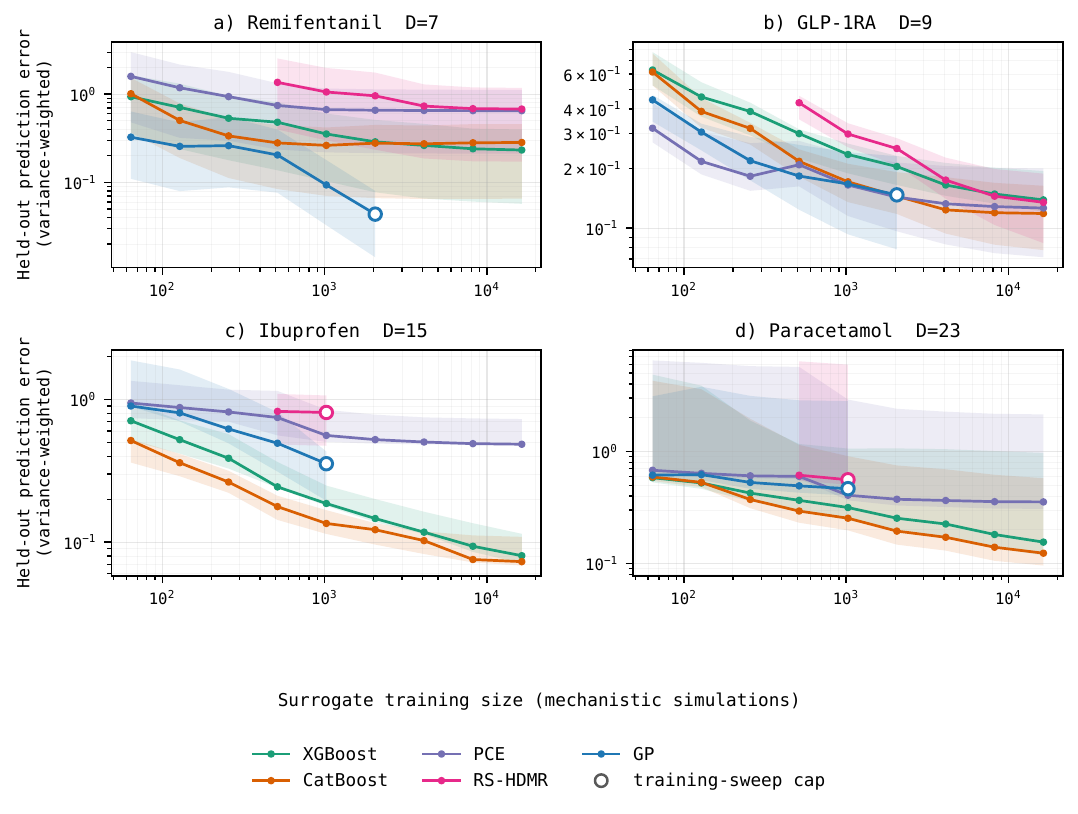}}
\caption{Surrogate fit quality against training size, for the five surrogates on the four models, on held-out sets drawn from each cohort's own distribution.
The score is the normalised root-mean-square error of \cref{eq:nrmse}, on a log axis, with the line the median across each model's cohorts and the band their 10th to 90th percentile.
}
\label{fig:fit}
\end{figure}

\cref{fig:fit} shows held-out fit against training size.
For one cohort, one replicate seed and one surrogate trained at a given size, the fit is scored by the normalised root-mean-square error of \cref{eq:nrmse}, in which $y_k(t)$ is the mechanistic output of the $k$th of the 1024 held-out draws of that cohort at timepoint $t$, $\hat{y}_k(t)$ is the surrogate's prediction for that same draw, and $\bar{y}(t)$ is the mean of the held-out draws at that timepoint.
\begin{equation}
\mathrm{NRMSE} = \sqrt{\frac{\sum_t \sum_k \left(\hat{y}_k(t) - y_k(t)\right)^{2}}
{\sum_t \sum_k \left(y_k(t) - \bar{y}(t)\right)^{2}}},
\label{eq:nrmse}
\end{equation}
Both sums run over every timepoint of the horizon.
The denominator is the cohort's own output spread about its mean trajectory, so the score is one minus a coefficient of determination, $R^{2} = 1 - \mathrm{NRMSE}^{2}$, and $\mathrm{NRMSE} = 1$ is a surrogate no better than predicting that cohort's mean trajectory.

Gaussian process and RS-HDMR training sweeps are limited: the Gaussian process is too slow to train on more than 2048 rows, and RS-HDMR exceeds available memory above 1024 training rows from fifteen inputs upwards, so those entries are not comparable with the tree entries at equal cost.

\begin{table}[htbp]
\centering
\caption{Held-out fit and ranking-tier training budget for the five surrogates, against the variance-based GSA.
The fit is the normalised root-mean-square error of \cref{eq:nrmse}, median over that model's cohorts, read at each surrogate's largest available training budget: 16{,}384 evaluations, except 2048 for the Gaussian process on the two lower-dimensional models and 1024 for the Gaussian process and RS-HDMR at fifteen inputs and above.}

\label{tab:surrogate-fit}
\scriptsize
\setlength{\tabcolsep}{2.5pt}
\begin{tabular}{@{}l rrc c rrc c rrc c rrc@{}}
\toprule
                     & \multicolumn{3}{c}{Remifentanil $D=7$} &                & \multicolumn{3}{c}{GLP-1RA $D=9$} &  & \multicolumn{3}{c}{Ibuprofen $D=15$} &                & \multicolumn{3}{c}{Paracetamol $D=23$}                                                                     \\
\cmidrule(lr){2-4}\cmidrule(lr){6-8}\cmidrule(lr){10-12}\cmidrule(l){14-16}
                     & NRMSE                                  & $N_\text{env}$ & $k$                               &  & NRMSE                                & $N_\text{env}$ & $k$                                    &  & NRMSE & $N_\text{env}$ & $k$ &  & NRMSE & $N_\text{env}$ & $k$ \\
\midrule
Gaussian process     & 0.04                                   & 64             & 5/5                               &  & 0.15                                 & 64             & 6/7                                    &  & 0.35  & 1024           & 4/5 &  & 0.47  & ---            & 0/4 \\
CatBoost             & 0.28                                   & 128            & 5/5                               &  & 0.12                                 & 128            & 7/7                                    &  & 0.07  & 64             & 5/5 &  & 0.12  & 64             & 3/4 \\
XGBoost              & 0.23                                   & 128            & 5/5                               &  & 0.14                                 & 128            & 7/7                                    &  & 0.08  & 64             & 5/5 &  & 0.16  & 128            & 3/4 \\
Polynomial chaos     & 0.65                                   & 1024           & 5/5                               &  & 0.13                                 & 128            & 7/7                                    &  & 0.49  & 512            & 3/5 &  & 0.35  & 64             & 3/4 \\
RS-HDMR              & 0.68                                   & 8192           & 4/5                               &  & 0.14                                 & 1024           & 6/7                                    &  & 0.81  & 512            & 3/5 &  & 0.56  & 1024           & 4/4 \\
\midrule
Mechanistic (direct) & \multicolumn{3}{c}{896}                &                & \multicolumn{3}{c}{1408}          &  & \multicolumn{3}{c}{2176}             &                & \multicolumn{3}{c}{3200}                                                                                   \\
\bottomrule
\end{tabular}
\end{table}

The five surrogates fit the four models to different accuracy (\cref{tab:surrogate-fit}).
On remifentanil the fit ordering and the budget ordering match exactly, the Gaussian process fitting best and reaching the ranking tier from 64 evaluations, RS-HDMR fitting worst and needing 8192.
On the ibuprofen model the two tree surrogates fit best and reach the tier from 64 evaluations, polynomial chaos and RS-HDMR fit worst and reach only three of that model's five cohorts, while the Gaussian process has the third best fit and needs the largest budget of the five.
All five fit the GLP-1RA model to between 0.12 and 0.15, so on that model there is no fit ordering for the cost to follow.

Fit and ranking recovery are only loosely coupled, as shown by the paracetamol model which exhibits a failure in each direction.
RS-HDMR reconstructs the ranking of the therapeutic cohort in all twelve seeds from 1024 evaluations, against a direct cost of 3072.

The Gaussian process fits three of the four paracetamol cohorts to between $0.40$ and $0.52$ and reconstructs the ranking on none of them.

The two directions fail for different reasons.
A sensitivity index is a share of variance, so an error that is systematic across the input space raises the fit score but largely cancels in the ratio of partial to total variance, which is how RS-HDMR ranks correctly while fitting worst in the study.
In the other direction a surrogate can carry a low error overall while flattening the response along an axis that is important but contributes little output variance.
The residual then sits exactly where the index is read, which is what happens to the Gaussian process on the paracetamol model.
Prediction accuracy is therefore neither necessary nor sufficient for index recovery, so a held-out fit score cannot serve as the acceptance test for a GSA surrogate.
That test has to be run on the indices themselves, against the ranking and absolute tiers of \cref{sec:gate}, at a cost of one converged mechanistic solution per model.

\subsection{Cost and reliability of surrogate GSA}
\label{sec:results-breakeven}

\begin{figure}[p]
\centering
\makebox[\textwidth][c]{\includegraphics[width=1.28\textwidth]{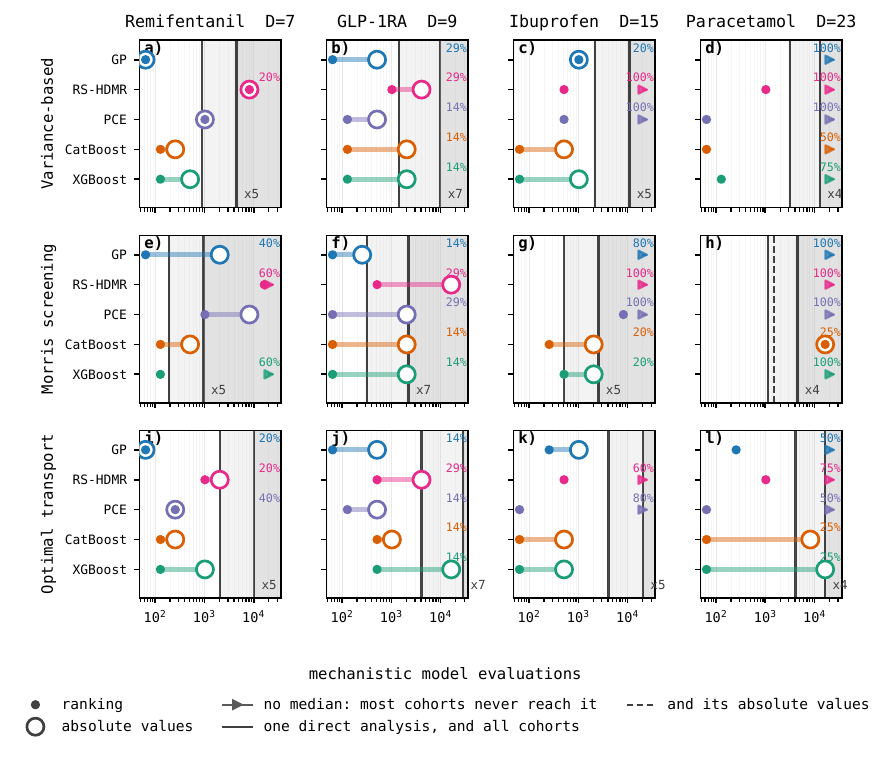}}
\caption{Ranking and absolute-tier budgets in mechanistic model evaluations, for three methods on four models.
Filled and hollow marks give surrogate budgets. Solid vertical lines mark the direct ranking cost for one cohort and for all cohorts; the two grey regions begin at those costs.
A dashed line marks the direct absolute-tier cost where it differs. Arrows indicate no median absolute-tier budget, and percentages give the share of cohorts that fail that tier within the measured sweep.
}
\label{fig:tier-gap}
\end{figure}

Two quantities are compared in this subsection.
The \emph{training budget} $N_\text{env}$ is the smallest training size at which a surrogate first clears a tier and stays clear at every larger size, incurred once for a model.
The \emph{break-even cohort count} $K^{*}$ expresses that budget in direct-cohort equivalents, $N_\text{env} / N_\text{direct}$ in \cref{eq:kstar}; it is a fractional threshold rather than necessarily an integer cohort count.
A $K^{*}$ below one means the surrogate costs fewer evaluations to train than one direct cohort analysis.
\cref{fig:tier-gap} gives $N_\text{env}$ and $N_\text{direct}$ for every surrogate, model and method at both tiers; \cref{fig:convergence} in the supplementary material shows the underlying curves from which each $N_\text{env}$ is assessed.

At the ranking tier a gradient-boosted tree costs fewer evaluations to train than one direct cohort analysis almost everywhere.
Against the variance-based GSA at seven inputs, CatBoost trains to the ranking on 128 evaluations against a direct cost of 896, so it breaks even at $0.14$ cohorts; at twenty-three inputs it needs 64 against 3200, so $0.02$.
Against optimal transport the same two training budgets stand against direct costs of 2048 and 4096, giving $0.06$ and $0.02$.
The one exception is Morris screening at twenty-three inputs, where CatBoost needs 16{,}384 training evaluations against a direct cost of 1120, so it breaks even at $14.63$ cohorts and the paracetamol model has four.

Which method is being reconstructed matters more than which surrogate is used.
Morris screening needs the fewest evaluations to run directly, 192 at seven inputs rising to 1120 at twenty-three, and optimal transport the most, 2048 rising to 4096, with the variance-based GSA between them at 896 and 3200.
The reduction in evaluations available to a surrogate is therefore smallest against Morris screening and largest against optimal transport.
On the ibuprofen model CatBoost trains to the ranking on 64 evaluations for the variance-based GSA and for optimal transport alike, against direct costs of 2176 and 4096, so $0.03$ and $0.02$ cohorts.
Morris screening costs it 256 evaluations against a direct cost of 512, so $0.50$ cohorts.
That is twenty-five times the ratio against optimal transport, on the same model with the same surrogate.

Within the variance-based GSA the two costs move in opposite directions as dimension grows.
The $(D+2)$ factor of \cref{eq:cost-sobol} raises the cost of the direct analysis.
The two gradient-boosted trees need 128 evaluations at seven and nine inputs and 64 at fifteen; at twenty-three CatBoost holds at 64 while XGBoost needs 96.
The surrogate therefore breaks even sooner the larger the model, contrary to the expectation that surrogates are least effective at high dimension.

The five surrogates differ more in reliability than in training cost.
CatBoost and XGBoost reach the ranking on 20 of the 21 cohorts.
The Gaussian process needs the fewest training evaluations of the five where it works, at 64 evaluations on the two smaller models, but reconstructs no paracetamol cohort at all.
RS-HDMR is the only surrogate that ever costs more to train than a direct variance-based GSA at this tier, needing 8192 evaluations on remifentanil against a direct cost of 896, so it breaks even at $9.14$ cohorts and that model has five.
Morris screening is also where surrogates fail most often: 6 of its 20 cells reach the ranking on no majority of cohorts, against 1 for the variance-based GSA and none at all for optimal transport.

Only the surrogate side costs more at the absolute tier.
The direct budget is unchanged on all four models, so a practitioner who has already run a mechanistic ranking obtains calibrated shares with no further evaluations, while the training budget rises by a factor of two to sixteen.
On the ibuprofen model the tier is carried by the two gradient-boosted trees at 512 and 1024 evaluations, with polynomial chaos and RS-HDMR reaching no cohort at all.
On the paracetamol model no surrogate reconstructs a majority of the four cohorts, CatBoost reaching two and XGBoost one, so that model allows for surrogate-derived index rankings, but not absolute index reconstruction.

On fit and cost taken together, CatBoost is the default surrogate.
It is best or joint-best on fit on three of the four models, reaches the ranking on 20 of the 21 cohorts, and holds the absolute tier furthest up the dimension range; XGBoost matches it on cohorts reconstructed but reaches the absolute tier at twice the training cost on remifentanil.
Below roughly ten inputs on a smooth response the Gaussian process is the most sample-efficient of the five and needs no tuning budget, though it reconstructs no paracetamol cohort, so that strength should not be generalised beyond the two lowest-dimensional models.
Polynomial chaos is the choice when the analytical decomposition itself is wanted, since it alone returns indices and higher-order terms directly from its coefficients, and RS-HDMR should be used only at main-effects order.
These rules may transfer beyond these four models because the smooth-basis methods incur a combinatorial cost in basis terms as dimension grows and the Gaussian process is capped by its cubic training cost, whereas tree-based surrogates have neither penalty.

\begin{figure}[p]
\centering
\makebox[\textwidth][c]{\includegraphics[width=1.35\textwidth]{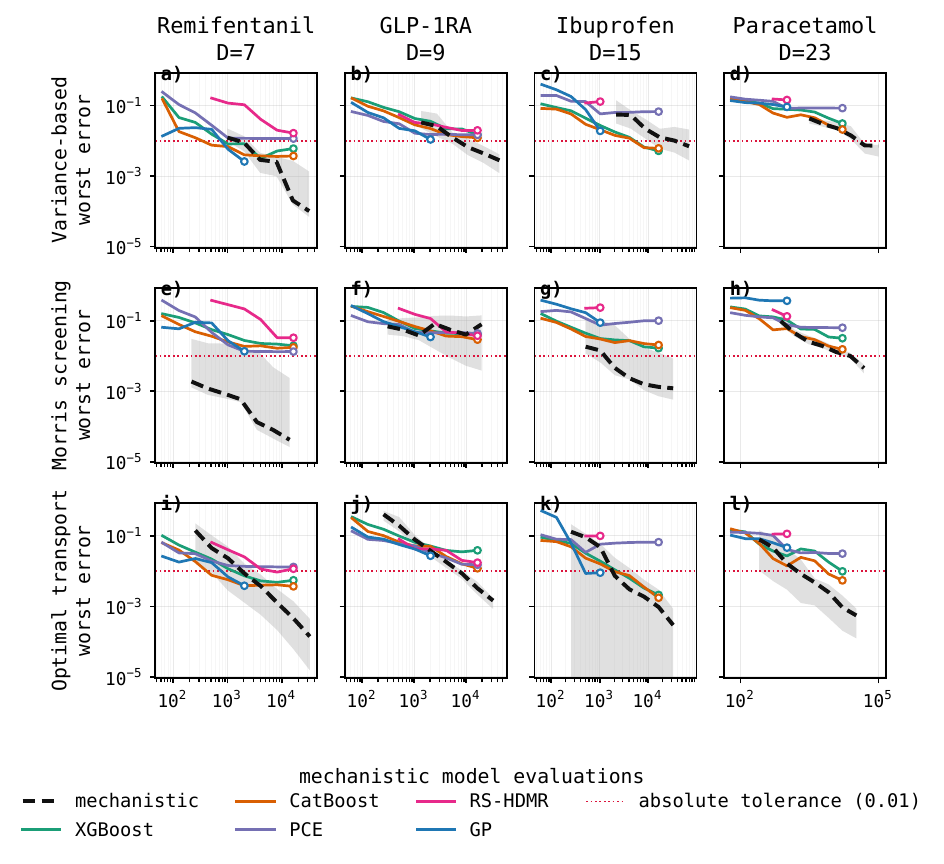}}
\caption{Index error against cost with the mechanistic ladder on the same axis, for the three methods (rows) on the four models in dimension order (columns).
}
\label{fig:cost-error}
\end{figure}

A surrogate training curve plateaus at an irreducible bias set by its hyperparameters, and past that point only the mechanistic route improves with more evaluations (\cref{fig:cost-error}).
Polynomial chaos on the ibuprofen model reaches $0.058$ at 1024 training rows and $0.068$ at 16{,}384, where CatBoost reaches $0.006$ and the direct analysis $0.007$.

\clearpage
\section{Expert-guided method tuning}
\label{sec:guidance}

\definecolor{surrogateGP}{HTML}{1F77B4}
\definecolor{surrogateCatBoost}{HTML}{D95F02}
\definecolor{surrogatePCE}{HTML}{7570B3}

\begin{figure}[H]
\centering
\begin{tikzpicture}[
  font=\footnotesize,
  band/.style={font=\footnotesize\bfseries, anchor=west},
  tablehead/.style={font=\scriptsize\bfseries, text=black!65, anchor=west},
  cell/.style={font=\footnotesize, anchor=west, align=left},
  card/.style 2 args={draw=#1, line width=0.8pt, fill=#1!5!white,
    rounded corners=2pt, text width=#2, minimum height=25mm, font=\scriptsize,
    align=center, inner sep=5pt, anchor=north},
  rule/.style={draw=black!24, line width=0.45pt}
]
\newlength{\figwidth}\setlength{\figwidth}{0.94\textwidth}

\node[band] at (0,0) {Step 1\quad Which route meets the required result?};
\node[tablehead] at (0.02\figwidth,-8mm) {Required result};
\node[tablehead] at (0.22\figwidth,-8mm) {Situation};
\node[tablehead] at (0.72\figwidth,-8mm) {Recommendation};
\draw[rule] (0,-12mm) -- (\figwidth,-12mm);

\node[cell] at (0.02\figwidth,-19mm) {Ranking};
\node[cell] at (0.22\figwidth,-19mm) {Variance-based or optimal-transport GSA};
\node[cell, font=\footnotesize\bfseries] at (0.72\figwidth,-19mm) {Build a surrogate};
\draw[rule] (0,-26mm) -- (\figwidth,-26mm);

\node[cell] at (0.02\figwidth,-33mm) {Ranking};
\node[cell] at (0.22\figwidth,-33mm) {Morris screening};
\node[cell, font=\footnotesize\bfseries] at (0.72\figwidth,-33mm) {Run direct};
\draw[rule] (0,-41mm) -- (\figwidth,-41mm);

\node[cell] at (0.02\figwidth,-48mm) {Index values};
\node[cell] at (0.22\figwidth,-48mm) {Up to 15 inputs in this study};
\node[cell, font=\footnotesize\bfseries] at (0.72\figwidth,-48mm) {Build a surrogate};
\draw[rule] (0,-56mm) -- (\figwidth,-56mm);

\node[cell] at (0.02\figwidth,-63mm) {Index values};
\node[cell] at (0.22\figwidth,-63mm) {23 inputs in this study};
\node[cell, font=\footnotesize\bfseries] at (0.72\figwidth,-63mm) {Run direct};
\draw[rule] (0,-70mm) -- (\figwidth,-70mm);

\node[band] at (0,-79mm) {Step 2\quad Which surrogate?};
\node[card={surrogateGP}{0.275\figwidth}] at (0.16\figwidth,-86mm)
  {\textbf{\footnotesize Gaussian process}\\[3pt]
   Smooth response; roughly nine inputs or fewer\\[3pt]
   Most sample-efficient};
\node[card={surrogateCatBoost}{0.275\figwidth}] at (0.50\figwidth,-86mm)
  {\textbf{\footnotesize CatBoost}\\[3pt]
   Default choice\\[3pt]
   Broadest cohort coverage};
\node[card={surrogatePCE}{0.275\figwidth}] at (0.84\figwidth,-86mm)
  {\textbf{\footnotesize Polynomial chaos}\\[3pt]
   Need coefficient decomposition\\[3pt]
   Returns higher-order terms};

\node[band] at (0,-125mm) {Step 3\quad Validate};
\node[draw=black!40, fill=black!3, rounded corners=2pt,
  text width=0.90\figwidth, minimum height=11mm, align=center,
  font=\footnotesize, anchor=north] at (0.50\figwidth,-131mm)
  {Validate against converged mechanistic indices at the required tier.};
\end{tikzpicture}
\caption{Choosing the direct or surrogate route, then selecting and validating a surrogate.
Direct Morris screening is preferable unless many cohorts are analysed; absolute-index surrogates may also need several cohorts to amortise.
Held-out prediction fit alone does not establish index recovery; a converged mechanistic reference is needed for validation.}
\label{fig:tree-surrogate}
\end{figure}

\cref{fig:tree-surrogate} summarises the route and surrogate choices supported by the results above.
Its branch conditions are the input dimension, the tier needed and the GSA method chosen; the following subsections explain the evidence and limits behind each recommendation.

\subsection{Choosing a method}
\label{sec:guidance-method}

Calibrated variance shares, and the separation of interaction from non-linearity through $\ST - \Sone$, come only from the variance-based GSA.
The shape or the tail of the output distribution is seen only by optimal transport, which on these models reported a substantial diffusive component without reordering the parameters that matter.
Effect direction comes only from Morris screening, through the sign of the mean elementary effect.
\cref{tab:guidance-method} compares GSA methods.

\begin{table}[htbp]
\centering
\caption{Method selection. Evaluation counts are the median direct budget per cohort in mechanistic model evaluations.}
\label{tab:guidance-method}
\scriptsize
\setlength{\tabcolsep}{4pt}
\renewcommand{\arraystretch}{1.15}
\begin{tabular}{@{}>{\raggedright\arraybackslash}p{0.225\textwidth}
>{\raggedright\arraybackslash}p{0.225\textwidth}
>{\raggedright\arraybackslash}p{0.195\textwidth}
>{\raggedright\arraybackslash}p{0.255\textwidth}@{}}
\toprule
                        & Variance-based                                                             & Morris screening & Optimal transport \\
\midrule
Answers
                        & how much output variance each input carries, alone and through interaction
                        & which inputs matter, and in which direction
                        & how much each input moves the whole output distribution                                                           \\
\addlinespace
Design cost             & $N(D+2)$                                                                   & $r(D+1)$         & $N$               \\
\addlinespace
Evaluations per cohort
                        & 896/1408/2176/3200
                        & 192/320/512/1120
                        & 2048/4096/4096/4096                                                                                               \\
\addlinespace
Gives effect direction  & no                                                                         & \textbf{yes}     & no                \\
Separates interaction from non-linearity
                        & \textbf{yes}                                                               & no               & no                \\
Sees distribution shape & no                                                                         & no               & \textbf{yes}      \\
\addlinespace
Reuses another method's design
                        & no
                        & structurally yes, untested here
                        & \textbf{yes}; the indices here are read from the variance-based $A$ matrix                                        \\
\bottomrule
\end{tabular}
\end{table}

\subsection{Choosing a surrogate}
\label{sec:guidance-surrogate}

A decision made from the ranking and ordering of GSA indices alone can be reached with any of the five surrogates on most model and cohort combinations, and with fewer simulations than the direct route needs.
Recovering accurate absolute index values requires the absolute tier, which four of the five reach on half their cells or fewer.

The best performing surrogate tested is CatBoost.
It reaches the ranking tier on all 12 model and method cells and the absolute tier on 11 of them, and on 11 of the 12 it costs less to train than one direct analysis, at a median break-even of $0.11$ cohorts.
XGBoost is the nearest to it, reaching the ranking on 11 cells and the absolute tier on 9, and the other three fall further behind.

The remaining three surrogates should be chosen on the dimensionality of the model and the problem rather than on accuracy.
Polynomial chaos is the choice when the analytical decomposition itself is wanted, since it returns the indices and the higher-order terms directly from its coefficients, but it reaches the absolute tier only on the lower-dimensional models, on all six of its cells at seven and nine inputs and none of the six above them.
The Gaussian process is the choice below roughly ten inputs on a smooth response, where it is the most sample-efficient of the five at a median break-even of $0.062$ cohorts and needs no tuning budget, fitting its hyperparameters by marginal likelihood; it cannot be trained on more than 2048 rows, its cost being cubic in the sample, and at twenty-three inputs it reaches neither tier on any cohort.
RS-HDMR can include pairwise terms and so return second-order indices, but the fit is memory-hungry and is capped at 1024 training rows above fourteen inputs, where it recovers the ordering on most cells and the values on none.
The Gaussian process therefore has the best median break-even of the five and the worst coverage, reaching a tier on 9 cells of 12 where CatBoost reaches all 12, so CatBoost is the default on coverage rather than on sample efficiency.

\subsubsection{How many cohorts before the surrogate route costs less}
\label{sec:guidance-kstar}

Against a variance-based GSA the two gradient-boosted trees and the Gaussian process cost less to train than a single direct analysis on every model they reach, and for the trees the margin grows with dimension, the break-even falling from $0.14$ cohorts at seven inputs to $0.03$ at fifteen and to $0.02$ for CatBoost and $0.04$ for XGBoost at twenty-three.
The other two do not.
Polynomial chaos needs 1024 training evaluations at seven inputs against a direct cost of 896, so $1.14$ cohorts, and RS-HDMR needs 8192 against the same 896, so $9.14$.
Across all three methods RS-HDMR's median break-even at this tier is $0.41$ cohorts with three cells above one.
The reason is arithmetic.
The $(D+2)$ factor inflates the direct cost as dimension grows, while the training budget needed for a correct ranking does not follow the same pattern.

If a practitioner wants to recover accurate index values, the guidance changes.
Through nine inputs the break-even sits between $0.07$ and $1.45$ cohorts, at fifteen it survives for the two gradient boosters and the Gaussian process between $0.24$ and $0.47$, and at twenty-three none of the five reaches the tier on a majority of cohorts.

Against Morris screening the comparison is much closer, because the direct screening design is itself inexpensive: twelve of the twenty cells fail to break even inside a single cohort, six of them because no training size measured here reaches the tier at all, and the worst finite break-even is $85.3$ for RS-HDMR at seven inputs.
If Morris screening answers the question, run it directly and build a surrogate only when the cohort count is large.

\section{Where the guidance stops applying}
\label{sec:guidance-scope}

The measured break-even counts and method recommendations apply within the input envelopes, patient cohorts, model responses and cost definition used here. The following conditions mark where that guidance needs to be reconsidered.

\begin{itemize}
\item \textbf{Cohorts outside the envelope.} Every surrogate result assumes that the cohort lies inside the input region used for training. A group outside that region would require an extended envelope and a new surrogate.
\item \textbf{Model scope.} The four models have seven to twenty-three free inputs and pharmacokinetic or pharmacokinetic--pharmacodynamic outputs. Performance above twenty-three inputs or for responses of different character was not measured.
\item \textbf{Small critical sets.} With fewer than three critical axes, the ranking tier uses exact agreement on set membership and the leading axis rather than a rank correlation. That criterion does not resolve the ordering of axes below the leader.
\item \textbf{Cost accounting.} Costs are counted in mechanistic model evaluations. When model solves are fast enough that surrogate fitting time is comparable, elapsed-time comparisons could differ.
\end{itemize}

A study limitation is the fixed training design: each surrogate was trained on a corpus drawn in advance. Adaptive refinement, which adds samples where the current surrogate is uncertain, was not tested. Its effect on the relative performance of the surrogates remains unknown.

\section{Conclusions}
\label{sec:conclusions}

Pharmacokinetic models are increasingly run as simulated clinical trials, where the unit of analysis is a virtual patient cohort and one study spans many cohorts and scenarios.
Global sensitivity analysis is then called on repeatedly over one model under a different input distribution each time, and under conventional accounting each call incurs the full cost of a fresh analysis.

One surrogate, trained once on a wide input envelope, reconstructs the sensitivity ranking under a variance-based GSA on 20 of the 21 patient cohorts of four PK and PK/PD models spanning seven to twenty-three free inputs.
That surrogate is CatBoost, and it costs fewer evaluations to train than one direct cohort analysis on every one of the four models, at break-even counts of $0.14$, $0.09$, $0.03$ and $0.02$ cohorts as the free dimension rises from seven to twenty-three.
Reconstructing the absolute index values is a harder and separate problem that narrows with dimension: gradient-boosted trees and the Gaussian process reconstruct absolute indices, while polynomial chaos and RS-HDMR fail at $D=15$.
All surrogates struggle at the highest dimension, none reaching the tier on a majority of the paracetamol cohorts even under a scale-aware tolerance.

The three methods are peers, and the useful question is what a second one adds for its cost.
Direct Morris screening is the cheapest route to a ranking at seven and nine inputs, holds only a marginal lead at fifteen and none at twenty-three, and a surrogate breaks even against it far more slowly than against a variance-based analysis, because the cost it is measured against is already small.
On ordering, the three recover similar trends: optimal transport reproduces the exact ordering of the variance-based important set on 18 of the 21 cohorts and Morris screening on 16.

Under that pattern of use the cost of one analysis is the wrong number to report; what matters is the count of cohorts at which the surrogate route has cost fewer evaluations than the direct one, and on all four models here that count is below one.

Two further extensions follow from this study.
Cohorts defined by covariate distributions rather than by fractions of an envelope bound would test the reordering finding properly.
And second-order indices can be read directly from the polynomial chaos coefficients already fitted here, and from the RS-HDMR coefficients on the two models where tuning selected pairwise terms, so the interaction structure would come from the surrogate already built, with no second mechanistic design.

\clearpage
\section*{Acknowledgements}
This work was supported by the Engineering and Physical Sciences Research Council
(EPSRC) for the Imperial College London Doctoral Training Partnership (DTP).

\section*{Code availability}
The variance-based, Morris screening and optimal transport estimators used in this study
are available in the open-source \texttt{jaxgsa} package \citep{pessina2026jaxgsa},
archived on Zenodo. Source code is available at
\href{https://github.com/DanielePessina/jaxgsa}{github.com/DanielePessina/jaxgsa}.

\clearpage

\bibliographystyle{elsarticle-harv}
\bibliography{references}

\begin{thebibliography}{81}
\expandafter\ifx\csname natexlab\endcsname\relax\def\natexlab#1{#1}\fi
\providecommand{\url}[1]{\texttt{#1}}
\providecommand{\href}[2]{#2}
\providecommand{\path}[1]{#1}
\providecommand{\DOIprefix}{doi:}
\providecommand{\ArXivprefix}{arXiv:}
\providecommand{\URLprefix}{URL: }
\providecommand{\Pubmedprefix}{pmid:}
\providecommand{\doi}[1]{\href{http://dx.doi.org/#1}{\path{#1}}}
\providecommand{\Pubmed}[1]{\href{pmid:#1}{\path{#1}}}
\providecommand{\bibinfo}[2]{#2}
\ifx\xfnm\relax \def\xfnm[#1]{\unskip,\space#1}\fi
%Type = Article
\bibitem[{Abbiati et~al.(2016)Abbiati, Lamberti, Grassi, Trotta and
  Manca}]{abbiati2016definition}
\bibinfo{author}{Abbiati, R.A.}, \bibinfo{author}{Lamberti, G.},
  \bibinfo{author}{Grassi, M.}, \bibinfo{author}{Trotta, F.},
  \bibinfo{author}{Manca, D.}, \bibinfo{year}{2016}.
\newblock \bibinfo{title}{Definition and validation of a patient-individualized
  physiologically-based pharmacokinetic model}.
\newblock \bibinfo{journal}{Computers \& Chemical Engineering}
  \bibinfo{volume}{84}, \bibinfo{pages}{394--408}.
\newblock \DOIprefix\doi{10.1016/j.compchemeng.2015.09.018}.
%Type = Incollection
\bibitem[{Arsène et~al.(2024)Arsène, Parès, Tixier, Granjeon-Noriot, Martin,
  Bruezière, Couty, Courcelles, Kahoul, Pitrat, Go, Monteiro,
  Kleine-Schultjann, Jemai, Pham, Boissel and
  Kulesza}]{arseneSilicoClinicalTrials2024}
\bibinfo{author}{Arsène, S.}, \bibinfo{author}{Parès, Y.},
  \bibinfo{author}{Tixier, E.}, \bibinfo{author}{Granjeon-Noriot, S.},
  \bibinfo{author}{Martin, B.}, \bibinfo{author}{Bruezière, L.},
  \bibinfo{author}{Couty, C.}, \bibinfo{author}{Courcelles, E.},
  \bibinfo{author}{Kahoul, R.}, \bibinfo{author}{Pitrat, J.},
  \bibinfo{author}{Go, N.}, \bibinfo{author}{Monteiro, C.},
  \bibinfo{author}{Kleine-Schultjann, J.}, \bibinfo{author}{Jemai, S.},
  \bibinfo{author}{Pham, E.}, \bibinfo{author}{Boissel, J.P.},
  \bibinfo{author}{Kulesza, A.}, \bibinfo{year}{2024}.
\newblock \bibinfo{title}{In {{Silico Clinical Trials}}: {{Is It Possible}}?},
  in: \bibinfo{editor}{Heifetz, A.} (Ed.), \bibinfo{booktitle}{High
  {{Performance Computing}} for {{Drug Discovery}} and {{Biomedicine}}}.
  \bibinfo{publisher}{Springer US}, pp. \bibinfo{pages}{51--99}.
\newblock \URLprefix \url{https://doi.org/10.1007/978-1-0716-3449-3_4},
  \DOIprefix\doi{10.1007/978-1-0716-3449-3_4}.
%Type = Article
\bibitem[{Borgonovo(2007)}]{borgonovo2007delta}
\bibinfo{author}{Borgonovo, E.}, \bibinfo{year}{2007}.
\newblock \bibinfo{title}{A new uncertainty importance measure}.
\newblock \bibinfo{journal}{Reliability Engineering \& System Safety}
  \bibinfo{volume}{92}, \bibinfo{pages}{771--784}.
\newblock \DOIprefix\doi{10.1016/j.ress.2006.04.015}.
%Type = Article
\bibitem[{Borgonovo et~al.(2025)Borgonovo, Figalli, Plischke and
  Savar{\'e}}]{borgonovo2025ot}
\bibinfo{author}{Borgonovo, E.}, \bibinfo{author}{Figalli, A.},
  \bibinfo{author}{Plischke, E.}, \bibinfo{author}{Savar{\'e}, G.},
  \bibinfo{year}{2025}.
\newblock \bibinfo{title}{Global sensitivity analysis via optimal transport}.
\newblock \bibinfo{journal}{Management Science} \bibinfo{volume}{71},
  \bibinfo{pages}{3809--3828}.
\newblock \DOIprefix\doi{10.1287/mnsc.2023.01796}. \bibinfo{note}{published
  online ahead of print in 2024; volume of record 2025}.
%Type = Article
\bibitem[{Bosch et~al.(2024)Bosch, Sijbrands and
  Snelder}]{bosch2024quantification}
\bibinfo{author}{Bosch, R.}, \bibinfo{author}{Sijbrands, E.J.G.},
  \bibinfo{author}{Snelder, N.}, \bibinfo{year}{2024}.
\newblock \bibinfo{title}{Quantification of the effect of {GLP-1R} agonists on
  body weight using in vitro efficacy information: An extension of the {Hall}
  body composition model}.
\newblock \bibinfo{journal}{CPT: Pharmacometrics \& Systems Pharmacology}
  \bibinfo{volume}{13}, \bibinfo{pages}{1488--1502}.
\newblock \DOIprefix\doi{10.1002/psp4.13183}. \bibinfo{note}{pMID: 38867373}.
%Type = Misc
\bibitem[{Boucharif et~al.(2025)Boucharif, Morio and Rochet}]{boucharif2025}
\bibinfo{author}{Boucharif, H.}, \bibinfo{author}{Morio, J.},
  \bibinfo{author}{Rochet, P.}, \bibinfo{year}{2025}.
\newblock \bibinfo{title}{Importance sampling for {Sobol'} indices estimation}.
\newblock \DOIprefix\doi{10.48550/arXiv.2507.05958},
  \href{http://arxiv.org/abs/2507.05958}{{\tt arXiv:2507.05958}}.
  \bibinfo{note}{submitted 8 July 2025; revised 1 March 2026}.
%Type = Misc
\bibitem[{Bradbury et~al.(2018)Bradbury, Frostig, Hawkins, Johnson, Leary,
  Maclaurin, Necula, Paszke, Vander{P}las, Wanderman-{M}ilne and
  Zhang}]{bradbury2018jax}
\bibinfo{author}{Bradbury, J.}, \bibinfo{author}{Frostig, R.},
  \bibinfo{author}{Hawkins, P.}, \bibinfo{author}{Johnson, M.J.},
  \bibinfo{author}{Leary, C.}, \bibinfo{author}{Maclaurin, D.},
  \bibinfo{author}{Necula, G.}, \bibinfo{author}{Paszke, A.},
  \bibinfo{author}{Vander{P}las, J.}, \bibinfo{author}{Wanderman-{M}ilne, S.},
  \bibinfo{author}{Zhang, Q.}, \bibinfo{year}{2018}.
\newblock \bibinfo{title}{{JAX}: composable transformations of {Python+NumPy}
  programs}.
\newblock \bibinfo{howpublished}{Software}.
\newblock \URLprefix \url{https://github.com/jax-ml/jax}.
%Type = Article
\bibitem[{Braniff et~al.(2025)Braniff, Joshi, Cassidy, Trogdon, Kumar, Poels,
  Allen, Musante and Shtylla}]{braniffIntegratedQuantitativeSystems2025}
\bibinfo{author}{Braniff, N.}, \bibinfo{author}{Joshi, T.},
  \bibinfo{author}{Cassidy, T.}, \bibinfo{author}{Trogdon, M.},
  \bibinfo{author}{Kumar, R.}, \bibinfo{author}{Poels, K.},
  \bibinfo{author}{Allen, R.}, \bibinfo{author}{Musante, C.J.},
  \bibinfo{author}{Shtylla, B.}, \bibinfo{year}{2025}.
\newblock \bibinfo{title}{An integrated quantitative systems pharmacology
  virtual population approach for calibration with oncology efficacy
  endpoints}.
\newblock \bibinfo{journal}{CPT: Pharmacometrics \& Systems Pharmacology}
  \bibinfo{volume}{14}, \bibinfo{pages}{268--278}.
\newblock \URLprefix
  \url{https://onlinelibrary.wiley.com/doi/abs/10.1002/psp4.13270},
  \DOIprefix\doi{10.1002/psp4.13270}.
%Type = Article
\bibitem[{Campolongo et~al.(2007)Campolongo, Cariboni and
  Saltelli}]{campolongo2007effective}
\bibinfo{author}{Campolongo, F.}, \bibinfo{author}{Cariboni, J.},
  \bibinfo{author}{Saltelli, A.}, \bibinfo{year}{2007}.
\newblock \bibinfo{title}{An effective screening design for sensitivity
  analysis of large models}.
\newblock \bibinfo{journal}{Environmental Modelling \& Software}
  \bibinfo{volume}{22}, \bibinfo{pages}{1509--1518}.
\newblock \DOIprefix\doi{10.1016/j.envsoft.2006.10.004}.
%Type = Article
\bibitem[{Campolongo et~al.(2011)Campolongo, Saltelli and
  Cariboni}]{campolongo2011unified}
\bibinfo{author}{Campolongo, F.}, \bibinfo{author}{Saltelli, A.},
  \bibinfo{author}{Cariboni, J.}, \bibinfo{year}{2011}.
\newblock \bibinfo{title}{From screening to quantitative sensitivity analysis.
  {A} unified approach}.
\newblock \bibinfo{journal}{Computer Physics Communications}
  \bibinfo{volume}{182}, \bibinfo{pages}{978--988}.
\newblock \DOIprefix\doi{10.1016/j.cpc.2010.12.039}.
%Type = Inproceedings
\bibitem[{Chen and Guestrin(2016)}]{chen2016xgboost}
\bibinfo{author}{Chen, T.}, \bibinfo{author}{Guestrin, C.},
  \bibinfo{year}{2016}.
\newblock \bibinfo{title}{{XGBoost}: a scalable tree boosting system}, in:
  \bibinfo{booktitle}{Proceedings of the 22nd ACM SIGKDD International
  Conference on Knowledge Discovery and Data Mining}, pp.
  \bibinfo{pages}{785--794}.
\newblock \DOIprefix\doi{10.1145/2939672.2939785}.
%Type = Incollection
\bibitem[{Cheng et~al.(2022)Cheng, Straube, Alnaif, Huang, Leil and
  Schmidt}]{chengVirtualPopulationsQuantitative2022a}
\bibinfo{author}{Cheng, Y.}, \bibinfo{author}{Straube, R.},
  \bibinfo{author}{Alnaif, A.E.}, \bibinfo{author}{Huang, L.},
  \bibinfo{author}{Leil, T.A.}, \bibinfo{author}{Schmidt, B.J.},
  \bibinfo{year}{2022}.
\newblock \bibinfo{title}{Virtual {{Populations}} for {{Quantitative Systems
  Pharmacology Models}}}, in: \bibinfo{editor}{Bai, J.P.},
  \bibinfo{editor}{Hur, J.} (Eds.), \bibinfo{booktitle}{Systems {{Medicine}}}.
  \bibinfo{publisher}{Springer US}, pp. \bibinfo{pages}{129--179}.
\newblock \URLprefix \url{https://doi.org/10.1007/978-1-0716-2265-0_8},
  \DOIprefix\doi{10.1007/978-1-0716-2265-0_8}.
%Type = Misc
\bibitem[{Chiani et~al.(2025)Chiani, Borgonovo, Plischke and
  Tavoni}]{chiani2025gsaot}
\bibinfo{author}{Chiani, L.}, \bibinfo{author}{Borgonovo, E.},
  \bibinfo{author}{Plischke, E.}, \bibinfo{author}{Tavoni, M.},
  \bibinfo{year}{2025}.
\newblock \bibinfo{title}{gsaot: {An} {R} package for optimal transport-based
  sensitivity analysis}.
\newblock \DOIprefix\doi{10.48550/arXiv.2507.18588},
  \href{http://arxiv.org/abs/2507.18588}{{\tt arXiv:2507.18588}}.
  \bibinfo{note}{arXiv preprint}.
%Type = Article
\bibitem[{Cortés-Ríos et~al.(2025)Cortés-Ríos, Magee, Sher, Jusko and
  Desikan}]{cortes-riosStepbyStepWorkflowPerforming2025}
\bibinfo{author}{Cortés-Ríos, J.}, \bibinfo{author}{Magee, M.},
  \bibinfo{author}{Sher, A.}, \bibinfo{author}{Jusko, W.J.},
  \bibinfo{author}{Desikan, R.}, \bibinfo{year}{2025}.
\newblock \bibinfo{title}{A {{Step-by-Step Workflow}} for {{Performing In
  Silico Clinical Trials With Nonlinear Mixed Effects Models}}}.
\newblock \bibinfo{journal}{CPT: Pharmacometrics \& Systems Pharmacology}
  \bibinfo{volume}{14}, \bibinfo{pages}{1949--1964}.
\newblock \URLprefix
  \url{https://onlinelibrary.wiley.com/doi/abs/10.1002/psp4.70122},
  \DOIprefix\doi{10.1002/psp4.70122}.
%Type = Article
\bibitem[{Da~Veiga(2015)}]{daveiga2015dependence}
\bibinfo{author}{Da~Veiga, S.}, \bibinfo{year}{2015}.
\newblock \bibinfo{title}{Global sensitivity analysis with dependence
  measures}.
\newblock \bibinfo{journal}{Journal of Statistical Computation and Simulation}
  \bibinfo{volume}{85}, \bibinfo{pages}{1283--1305}.
\newblock \DOIprefix\doi{10.1080/00949655.2014.945932}.
%Type = Article
\bibitem[{El~Amri and Marrel(2022)}]{elamri2022optimized}
\bibinfo{author}{El~Amri, M.R.}, \bibinfo{author}{Marrel, A.},
  \bibinfo{year}{2022}.
\newblock \bibinfo{title}{Optimized {HSIC}-based tests for sensitivity
  analysis: application to thermalhydraulic simulation of accidental scenario
  on nuclear reactor}.
\newblock \bibinfo{journal}{Quality and Reliability Engineering International}
  \bibinfo{volume}{38}, \bibinfo{pages}{1386--1403}.
\newblock \DOIprefix\doi{10.1002/qre.2954}.
%Type = Article
\bibitem[{Feng et~al.(2019)Feng, Lu and Yang}]{feng2019enhanced}
\bibinfo{author}{Feng, K.}, \bibinfo{author}{Lu, Z.}, \bibinfo{author}{Yang,
  C.}, \bibinfo{year}{2019}.
\newblock \bibinfo{title}{Enhanced {Morris} method for global sensitivity
  analysis: good proxy of {Sobol'} index}.
\newblock \bibinfo{journal}{Structural and Multidisciplinary Optimization}
  \bibinfo{volume}{59}, \bibinfo{pages}{373--387}.
\newblock \DOIprefix\doi{10.1007/s00158-018-2071-7}.
%Type = Article
\bibitem[{Friedman(2001)}]{friedman2001}
\bibinfo{author}{Friedman, J.H.}, \bibinfo{year}{2001}.
\newblock \bibinfo{title}{Greedy function approximation: a gradient boosting
  machine}.
\newblock \bibinfo{journal}{The Annals of Statistics} \bibinfo{volume}{29},
  \bibinfo{pages}{1189--1232}.
\newblock \DOIprefix\doi{10.1214/aos/1013203451}.
%Type = Article
\bibitem[{Gamboa et~al.(2014)Gamboa, Janon, Klein and
  Lagnoux}]{gamboa2014multivariate}
\bibinfo{author}{Gamboa, F.}, \bibinfo{author}{Janon, A.},
  \bibinfo{author}{Klein, T.}, \bibinfo{author}{Lagnoux, A.},
  \bibinfo{year}{2014}.
\newblock \bibinfo{title}{Sensitivity analysis for multidimensional and
  functional outputs}.
\newblock \bibinfo{journal}{Electronic Journal of Statistics}
  \bibinfo{volume}{8}, \bibinfo{pages}{575--603}.
\newblock \DOIprefix\doi{10.1214/14-EJS895}. \bibinfo{note}{page range taken
  from the Project Euclid record; the Crossref record has a null page field}.
%Type = Article
\bibitem[{Gilboa et~al.(2015)Gilboa, Saat{\c{c}}i and
  Cunningham}]{gilboa2015scaling}
\bibinfo{author}{Gilboa, E.}, \bibinfo{author}{Saat{\c{c}}i, Y.},
  \bibinfo{author}{Cunningham, J.P.}, \bibinfo{year}{2015}.
\newblock \bibinfo{title}{Scaling multidimensional inference for structured
  {Gaussian} processes}.
\newblock \bibinfo{journal}{IEEE Transactions on Pattern Analysis and Machine
  Intelligence} \bibinfo{volume}{37}, \bibinfo{pages}{424--436}.
\newblock \DOIprefix\doi{10.1109/TPAMI.2013.192}.
%Type = Article
\bibitem[{Hall et~al.(2011)Hall, Sacks, Chandramohan, Chow, Wang, Gortmaker and
  Swinburn}]{hall2011quantification}
\bibinfo{author}{Hall, K.D.}, \bibinfo{author}{Sacks, G.},
  \bibinfo{author}{Chandramohan, D.}, \bibinfo{author}{Chow, C.C.},
  \bibinfo{author}{Wang, Y.C.}, \bibinfo{author}{Gortmaker, S.L.},
  \bibinfo{author}{Swinburn, B.A.}, \bibinfo{year}{2011}.
\newblock \bibinfo{title}{Quantification of the effect of energy imbalance on
  bodyweight}.
\newblock \bibinfo{journal}{The Lancet} \bibinfo{volume}{378},
  \bibinfo{pages}{826--837}.
\newblock \DOIprefix\doi{10.1016/S0140-6736(11)60812-X}. \bibinfo{note}{pMID:
  21872751}.
%Type = Article
\bibitem[{Hilhorst et~al.(2025)Hilhorst, {van de Wouw}, Zajac, {van 't Veer},
  Tonino, {van de Vosse} and Huberts}]{hilhorstSensitivityAnalysisExploring}
\bibinfo{author}{Hilhorst, P.}, \bibinfo{author}{{van de Wouw}, B.},
  \bibinfo{author}{Zajac, K.}, \bibinfo{author}{{van 't Veer}, M.},
  \bibinfo{author}{Tonino, P.}, \bibinfo{author}{{van de Vosse}, F.},
  \bibinfo{author}{Huberts, W.}, \bibinfo{year}{2025}.
\newblock \bibinfo{title}{Sensitivity analysis for exploring the variability
  and parameter landscape in virtual patient cohorts of multi-vessel coronary
  artery disease}.
\newblock \bibinfo{journal}{Philosophical transactions. Series A, Mathematical,
  physical, and engineering sciences} \bibinfo{volume}{383},
  \bibinfo{pages}{20240230}.
\newblock \URLprefix \url{https://pmc.ncbi.nlm.nih.gov/articles/PMC11966647/},
  \DOIprefix\doi{10.1098/rsta.2024.0230},
  \href{http://arxiv.org/abs/40172563}{{\tt arXiv:40172563}}.
%Type = Article
\bibitem[{Homma and Saltelli(1996)}]{homma1996}
\bibinfo{author}{Homma, T.}, \bibinfo{author}{Saltelli, A.},
  \bibinfo{year}{1996}.
\newblock \bibinfo{title}{Importance measures in global sensitivity analysis of
  nonlinear models}.
\newblock \bibinfo{journal}{Reliability Engineering \& System Safety}
  \bibinfo{volume}{52}, \bibinfo{pages}{1--17}.
\newblock \DOIprefix\doi{10.1016/0951-8320(96)00002-6}.
%Type = Article
\bibitem[{Hsieh et~al.(2018)Hsieh, Reisfeld, Bois and Chiu}]{hsieh2018applying}
\bibinfo{author}{Hsieh, N.H.}, \bibinfo{author}{Reisfeld, B.},
  \bibinfo{author}{Bois, F.Y.}, \bibinfo{author}{Chiu, W.A.},
  \bibinfo{year}{2018}.
\newblock \bibinfo{title}{Applying a global sensitivity analysis workflow to
  improve the computational efficiencies in physiologically-based
  pharmacokinetic modeling}.
\newblock \bibinfo{journal}{Frontiers in Pharmacology} \bibinfo{volume}{9},
  \bibinfo{pages}{588}.
\newblock \DOIprefix\doi{10.3389/fphar.2018.00588}. \bibinfo{note}{pMID:
  29937730}.
%Type = Article
\bibitem[{Hsieh et~al.(2020)Hsieh, Reisfeld and Chiu}]{hsieh2020pksensi}
\bibinfo{author}{Hsieh, N.H.}, \bibinfo{author}{Reisfeld, B.},
  \bibinfo{author}{Chiu, W.A.}, \bibinfo{year}{2020}.
\newblock \bibinfo{title}{pksensi: An {R} package to apply global sensitivity
  analysis in physiologically based kinetic modeling}.
\newblock \bibinfo{journal}{SoftwareX} \bibinfo{volume}{12},
  \bibinfo{pages}{100609}.
\newblock \DOIprefix\doi{10.1016/j.softx.2020.100609}. \bibinfo{note}{pMID:
  33426260}.
%Type = Incollection
\bibitem[{Iooss and Lema{\^i}tre(2015)}]{iooss2015}
\bibinfo{author}{Iooss, B.}, \bibinfo{author}{Lema{\^i}tre, P.},
  \bibinfo{year}{2015}.
\newblock \bibinfo{title}{A review on global sensitivity analysis methods}, in:
  \bibinfo{booktitle}{Uncertainty Management in Simulation-Optimization of
  Complex Systems}. \bibinfo{publisher}{Springer}, pp.
  \bibinfo{pages}{101--122}.
\newblock \DOIprefix\doi{10.1007/978-1-4899-7547-8_5}.
%Type = Article
\bibitem[{Jansen(1999)}]{jansen1999}
\bibinfo{author}{Jansen, M.J.W.}, \bibinfo{year}{1999}.
\newblock \bibinfo{title}{Analysis of variance designs for model output}.
\newblock \bibinfo{journal}{Computer Physics Communications}
  \bibinfo{volume}{117}, \bibinfo{pages}{35--43}.
\newblock \DOIprefix\doi{10.1016/S0010-4655(98)00154-4}.
%Type = Article
\bibitem[{Khorashadi~Zadeh et~al.(2017)Khorashadi~Zadeh, Nossent, Sarrazin,
  Pianosi, van Griensven, Wagener and Bauwens}]{khorashadizadeh2017comparison}
\bibinfo{author}{Khorashadi~Zadeh, F.}, \bibinfo{author}{Nossent, J.},
  \bibinfo{author}{Sarrazin, F.}, \bibinfo{author}{Pianosi, F.},
  \bibinfo{author}{van Griensven, A.}, \bibinfo{author}{Wagener, T.},
  \bibinfo{author}{Bauwens, W.}, \bibinfo{year}{2017}.
\newblock \bibinfo{title}{Comparison of variance-based and moment-independent
  global sensitivity analysis approaches by application to the {SWAT} model}.
\newblock \bibinfo{journal}{Environmental Modelling \& Software}
  \bibinfo{volume}{91}, \bibinfo{pages}{210--222}.
\newblock \DOIprefix\doi{10.1016/j.envsoft.2017.02.001}.
%Type = Phdthesis
\bibitem[{Kidger(2022)}]{kidger2022}
\bibinfo{author}{Kidger, P.}, \bibinfo{year}{2022}.
\newblock \bibinfo{title}{On Neural Differential Equations}.
\newblock Ph.D. thesis. University of Oxford.
\newblock \URLprefix \url{https://arxiv.org/abs/2202.02435},
  \href{http://arxiv.org/abs/2202.02435}{{\tt arXiv:2202.02435}}.
  \bibinfo{note}{reference for the Diffrax differentiable ODE-solver library}.
%Type = Article
\bibitem[{Kolesova et~al.(2022)Kolesova, Stepanov, Lebedeva and
  Demin}]{kolesovaApplicationDifferentApproaches2022}
\bibinfo{author}{Kolesova, G.}, \bibinfo{author}{Stepanov, A.},
  \bibinfo{author}{Lebedeva, G.}, \bibinfo{author}{Demin, O.},
  \bibinfo{year}{2022}.
\newblock \bibinfo{title}{Application of different approaches to generate
  virtual patient populations for the quantitative systems pharmacology model
  of erythropoiesis}.
\newblock \bibinfo{journal}{Journal of Pharmacokinetics and Pharmacodynamics}
  \bibinfo{volume}{49}, \bibinfo{pages}{511--524}.
\newblock \URLprefix \url{https://doi.org/10.1007/s10928-022-09814-y},
  \DOIprefix\doi{10.1007/s10928-022-09814-y}.
%Type = Article
\bibitem[{Lamboni et~al.(2011)Lamboni, Monod and
  Makowski}]{lamboni2011multivariate}
\bibinfo{author}{Lamboni, M.}, \bibinfo{author}{Monod, H.},
  \bibinfo{author}{Makowski, D.}, \bibinfo{year}{2011}.
\newblock \bibinfo{title}{Multivariate sensitivity analysis to measure global
  contribution of input factors in dynamic models}.
\newblock \bibinfo{journal}{Reliability Engineering \& System Safety}
  \bibinfo{volume}{96}, \bibinfo{pages}{450--459}.
\newblock \DOIprefix\doi{10.1016/j.ress.2010.12.002}.
%Type = Incollection
\bibitem[{Le~Gratiet et~al.(2017)Le~Gratiet, Marelli and
  Sudret}]{legratiet2017}
\bibinfo{author}{Le~Gratiet, L.}, \bibinfo{author}{Marelli, S.},
  \bibinfo{author}{Sudret, B.}, \bibinfo{year}{2017}.
\newblock \bibinfo{title}{Metamodel-based sensitivity analysis: polynomial
  chaos expansions and {Gaussian} processes}, in: \bibinfo{booktitle}{Handbook
  of Uncertainty Quantification}. \bibinfo{publisher}{Springer},
  \bibinfo{address}{Cham}, pp. \bibinfo{pages}{1289--1325}.
\newblock \DOIprefix\doi{10.1007/978-3-319-12385-1_38}.
%Type = Article
\bibitem[{Lemaire et~al.(2023)Lemaire, Bassen, Reed, Song, Khalili, Lien,
  Huang, Singh, Stamatelos, Bottino and Hua}]{lemaireColdHotChanging2023}
\bibinfo{author}{Lemaire, V.}, \bibinfo{author}{Bassen, D.},
  \bibinfo{author}{Reed, M.}, \bibinfo{author}{Song, R.},
  \bibinfo{author}{Khalili, S.}, \bibinfo{author}{Lien, Y.T.K.},
  \bibinfo{author}{Huang, L.}, \bibinfo{author}{Singh, A.P.},
  \bibinfo{author}{Stamatelos, S.}, \bibinfo{author}{Bottino, D.},
  \bibinfo{author}{Hua, F.}, \bibinfo{year}{2023}.
\newblock \bibinfo{title}{From {{Cold}} to {{Hot}}: {{Changing Perceptions}}
  and {{Future Opportunities}} for {{Quantitative Systems Pharmacology
  Modeling}} in {{Cancer Immunotherapy}}}.
\newblock \bibinfo{journal}{Clinical Pharmacology \& Therapeutics}
  \bibinfo{volume}{113}, \bibinfo{pages}{963--972}.
\newblock \URLprefix
  \url{https://onlinelibrary.wiley.com/doi/abs/10.1002/cpt.2770},
  \DOIprefix\doi{10.1002/cpt.2770}.
%Type = Article
\bibitem[{Li et~al.(2002)Li, Wang and Rabitz}]{li2002practical}
\bibinfo{author}{Li, G.}, \bibinfo{author}{Wang, S.W.},
  \bibinfo{author}{Rabitz, H.}, \bibinfo{year}{2002}.
\newblock \bibinfo{title}{Practical approaches to construct {RS-HDMR} component
  functions}.
\newblock \bibinfo{journal}{The Journal of Physical Chemistry A}
  \bibinfo{volume}{106}, \bibinfo{pages}{8721--8733}.
\newblock \DOIprefix\doi{10.1021/jp014567t}.
%Type = Article
\bibitem[{Loizou et~al.(2015)Loizou, McNally, Jones and
  Cocker}]{loizou2015application}
\bibinfo{author}{Loizou, G.D.}, \bibinfo{author}{McNally, K.},
  \bibinfo{author}{Jones, K.}, \bibinfo{author}{Cocker, J.},
  \bibinfo{year}{2015}.
\newblock \bibinfo{title}{The application of global sensitivity analysis in the
  development of a physiologically based pharmacokinetic model for m-xylene and
  ethanol co-exposure in humans}.
\newblock \bibinfo{journal}{Frontiers in Pharmacology} \bibinfo{volume}{6},
  \bibinfo{pages}{135}.
\newblock \DOIprefix\doi{10.3389/fphar.2015.00135}. \bibinfo{note}{pMID:
  26175688}.
%Type = Article
\bibitem[{Marino et~al.(2008)Marino, Hogue, Ray and
  Kirschner}]{marino2008methodology}
\bibinfo{author}{Marino, S.}, \bibinfo{author}{Hogue, I.B.},
  \bibinfo{author}{Ray, C.J.}, \bibinfo{author}{Kirschner, D.E.},
  \bibinfo{year}{2008}.
\newblock \bibinfo{title}{A methodology for performing global uncertainty and
  sensitivity analysis in systems biology}.
\newblock \bibinfo{journal}{Journal of Theoretical Biology}
  \bibinfo{volume}{254}, \bibinfo{pages}{178--196}.
\newblock \DOIprefix\doi{10.1016/j.jtbi.2008.04.011}. \bibinfo{note}{pMID:
  18572196}.
%Type = Article
\bibitem[{McNally et~al.(2011)McNally, Cotton and Loizou}]{mcnally2011workflow}
\bibinfo{author}{McNally, K.}, \bibinfo{author}{Cotton, R.},
  \bibinfo{author}{Loizou, G.D.}, \bibinfo{year}{2011}.
\newblock \bibinfo{title}{A workflow for global sensitivity analysis of {PBPK}
  models}.
\newblock \bibinfo{journal}{Frontiers in Pharmacology} \bibinfo{volume}{2},
  \bibinfo{pages}{31}.
\newblock \DOIprefix\doi{10.3389/fphar.2011.00031}. \bibinfo{note}{pMID:
  21772819}.
%Type = Article
\bibitem[{Melillo et~al.(2019)Melillo, Aarons, Magni and
  Darwich}]{melillo2019variance}
\bibinfo{author}{Melillo, N.}, \bibinfo{author}{Aarons, L.},
  \bibinfo{author}{Magni, P.}, \bibinfo{author}{Darwich, A.S.},
  \bibinfo{year}{2019}.
\newblock \bibinfo{title}{Variance based global sensitivity analysis of
  physiologically based pharmacokinetic absorption models for {BCS} {I}--{IV}
  drugs}.
\newblock \bibinfo{journal}{Journal of Pharmacokinetics and Pharmacodynamics}
  \bibinfo{volume}{46}, \bibinfo{pages}{27--42}.
\newblock \DOIprefix\doi{10.1007/s10928-018-9615-8}. \bibinfo{note}{pMID:
  30552544}.
%Type = Article
\bibitem[{Meyer et~al.(2026)Meyer, Friede, Grieve, Jennison, Krams, Lorenzi,
  Manji, Mielke, Overbey, Viele, Walton and
  König}]{meyerDesignSimulateRefine2026}
\bibinfo{author}{Meyer, E.L.}, \bibinfo{author}{Friede, T.},
  \bibinfo{author}{Grieve, A.P.}, \bibinfo{author}{Jennison, C.},
  \bibinfo{author}{Krams, M.}, \bibinfo{author}{Lorenzi, E.},
  \bibinfo{author}{Manji, H.K.}, \bibinfo{author}{Mielke, T.},
  \bibinfo{author}{Overbey, J.R.}, \bibinfo{author}{Viele, K.},
  \bibinfo{author}{Walton, M.K.}, \bibinfo{author}{König, F.},
  \bibinfo{year}{2026}.
\newblock \bibinfo{title}{Design, simulate, refine: Simulation-guided clinical
  trials for accelerated drug development}.
\newblock \bibinfo{journal}{Nature Reviews Drug Discovery} \URLprefix
  \url{https://doi.org/10.1038/s41573-026-01481-9},
  \DOIprefix\doi{10.1038/s41573-026-01481-9}.
%Type = Article
\bibitem[{Miller et~al.(1976)Miller, Roberts and
  Fischer}]{miller1976acetaminophen}
\bibinfo{author}{Miller, R.P.}, \bibinfo{author}{Roberts, R.J.},
  \bibinfo{author}{Fischer, L.J.}, \bibinfo{year}{1976}.
\newblock \bibinfo{title}{Acetaminophen elimination kinetics in neonates,
  children, and adults}.
\newblock \bibinfo{journal}{Clinical Pharmacology and Therapeutics}
  \bibinfo{volume}{19}, \bibinfo{pages}{284--294}.
\newblock \DOIprefix\doi{10.1002/cpt1976193284}. \bibinfo{note}{pMID: 1261167.
  Neonates and young children route more of the dose to sulfate, compensating
  for immature glucuronidation}.
%Type = Article
\bibitem[{Minto et~al.(1997a)Minto, Schnider, Egan, Youngs, Lemmens, Gambus,
  Billard, Hoke, Moore, Hermann, Muir, Mandema and Shafer}]{minto1997influence}
\bibinfo{author}{Minto, C.F.}, \bibinfo{author}{Schnider, T.W.},
  \bibinfo{author}{Egan, T.D.}, \bibinfo{author}{Youngs, E.},
  \bibinfo{author}{Lemmens, H.J.M.}, \bibinfo{author}{Gambus, P.L.},
  \bibinfo{author}{Billard, V.}, \bibinfo{author}{Hoke, J.F.},
  \bibinfo{author}{Moore, K.H.P.}, \bibinfo{author}{Hermann, D.J.},
  \bibinfo{author}{Muir, K.T.}, \bibinfo{author}{Mandema, J.W.},
  \bibinfo{author}{Shafer, S.L.}, \bibinfo{year}{1997}a.
\newblock \bibinfo{title}{Influence of age and gender on the pharmacokinetics
  and pharmacodynamics of remifentanil: {I}. model development}.
\newblock \bibinfo{journal}{Anesthesiology} \bibinfo{volume}{86},
  \bibinfo{pages}{10--23}.
\newblock \DOIprefix\doi{10.1097/00000542-199701000-00004}.
  \bibinfo{note}{pMID: 9009935}.
%Type = Article
\bibitem[{Minto et~al.(1997b)Minto, Schnider and
  Shafer}]{minto1997pharmacokinetics}
\bibinfo{author}{Minto, C.F.}, \bibinfo{author}{Schnider, T.W.},
  \bibinfo{author}{Shafer, S.L.}, \bibinfo{year}{1997}b.
\newblock \bibinfo{title}{Pharmacokinetics and pharmacodynamics of
  remifentanil: {II}. model application}.
\newblock \bibinfo{journal}{Anesthesiology} \bibinfo{volume}{86},
  \bibinfo{pages}{24--33}.
\newblock \DOIprefix\doi{10.1097/00000542-199701000-00005}.
  \bibinfo{note}{pMID: 9009936}.
%Type = Article
\bibitem[{Morris(1991)}]{morris1991factorial}
\bibinfo{author}{Morris, M.D.}, \bibinfo{year}{1991}.
\newblock \bibinfo{title}{Factorial sampling plans for preliminary
  computational experiments}.
\newblock \bibinfo{journal}{Technometrics} \bibinfo{volume}{33},
  \bibinfo{pages}{161--174}.
\newblock \DOIprefix\doi{10.1080/00401706.1991.10484804}.
%Type = Article
\bibitem[{Nestorov(1999)}]{nestorov1999sensitivity}
\bibinfo{author}{Nestorov, I.A.}, \bibinfo{year}{1999}.
\newblock \bibinfo{title}{Sensitivity analysis of pharmacokinetic and
  pharmacodynamic systems: {I}. a structural approach to sensitivity analysis
  of physiologically based pharmacokinetic models}.
\newblock \bibinfo{journal}{Journal of Pharmacokinetics and Biopharmaceutics}
  \bibinfo{volume}{27}, \bibinfo{pages}{577--596}.
\newblock \DOIprefix\doi{10.1023/A:1020926525495}. \bibinfo{note}{pMID:
  11153447}.
%Type = Article
\bibitem[{Owen(1997)}]{owen1997}
\bibinfo{author}{Owen, A.B.}, \bibinfo{year}{1997}.
\newblock \bibinfo{title}{Scrambled net variance for integrals of smooth
  functions}.
\newblock \bibinfo{journal}{The Annals of Statistics} \bibinfo{volume}{25},
  \bibinfo{pages}{1541--1562}.
\newblock \DOIprefix\doi{10.1214/aos/1031594731}.
%Type = Article
\bibitem[{Owen(1998)}]{owen1998}
\bibinfo{author}{Owen, A.B.}, \bibinfo{year}{1998}.
\newblock \bibinfo{title}{Scrambling {Sobol'} and {Niederreiter--Xing} points}.
\newblock \bibinfo{journal}{Journal of Complexity} \bibinfo{volume}{14},
  \bibinfo{pages}{466--489}.
\newblock \DOIprefix\doi{10.1006/jcom.1998.0487}.
%Type = Article
\bibitem[{Owen(2013)}]{owen2013}
\bibinfo{author}{Owen, A.B.}, \bibinfo{year}{2013}.
\newblock \bibinfo{title}{Better estimation of small {Sobol'} sensitivity
  indices}.
\newblock \bibinfo{journal}{ACM Transactions on Modeling and Computer
  Simulation} \bibinfo{volume}{23}, \bibinfo{pages}{1--17}.
\newblock \DOIprefix\doi{10.1145/2457459.2457460}. \bibinfo{note}{article 11}.
%Type = Article
\bibitem[{Pearce et~al.(2017)Pearce, Setzer, Strope, Sipes and
  Wambaugh}]{pearce2017httk}
\bibinfo{author}{Pearce, R.G.}, \bibinfo{author}{Setzer, R.W.},
  \bibinfo{author}{Strope, C.L.}, \bibinfo{author}{Sipes, N.S.},
  \bibinfo{author}{Wambaugh, J.F.}, \bibinfo{year}{2017}.
\newblock \bibinfo{title}{httk: {R} package for high-throughput
  toxicokinetics}.
\newblock \bibinfo{journal}{Journal of Statistical Software}
  \bibinfo{volume}{79}, \bibinfo{pages}{1--26}.
\newblock \DOIprefix\doi{10.18637/jss.v079.i04}. \bibinfo{note}{the primary
  software citation for the httk package}.
%Type = Misc
\bibitem[{Pessina and Papathanasiou(2026)}]{pessina2026jaxgsa}
\bibinfo{author}{Pessina, D.}, \bibinfo{author}{Papathanasiou, M.M.},
  \bibinfo{year}{2026}.
\newblock \bibinfo{title}{{jaxgsa}: Global sensitivity analysis in {JAX}}.
\newblock \bibinfo{howpublished}{Computer software, Zenodo}.
\newblock \URLprefix \url{https://github.com/DanielePessina/jaxgsa},
  \DOIprefix\doi{10.5281/zenodo.22099117}.
%Type = Article
\bibitem[{Peyr{\'e} and Cuturi(2019)}]{peyre2019computational}
\bibinfo{author}{Peyr{\'e}, G.}, \bibinfo{author}{Cuturi, M.},
  \bibinfo{year}{2019}.
\newblock \bibinfo{title}{Computational optimal transport: With applications to
  data science}.
\newblock \bibinfo{journal}{Foundations and Trends in Machine Learning}
  \bibinfo{volume}{11}, \bibinfo{pages}{355--607}.
\newblock \DOIprefix\doi{10.1561/2200000073}. \bibinfo{note}{the reference
  cited by the gsaot paper for the closed-form one-dimensional empirical
  solution; a monograph edition exists at doi 10.1561/9781680835519}.
%Type = Article
\bibitem[{Pianosi and Wagener(2015)}]{pianosi2015pawn}
\bibinfo{author}{Pianosi, F.}, \bibinfo{author}{Wagener, T.},
  \bibinfo{year}{2015}.
\newblock \bibinfo{title}{A simple and efficient method for global sensitivity
  analysis based on cumulative distribution functions}.
\newblock \bibinfo{journal}{Environmental Modelling \& Software}
  \bibinfo{volume}{67}, \bibinfo{pages}{1--11}.
\newblock \DOIprefix\doi{10.1016/j.envsoft.2015.01.004}.
%Type = Book
\bibitem[{Pinheiro and Bates(2000)}]{pinheirobates2000}
\bibinfo{author}{Pinheiro, J.C.}, \bibinfo{author}{Bates, D.M.},
  \bibinfo{year}{2000}.
\newblock \bibinfo{title}{Mixed-Effects Models in {S} and {S-PLUS}}.
\newblock \bibinfo{publisher}{Springer}, \bibinfo{address}{New York}.
\newblock \DOIprefix\doi{10.1007/b98882}.
%Type = Article
\bibitem[{Plischke et~al.(2013)Plischke, Borgonovo and
  Smith}]{plischke2013givendata}
\bibinfo{author}{Plischke, E.}, \bibinfo{author}{Borgonovo, E.},
  \bibinfo{author}{Smith, C.L.}, \bibinfo{year}{2013}.
\newblock \bibinfo{title}{Global sensitivity measures from given data}.
\newblock \bibinfo{journal}{European Journal of Operational Research}
  \bibinfo{volume}{226}, \bibinfo{pages}{536--550}.
\newblock \DOIprefix\doi{10.1016/j.ejor.2012.11.047}.
%Type = Inproceedings
\bibitem[{Prokhorenkova et~al.(2018)Prokhorenkova, Gusev, Vorobev, Dorogush and
  Gulin}]{prokhorenkova2018catboost}
\bibinfo{author}{Prokhorenkova, L.}, \bibinfo{author}{Gusev, G.},
  \bibinfo{author}{Vorobev, A.}, \bibinfo{author}{Dorogush, A.V.},
  \bibinfo{author}{Gulin, A.}, \bibinfo{year}{2018}.
\newblock \bibinfo{title}{{CatBoost}: unbiased boosting with categorical
  features}, in: \bibinfo{booktitle}{Advances in Neural Information Processing
  Systems}, \bibinfo{publisher}{Curran Associates, Inc.}. pp.
  \bibinfo{pages}{6639--6649}.
\newblock \URLprefix
  \url{https://proceedings.neurips.cc/paper_files/paper/2018/hash/14491b756b3a51daac41c24863285549-Abstract.html},
  \href{http://arxiv.org/abs/1706.09516}{{\tt arXiv:1706.09516}}.
  \bibinfo{note}{introduces ordered boosting and ordered target statistics to
  remove the prediction shift caused by target leakage in classical gradient
  boosting}.
%Type = Book
\bibitem[{Rasmussen and Williams(2006)}]{rasmussen2006}
\bibinfo{author}{Rasmussen, C.E.}, \bibinfo{author}{Williams, C.K.I.},
  \bibinfo{year}{2006}.
\newblock \bibinfo{title}{Gaussian Processes for Machine Learning}.
\newblock \bibinfo{publisher}{MIT Press}, \bibinfo{address}{Cambridge, MA}.
\newblock \DOIprefix\doi{10.7551/mitpress/3206.001.0001}.
%Type = Article
\bibitem[{Ring et~al.(2017)Ring, Pearce, Setzer, Wetmore and
  Wambaugh}]{ring2017identifying}
\bibinfo{author}{Ring, C.L.}, \bibinfo{author}{Pearce, R.G.},
  \bibinfo{author}{Setzer, R.W.}, \bibinfo{author}{Wetmore, B.A.},
  \bibinfo{author}{Wambaugh, J.F.}, \bibinfo{year}{2017}.
\newblock \bibinfo{title}{Identifying populations sensitive to environmental
  chemicals by simulating toxicokinetic variability}.
\newblock \bibinfo{journal}{Environment International} \bibinfo{volume}{106},
  \bibinfo{pages}{105--118}.
\newblock \DOIprefix\doi{10.1016/j.envint.2017.06.004}. \bibinfo{note}{pMID:
  28628784. The httk-pop virtual population generator; NHANES-based,
  correlation-preserving Monte Carlo}.
%Type = Article
\bibitem[{Rougier(2008)}]{rougier2008outerproduct}
\bibinfo{author}{Rougier, J.}, \bibinfo{year}{2008}.
\newblock \bibinfo{title}{Efficient emulators for multivariate deterministic
  functions}.
\newblock \bibinfo{journal}{Journal of Computational and Graphical Statistics}
  \bibinfo{volume}{17}, \bibinfo{pages}{827--843}.
\newblock \DOIprefix\doi{10.1198/106186008X384032}.
%Type = Article
\bibitem[{Ruano et~al.(2012)Ruano, Ribes, Seco and Ferrer}]{ruano2012improved}
\bibinfo{author}{Ruano, M.V.}, \bibinfo{author}{Ribes, J.},
  \bibinfo{author}{Seco, A.}, \bibinfo{author}{Ferrer, J.},
  \bibinfo{year}{2012}.
\newblock \bibinfo{title}{An improved sampling strategy based on trajectory
  design for application of the {Morris} method to systems with many input
  factors}.
\newblock \bibinfo{journal}{Environmental Modelling \& Software}
  \bibinfo{volume}{37}, \bibinfo{pages}{103--109}.
\newblock \DOIprefix\doi{10.1016/j.envsoft.2012.03.008}.
%Type = Phdthesis
\bibitem[{Saat{\c{c}}i(2011)}]{saatci2011structured}
\bibinfo{author}{Saat{\c{c}}i, Y.}, \bibinfo{year}{2011}.
\newblock \bibinfo{title}{Scalable Inference for Structured {Gaussian} Process
  Models}.
\newblock Ph.D. thesis. University of Cambridge.
\newblock \URLprefix \url{https://mlg.eng.cam.ac.uk/pub/pdf/Saa11.pdf}.
%Type = Article
\bibitem[{Saltelli et~al.(2010)Saltelli, Annoni, Azzini, Campolongo, Ratto and
  Tarantola}]{saltelli2010}
\bibinfo{author}{Saltelli, A.}, \bibinfo{author}{Annoni, P.},
  \bibinfo{author}{Azzini, I.}, \bibinfo{author}{Campolongo, F.},
  \bibinfo{author}{Ratto, M.}, \bibinfo{author}{Tarantola, S.},
  \bibinfo{year}{2010}.
\newblock \bibinfo{title}{Variance based sensitivity analysis of model output.
  {Design} and estimator for the total sensitivity index}.
\newblock \bibinfo{journal}{Computer Physics Communications}
  \bibinfo{volume}{181}, \bibinfo{pages}{259--270}.
\newblock \DOIprefix\doi{10.1016/j.cpc.2009.09.018}.
%Type = Book
\bibitem[{Saltelli et~al.(2008)Saltelli, Ratto, Andres, Campolongo, Cariboni,
  Gatelli, Saisana and Tarantola}]{saltelli2008primer}
\bibinfo{author}{Saltelli, A.}, \bibinfo{author}{Ratto, M.},
  \bibinfo{author}{Andres, T.}, \bibinfo{author}{Campolongo, F.},
  \bibinfo{author}{Cariboni, J.}, \bibinfo{author}{Gatelli, D.},
  \bibinfo{author}{Saisana, M.}, \bibinfo{author}{Tarantola, S.},
  \bibinfo{year}{2008}.
\newblock \bibinfo{title}{Global Sensitivity Analysis: The Primer}.
\newblock \bibinfo{publisher}{Wiley}.
\newblock \DOIprefix\doi{10.1002/9780470725184}.
%Type = Article
\bibitem[{Saltelli et~al.(1999)Saltelli, Tarantola and Chan}]{saltelli1999fast}
\bibinfo{author}{Saltelli, A.}, \bibinfo{author}{Tarantola, S.},
  \bibinfo{author}{Chan, K.P.S.}, \bibinfo{year}{1999}.
\newblock \bibinfo{title}{A quantitative model-independent method for global
  sensitivity analysis of model output}.
\newblock \bibinfo{journal}{Technometrics} \bibinfo{volume}{41},
  \bibinfo{pages}{39--56}.
\newblock \DOIprefix\doi{10.1080/00401706.1999.10485594}.
%Type = Article
\bibitem[{S\'{a}nchez~Restrepo and
  Hern\'{a}ndez~Valdivieso(2022)}]{sanchezrestrepo2022global}
\bibinfo{author}{S\'{a}nchez~Restrepo, J.D.},
  \bibinfo{author}{Hern\'{a}ndez~Valdivieso, A.M.}, \bibinfo{year}{2022}.
\newblock \bibinfo{title}{Global sensitivity analysis in physiologically-based
  pharmacokinetic/pharmacodynamic models of inhaled and opioids anesthetics and
  its application to generate virtual populations}.
\newblock \bibinfo{journal}{Journal of Pharmacokinetics and Pharmacodynamics}
  \bibinfo{volume}{49}, \bibinfo{pages}{411--428}.
\newblock \DOIprefix\doi{10.1007/s10928-022-09810-2}. \bibinfo{note}{pMID:
  35616803}.
%Type = Book
\bibitem[{Santambrogio(2015)}]{santambrogio2015ot}
\bibinfo{author}{Santambrogio, F.}, \bibinfo{year}{2015}.
\newblock \bibinfo{title}{Optimal Transport for Applied Mathematicians}.
  volume~\bibinfo{volume}{87} of \textit{\bibinfo{series}{Progress in Nonlinear
  Differential Equations and Their Applications}}.
\newblock \bibinfo{publisher}{Birkh{\"a}user}, \bibinfo{address}{Cham}.
\newblock \DOIprefix\doi{10.1007/978-3-319-20828-2}. \bibinfo{note}{chapter 2
  gives the one-dimensional case: for a convex cost the optimal map is the
  monotone (quantile-matching) rearrangement}.
%Type = Article
\bibitem[{Sarrazin et~al.(2016)Sarrazin, Pianosi and Wagener}]{sarrazin2016}
\bibinfo{author}{Sarrazin, F.}, \bibinfo{author}{Pianosi, F.},
  \bibinfo{author}{Wagener, T.}, \bibinfo{year}{2016}.
\newblock \bibinfo{title}{Global sensitivity analysis of environmental models:
  convergence and validation}.
\newblock \bibinfo{journal}{Environmental Modelling \& Software}
  \bibinfo{volume}{79}, \bibinfo{pages}{135--152}.
\newblock \DOIprefix\doi{10.1016/j.envsoft.2016.02.005}.
%Type = Inproceedings
\bibitem[{Savy et~al.(2018)Savy, Savy, Andrieu and
  Marque}]{savySimulatedClinicalTrials2018}
\bibinfo{author}{Savy, N.}, \bibinfo{author}{Savy, S.},
  \bibinfo{author}{Andrieu, S.}, \bibinfo{author}{Marque, S.},
  \bibinfo{year}{2018}.
\newblock \bibinfo{title}{Simulated {{Clinical Trials}}: {{Principle}}, {{Good
  Practices}}, and {{Focus}} on {{Virtual Patients Generation}}}, in:
  \bibinfo{editor}{Pilz, J.}, \bibinfo{editor}{Rasch, D.},
  \bibinfo{editor}{Melas, V.B.}, \bibinfo{editor}{Moder, K.} (Eds.),
  \bibinfo{booktitle}{Statistics and {{Simulation}}},
  \bibinfo{publisher}{Springer International Publishing}. pp.
  \bibinfo{pages}{301--314}.
\newblock \DOIprefix\doi{10.1007/978-3-319-76035-3_21}.
%Type = Article
\bibitem[{Scannell et~al.(2012)Scannell, Blanckley, Boldon and
  Warrington}]{scannellDiagnosingDeclinePharmaceutical2012}
\bibinfo{author}{Scannell, J.W.}, \bibinfo{author}{Blanckley, A.},
  \bibinfo{author}{Boldon, H.}, \bibinfo{author}{Warrington, B.},
  \bibinfo{year}{2012}.
\newblock \bibinfo{title}{Diagnosing the decline in pharmaceutical {{R}}\&{{D}}
  efficiency}.
\newblock \bibinfo{journal}{Nature Reviews Drug Discovery}
  \bibinfo{volume}{11}, \bibinfo{pages}{191--200}.
\newblock \URLprefix \url{https://www.nature.com/articles/nrd3681},
  \DOIprefix\doi{10.1038/nrd3681}.
%Type = Article
\bibitem[{Sch{\"a}fer et~al.(2024)Sch{\"a}fer, Schiavazzi, Hellevik and
  Sturdy}]{schafer2024multifidelity}
\bibinfo{author}{Sch{\"a}fer, F.}, \bibinfo{author}{Schiavazzi, D.E.},
  \bibinfo{author}{Hellevik, L.R.}, \bibinfo{author}{Sturdy, J.},
  \bibinfo{year}{2024}.
\newblock \bibinfo{title}{Global sensitivity analysis with multifidelity {Monte
  Carlo} and polynomial chaos expansion for vascular haemodynamics}.
\newblock \bibinfo{journal}{International Journal for Numerical Methods in
  Biomedical Engineering} \bibinfo{volume}{40}, \bibinfo{pages}{e3836}.
\newblock \DOIprefix\doi{10.1002/cnm.3836}.
%Type = Article
\bibitem[{Sobol'(1967)}]{sobol1967}
\bibinfo{author}{Sobol', I.M.}, \bibinfo{year}{1967}.
\newblock \bibinfo{title}{On the distribution of points in a cube and the
  approximate evaluation of integrals}.
\newblock \bibinfo{journal}{USSR Computational Mathematics and Mathematical
  Physics} \bibinfo{volume}{7}, \bibinfo{pages}{86--112}.
\newblock \DOIprefix\doi{10.1016/0041-5553(67)90144-9}.
%Type = Article
\bibitem[{Sobol'(1993)}]{sobol1993}
\bibinfo{author}{Sobol', I.M.}, \bibinfo{year}{1993}.
\newblock \bibinfo{title}{Sensitivity estimates for nonlinear mathematical
  models}.
\newblock \bibinfo{journal}{Mathematical Modelling and Computational
  Experiments} \bibinfo{volume}{1}, \bibinfo{pages}{407--414}.
\newblock \bibinfo{note}{English translation of the 1990 Russian original;
  journal predates DOI registration}.
%Type = Article
\bibitem[{Sobol'(2001)}]{sobol2001}
\bibinfo{author}{Sobol', I.M.}, \bibinfo{year}{2001}.
\newblock \bibinfo{title}{Global sensitivity indices for nonlinear mathematical
  models and their {Monte Carlo} estimates}.
\newblock \bibinfo{journal}{Mathematics and Computers in Simulation}
  \bibinfo{volume}{55}, \bibinfo{pages}{271--280}.
\newblock \DOIprefix\doi{10.1016/S0378-4754(00)00270-6}.
%Type = Article
\bibitem[{Sobol' and Kucherenko(2009)}]{sobol2009derivative}
\bibinfo{author}{Sobol', I.M.}, \bibinfo{author}{Kucherenko, S.},
  \bibinfo{year}{2009}.
\newblock \bibinfo{title}{Derivative based global sensitivity measures and
  their link with global sensitivity indices}.
\newblock \bibinfo{journal}{Mathematics and Computers in Simulation}
  \bibinfo{volume}{79}, \bibinfo{pages}{3009--3017}.
\newblock \DOIprefix\doi{10.1016/j.matcom.2009.01.023}.
%Type = Article
\bibitem[{Strong and Oakley(2013)}]{strong2013evppi}
\bibinfo{author}{Strong, M.}, \bibinfo{author}{Oakley, J.E.},
  \bibinfo{year}{2013}.
\newblock \bibinfo{title}{An efficient method for computing single-parameter
  partial expected value of perfect information}.
\newblock \bibinfo{journal}{Medical Decision Making} \bibinfo{volume}{33},
  \bibinfo{pages}{755--766}.
\newblock \DOIprefix\doi{10.1177/0272989X12465123}. \bibinfo{note}{online-first
  December 2012; issue of record August 2013. Cited by the gsaot paper for the
  plateau behaviour of given-data estimates with respect to the number of
  conditioning classes}.
%Type = Article
\bibitem[{Sudret(2008)}]{sudret2008global}
\bibinfo{author}{Sudret, B.}, \bibinfo{year}{2008}.
\newblock \bibinfo{title}{Global sensitivity analysis using polynomial chaos
  expansions}.
\newblock \bibinfo{journal}{Reliability Engineering \& System Safety}
  \bibinfo{volume}{93}, \bibinfo{pages}{964--979}.
\newblock \DOIprefix\doi{10.1016/j.ress.2007.04.002}.
%Type = Book
\bibitem[{Villani(2009)}]{villani2009optimal}
\bibinfo{author}{Villani, C.}, \bibinfo{year}{2009}.
\newblock \bibinfo{title}{Optimal Transport: Old and New}. volume
  \bibinfo{volume}{338} of \textit{\bibinfo{series}{Grundlehren der
  mathematischen Wissenschaften}}.
\newblock \bibinfo{publisher}{Springer}, \bibinfo{address}{Berlin, Heidelberg}.
\newblock \DOIprefix\doi{10.1007/978-3-540-71050-9}.
%Type = Article
\bibitem[{Wagner(1995)}]{wagner1995global}
\bibinfo{author}{Wagner, H.M.}, \bibinfo{year}{1995}.
\newblock \bibinfo{title}{Global sensitivity analysis}.
\newblock \bibinfo{journal}{Operations Research} \bibinfo{volume}{43},
  \bibinfo{pages}{948--969}.
\newblock \DOIprefix\doi{10.1287/opre.43.6.948}. \bibinfo{note}{the univariate
  variance-based measure that the advective part of the optimal-transport index
  reproduces}.
%Type = Article
\bibitem[{Wambaugh et~al.(2015)Wambaugh, Wetmore, Pearce, Strope, Goldsmith,
  Sluka, Sedykh, Tropsha, Bosgra, Shah, Judson, Thomas and
  Setzer}]{wambaugh2015toxicokinetic}
\bibinfo{author}{Wambaugh, J.F.}, \bibinfo{author}{Wetmore, B.A.},
  \bibinfo{author}{Pearce, R.}, \bibinfo{author}{Strope, C.},
  \bibinfo{author}{Goldsmith, R.}, \bibinfo{author}{Sluka, J.P.},
  \bibinfo{author}{Sedykh, A.}, \bibinfo{author}{Tropsha, A.},
  \bibinfo{author}{Bosgra, S.}, \bibinfo{author}{Shah, I.},
  \bibinfo{author}{Judson, R.}, \bibinfo{author}{Thomas, R.S.},
  \bibinfo{author}{Setzer, R.W.}, \bibinfo{year}{2015}.
\newblock \bibinfo{title}{Toxicokinetic triage for environmental chemicals}.
\newblock \bibinfo{journal}{Toxicological Sciences} \bibinfo{volume}{147},
  \bibinfo{pages}{55--67}.
\newblock \DOIprefix\doi{10.1093/toxsci/kfv118}.
%Type = Article
\bibitem[{Wilding et~al.(2021)Wilding, Batterham, Calanna, Davies, Van~Gaal,
  Lingvay, McGowan, Rosenstock, Tran, Wadden, Wharton, Yokote, Zeuthen and
  Kushner}]{wilding2021onceweekly}
\bibinfo{author}{Wilding, J.P.H.}, \bibinfo{author}{Batterham, R.L.},
  \bibinfo{author}{Calanna, S.}, \bibinfo{author}{Davies, M.},
  \bibinfo{author}{Van~Gaal, L.F.}, \bibinfo{author}{Lingvay, I.},
  \bibinfo{author}{McGowan, B.M.}, \bibinfo{author}{Rosenstock, J.},
  \bibinfo{author}{Tran, M.T.D.}, \bibinfo{author}{Wadden, T.A.},
  \bibinfo{author}{Wharton, S.}, \bibinfo{author}{Yokote, K.},
  \bibinfo{author}{Zeuthen, N.}, \bibinfo{author}{Kushner, R.F.},
  \bibinfo{year}{2021}.
\newblock \bibinfo{title}{Once-weekly semaglutide in adults with overweight or
  obesity}.
\newblock \bibinfo{journal}{New England Journal of Medicine}
  \bibinfo{volume}{384}, \bibinfo{pages}{989--1002}.
\newblock \DOIprefix\doi{10.1056/NEJMoa2032183}. \bibinfo{note}{sTEP 1 trial.
  PMID: 33567185}.
%Type = Article
\bibitem[{Xiu and Karniadakis(2002)}]{xiu2002}
\bibinfo{author}{Xiu, D.}, \bibinfo{author}{Karniadakis, G.E.},
  \bibinfo{year}{2002}.
\newblock \bibinfo{title}{The {Wiener--Askey} polynomial chaos for stochastic
  differential equations}.
\newblock \bibinfo{journal}{SIAM Journal on Scientific Computing}
  \bibinfo{volume}{24}, \bibinfo{pages}{619--644}.
\newblock \DOIprefix\doi{10.1137/S1064827501387826}.
%Type = Article
\bibitem[{Zhang et~al.(2015)Zhang, Trame, Lesko and Schmidt}]{zhang2015sobol}
\bibinfo{author}{Zhang, X.Y.}, \bibinfo{author}{Trame, M.N.},
  \bibinfo{author}{Lesko, L.J.}, \bibinfo{author}{Schmidt, S.},
  \bibinfo{year}{2015}.
\newblock \bibinfo{title}{{Sobol} sensitivity analysis: A tool to guide the
  development and evaluation of systems pharmacology models}.
\newblock \bibinfo{journal}{CPT: Pharmacometrics \& Systems Pharmacology}
  \bibinfo{volume}{4}, \bibinfo{pages}{69--79}.
\newblock \DOIprefix\doi{10.1002/psp4.6}. \bibinfo{note}{pMID: 27548289}.
%Type = Article
\bibitem[{Zurlinden and Reisfeld(2016)}]{zurlinden2016physiologically}
\bibinfo{author}{Zurlinden, T.J.}, \bibinfo{author}{Reisfeld, B.},
  \bibinfo{year}{2016}.
\newblock \bibinfo{title}{Physiologically based modeling of the
  pharmacokinetics of acetaminophen and its major metabolites in humans using a
  {Bayesian} population approach}.
\newblock \bibinfo{journal}{European Journal of Drug Metabolism and
  Pharmacokinetics} \bibinfo{volume}{41}, \bibinfo{pages}{267--280}.
\newblock \DOIprefix\doi{10.1007/s13318-015-0253-x}. \bibinfo{note}{pMID:
  25636597. THE source model behind the pksensi acetaminophen example}.

\end{thebibliography}

\clearpage
\setcounter{section}{0}\setcounter{figure}{0}\setcounter{table}{0}\setcounter{equation}{0}
\renewcommand{\thesection}{S\arabic{section}}
\renewcommand{\thesubsection}{S\arabic{section}.\arabic{subsection}}
\renewcommand{\thefigure}{S\arabic{figure}}
\renewcommand{\thetable}{S\arabic{table}}
\renewcommand{\theequation}{S\arabic{equation}}
\renewcommand{\theHsection}{S\arabic{section}}
\renewcommand{\theHsubsection}{S\arabic{section}.\arabic{subsection}}
\renewcommand{\theHfigure}{S\arabic{figure}}
\renewcommand{\theHtable}{S\arabic{table}}
\renewcommand{\theHequation}{S\arabic{equation}}
\begin{center}
{\LARGE\bfseries Supplementary material}
\end{center}
\vspace{1em}
\section{Nomenclature}
\label{app:nomenclature}

Inside the defining equations of \cref{sec:methods-gsa} the sensitivity indices are written in per-input form, $\Si$ and $\STi$; everywhere else they are written in order-labelled form, $\Sone$ and $\ST$.
The two forms denote the same quantities.

\begin{longtable}{@{}p{0.21\textwidth}p{0.71\textwidth}@{}}
\caption{Model, output and sensitivity index symbols.}\label{tab:nom-indices}                                                          \\
\toprule
Symbol                                      & Meaning                                                                                  \\
\midrule
\endfirsthead
\caption[]{\emph{continued}}                                                                                                           \\
\toprule
Symbol                                      & Meaning                                                                                  \\
\midrule
\endhead
\bottomrule
\endfoot
$f$                                         & the mechanistic model map, $Y = f(X_1,\dots,X_D)$                                        \\
$Y$                                         & the scalar model output being analysed, assumed square-integrable                        \\
$X_i$                                       & the $i$th model input, treated as a random variable                                      \\
$\bm{X}_{\sim i}$                           & every input except $X_i$                                                                 \\
$D$                                         & the number of free inputs of a design, that is the input dimension after any pinned axis
has been removed                                                                                                                       \\
$\E$, $\Var$                                & expectation and variance                                                                 \\
$\Si$, $\Sone$                              & first-order index of an input: the share of output variance removed by learning
that input alone                                                                                                                       \\
$\STi$, $\ST$                               & total-effect index of an input: the share of output variance remaining when every
other input is fixed                                                                                                                   \\
$\ST - \Sone$                               & interaction gap, the part of an input's influence that acts only through
interaction, and zero for an additive model                                                                                            \\
$\mathrm{EE}_i(\bm{x})$                     & elementary effect of input $i$ at the point $\bm{x}$, a finite-difference
slope over the step $\Delta_i$                                                                                                         \\
$\bm{x}$, $\bm{e}_i$                        & a point in the unit-cube input space, and the unit vector along
coordinate $i$                                                                                                                         \\
$\Delta_i$                                  & the perturbation step taken in coordinate $i$                                            \\
$\mustar_i$                                 & mean absolute elementary effect, the Morris screening importance measure; it carries the
units of the output and is not a variance share                                                                                        \\
$\mu_i$                                     & signed mean elementary effect, giving the average direction of an effect                 \\
$\sigma_i$                                  & standard deviation of the elementary effects of input $i$, which flags non-linearity
or interaction without separating them                                                                                                 \\
$\mathcal{L}(Y)$, $\mathcal{L}(Y \mid X_i)$ & the unconditional and conditional laws of the output                                     \\
$W_2^{2}$                                   & squared $2$-Wasserstein distance between two output distributions                        \\
$\mathrm{OT}_i$                             & optimal transport index of input $i$, the expected squared $2$-Wasserstein
distance between the conditional and the unconditional output law, normalised to $[0,1]$                                               \\
\end{longtable}

\begin{longtable}{@{}p{0.21\textwidth}p{0.71\textwidth}@{}}
\caption{Design, cost and budget symbols.
Every cost is quoted in mechanistic model evaluations.
}\label{tab:nom-cost}                                                                  \\
\toprule
Symbol                  & Meaning                                                                                      \\
\midrule
\endfirsthead
\caption[]{\emph{continued}}                                                                                           \\
\toprule
Symbol                  & Meaning                                                                                      \\
\midrule
\endhead
\bottomrule
\endfoot
$A$, $B$                & the sample and re-sample matrices of the hybrid-matrix design, $N$ rows each                 \\
$AB_i$                  & the hybrid matrix taking column $i$ from $B$ and every other column from $A$                 \\
$N$                     & base sample size: rows per point set in a variance-based design, and the whole sample for an
optimal transport analysis                                                                                             \\
$N'$                    & the length of a prefix of the scrambled sequence, itself a valid design of smaller size      \\
$r$                     & the number of radial Morris trajectories                                                     \\
$N(D+2)$, $r(D+1)$, $N$ & the evaluation cost of the variance-based, Morris screening and optimal
transport designs respectively, from
\cref{eq:cost-sobol,eq:cost-morris,eq:cost-ot}                                                                         \\
$N_\text{env}$          & the one-time cost of training a surrogate on the envelope corpus to a given tier             \\
$N_\text{direct}$       & the recurring cost of one direct mechanistic analysis of one cohort, which for
the variance-based method is $N(D+2)$ at the converged sample size                                                     \\
$K$                     & the number of cohorts analysed                                                               \\
$K^{*}$                 & the break-even cohort count $N_\text{env} / N_\text{direct}$, above which the surrogate
route is the cheaper one                                                                                               \\
$n$, $T$                & Gaussian process training-set size and number of trajectory timepoints, appearing only
in its training cost                                                                                                   \\
\end{longtable}

\begin{longtable}{@{}p{0.21\textwidth}p{0.71\textwidth}@{}}
\caption{Scoring and convergence symbols and thresholds.
All four numeric thresholds are fixed for the whole study.
}\label{tab:nom-gate}                                                                              \\
\toprule
Symbol                          & Meaning                                                                                              \\
\midrule
\endfirsthead
\caption[]{\emph{continued}}                                                                                                           \\
\toprule
Symbol                          & Meaning                                                                                              \\
\midrule
\endhead
\bottomrule
\endfoot
$\bm{s}$                        & the converged solution index vector: the element-wise median over the twelve replicate seeds at base
sample 8192, computed for the same method as the estimate it scores                                                                    \\
$\hat{\bm{s}}$, $\hat{s}_i$     & the estimated index vector at a given budget, and one of its entries                                 \\
$\varepsilon_\text{abs} = 0.01$ & the absolute floor of the mixed convergence tolerance,
\cref{eq:tolerance}                                                                                                                    \\
$\varepsilon_\text{rel} = 0.10$ & the relative term of the same tolerance                                                              \\
$0.05$                          & the importance threshold defining the important set, taken on $\ST$ for the variance-based
method and relative to the converged solution's own maximum for Morris screening and optimal transport                                \\
$0.90$                          & the minimum Spearman rank correlation over the important set when it has at least three axes;
smaller sets use exact membership and leading-axis agreement                                                                         \\
$0.95$                          & the minimum recall of the unimportant set, so that no inert axis is promoted                         \\
$\mathrm{NRMSE}$                & the held-out fit score of \cref{eq:nrmse}, the residual and total sums of squares pooled
over the horizon and square-rooted, so that $1$ is a surrogate no better than the cohort mean                                          \\
\end{longtable}

\begin{longtable}{@{}p{0.21\textwidth}p{0.71\textwidth}@{}}
\caption{Abbreviations.
Surrogates are named by their display name in the results and by their implementation name in \cref{sec:surrogates}; both are given here.
}\label{tab:nom-abbrev}    \\
\toprule
Abbreviation                                             & Meaning                                                                          \\
\midrule
\endfirsthead
\caption[]{\emph{continued}}                                                                                                                \\
\toprule
Abbreviation                                             & Meaning                                                                          \\
\midrule
\endhead
\bottomrule
\endfoot
GSA                                                      & global sensitivity analysis                                                      \\
PK                                                       & pharmacokinetic                                                                  \\
PBPK                                                     & physiologically based pharmacokinetic                                            \\
GLP-1RA                                                  & glucagon-like peptide-1 receptor agonist                                         \\
NAPQI                                                    & \emph{N}-acetyl-\emph{p}-benzoquinone imine, the reactive paracetamol metabolite \\
NRMSE                                                    & envelope-normalised root-mean-square error                                       \\
PCE (\texttt{pce})                                       & polynomial chaos expansion                                                       \\
RS-HDMR (\texttt{hdmr})                                  & random-sampling high-dimensional model representation                            \\
GP (\texttt{gp})                                         & Gaussian process                                                                 \\
XGBoost (\texttt{xgboost}), CatBoost (\texttt{catboost}) & the two gradient-boosted-tree
surrogates                                                                                                                                  \\
\end{longtable}

\section{Sensitivity analysis methods in detail}
\label{app:gsa-methods}

This appendix explains each estimator method uses for the three GSA methods.
\subsection{The variance decomposition and the hybrid-matrix estimator}
\label{app:gsa-sobol}

Let $Y = f(X_1,\dots,X_D)$ be a square-integrable output of $D$ mutually independent inputs on the unit hypercube.
The function admits a unique decomposition into a mean, main-effect terms and interaction terms of increasing order, all mutually orthogonal, so that the output variance decomposes additively into partial variances \citep{sobol1993,sobol2001}.
This is the functional analysis-of-variance identity, and it is what gives the indices of \cref{eq:s1,eq:st} their interpretation as a variance budget.
The two answer different questions: the first-order index answers what an input explains on its own, and the total-effect index answers what would still be unknown after fixing everything else.

The total-effect index has the equivalent form of \cref{eq:st-alt}, which makes the complement structure explicit \citep{homma1996}.
\begin{equation}
\STi = 1 - \frac{\Var\!\left(\E\!\left[Y \mid \bm{X}_{\sim i}\right]\right)}{\Var(Y)},
\label{eq:st-alt}
\end{equation}

Both indices are estimated here on a hybrid-matrix design.
Two independent point sets $A$ and $B$ of $N$ rows each are drawn, the sample and the re-sample matrix, and for each input $i$ a hybrid matrix $AB_i$ is built by taking column $i$ from $B$ and every other column from $A$.
Comparing $A$ with $AB_i$ isolates the effect of input $i$, because the two sets differ in that coordinate alone.
First-order indices use the Saltelli estimator and total-effect indices the Jansen estimator \citep{saltelli2010,jansen1999}, giving the combined cost of \cref{eq:cost-sobol}.

Points are drawn from Owen-scrambled Sobol sequences, that is by randomised quasi-Monte Carlo, which improves the convergence rate over plain Monte Carlo while retaining an unbiased estimator and a usable error estimate across independent scrambles \citep{owen1997,owen1998,sobol1967}.
Every $N$ in the resulting convergence sweep gives a valid design in its own right rather than a subsample of a larger one.

\subsection{Radial Morris screening designs}
\label{app:gsa-morris}

Plotting $\mustar$ against $\sigma$ separates inputs into three groups: negligible; important and close to linear; and important with a state-dependent effect \citep{campolongo2011unified}.
The elementary effects of \cref{eq:ee} are computed here on radial star trajectories built from scrambled Sobol point pairs rather than on the original winding trajectories through a discretised grid \citep{ruano2012improved}.
A radial design draws a base point and a partner point, then perturbs the base towards the partner one coordinate at a time, giving the cost of \cref{eq:cost-morris}.
Effects are taken in unit-cube coordinates, which makes $\mustar$ comparable across inputs carrying different physical units; it does not make a truncated-Gaussian input uniform, which is the substance of the caveat in \cref{sec:morris}.

\subsection{The optimal transport index and its decomposition}
\label{app:gsa-ot}

The distance used to detect a change in the output distribution is the Wasserstein distance, which measures the work needed to transport one distribution onto another \citep{villani2009optimal,santambrogio2015ot}.
Informally, it is the mass moved multiplied by the distance moved, minimised over all ways of moving it.
For scalar outputs the optimal plan matches quantiles, so sorting both samples and pairing them in order is sufficient \citep{peyre2019computational}, which is the closed form of \cref{eq:w2} behind the index of \cref{eq:ot}.

Twenty-five equal-frequency rank classes are used for the conditioning.
The guidance is that a class should hold at least on the order of a hundred points before its conditional distribution is resolved \citep{strong2013evppi}, which places twenty-five classes inside the recommended regime for the sample sizes of \cref{sec:designs}.

Under the decomposition of \citet{borgonovo2025ot} the index separates into an advective term, a diffusive term and a non-negative remainder.
The advective and diffusive terms together form the Wasserstein--Bures index, which equals the whole index only when the conditional and unconditional laws are elliptical with a common characteristic generator.

The advective term carries a further condition that governs how it should be read.
In general it equals a sum of univariate variance-based measures in the sense of \citet{wagner1995global}.
Only when the inputs are independent does it equal exactly one half of the given-data first-order index, generalising the multivariate results of \citet{lamboni2011multivariate} and \citet{gamboa2014multivariate}.
Where cohort priors are correlated this identity no longer holds exactly, and the comparison in the main text is reported accordingly.

\section{Retained sample when an envelope draw is reweighted to a cohort}
\label{app:reweighting}

\cref{sec:introduction} rules out reweighting a single envelope sample to serve every cohort.
The per-cohort figures behind that are given here.

Importance sampling weights each envelope draw by $w(\bm{x}) = p_\text{c}(\bm{x}) / p_\text{env}(\bm{x})$, and the usable size of the weighted sample is its effective sample size $\mathrm{ESS} = (\sum_j w_j)^2 / \sum_j w_j^2$, whose expected value relative to the drawn size $n$ is given by \cref{eq:ess}.
\begin{equation}
\frac{\mathrm{ESS}}{n} \;=\; \frac{\left(\E_\text{env}[w]\right)^2}{\E_\text{env}[w^2]} ,
\label{eq:ess}
\end{equation}
Every axis is sampled independently, so $w$ factorises and \cref{eq:ess} becomes a product of per-axis factors, each at most one: an axis the cohort leaves alone contributes one, a narrowed uniform axis occupying a fraction $f$ of the envelope range contributes exactly $f$, and a truncated Gaussian axis with density $p_\text{G}$ on a range of width $L$ contributes $(L \int p_\text{G}^2)^{-1}$.
\cref{tab:reweighting} evaluates the product for the nine cohorts on which the estimator is defined at all.
For the eight cohorts that pin an axis it does not exist, the pinned value being a set of measure zero under the envelope's uniform marginal, so no density ratio represents it at any sample size.

\begin{table}[htbp]
\centering
\caption{Geometric support volume and retained fraction $\mathrm{ESS}/n$ when one envelope sample is reweighted to a cohort, for
the nine cohorts that override at least one axis without pinning any.
The support volume $v_c$ is the cohort support divided by the full envelope support, so it is the product of the per-axis width fractions; it is one for the two Gaussian cohorts because their truncation leaves the full support unchanged.
The last column is the envelope sample needed to leave an effective 1024 points, against the 1024 evaluations a direct design of that size costs outright.
}
\label{tab:reweighting}
\footnotesize
\setlength{\tabcolsep}{5pt}
\begin{tabular}{@{}llccrrr@{}}
\toprule
Model        & Cohort           & Narrowed & Gaussian & $v_c$ & $\mathrm{ESS}/n$    & Points for 1024 \\
\midrule
Remifentanil & Gauss            & 0        & 5        & 1     & 0.0262              & 39,132          \\
\addlinespace
GLP-1RA      & Gauss + obese    & 0        & 9        & 1     & $4.38\times10^{-4}$ & 2,339,648       \\
             & Strong lifestyle & 1        & 0        & $1/3$ & 0.3333              & 3,073           \\
             & Weak lifestyle   & 1        & 0        & $1/3$ & 0.3333              & 3,072           \\
\addlinespace
Ibuprofen    & Elderly          & 2        & 0        & $1/9$ & 0.1111              & 9,216           \\
             & Hepatic impair.  & 2        & 0        & $1/9$ & 0.1111              & 9,216           \\
             & Enzyme induct.   & 1        & 0        & $1/3$ & 0.3333              & 3,073           \\
             & Hypoalbumin.     & 1        & 0        & $1/3$ & 0.3333              & 3,072           \\
\addlinespace
Paracetamol  & Paediatric conj. & 2        & 0        & $1/9$ & 0.1111              & 9,216           \\
\bottomrule
\end{tabular}
\end{table}

The median retained fraction over those nine is $0.1111$, so a typical cohort would need 9216 envelope evaluations to attain the effective sample size a direct design attains with 1024, roughly nine times the cost of the route it was meant to replace.
For the uniform narrowed cohorts, $v_c=\mathrm{ESS}/n$ because the envelope density is constant.
The two Gaussian cohorts have $v_c=1$ but retain only $0.0262$ and $4.38\times10^{-4}$ of the envelope sample in effective size, the latter being 2.34 million envelope evaluations for one cohort analysis.

\section{Input distributions and their provenance}
\label{app:provenance}

This appendix separates the source of an input envelope from the construction of a cohort.
The envelope is one wide box over which the model is sampled, while a cohort is a restriction of that box that asks a population-specific question.
The two are related, but they are not the same kind of evidence: an envelope may be fitted or population-derived, whereas a cohort may be a deliberately constructed simulation scenario.

The four models have $54$ sampled axes in total: $D=7$ for remifentanil, $D=9$ for the GLP-1RA model, $D=15$ for ibuprofen and $D=23$ for paracetamol.
Every envelope marginal is uniform and every axis is sampled independently, because the envelope must cover the corners that any later cohort can reach and the variance-based analysis assumes independent inputs.
The paper reports $21$ cohorts, five, seven, five and four in model order.
The tables below describe that reported set; additional variants retained in the analysis registry are not included in the reported aggregates.

\begin{table}[htbp]
\centering
\caption{Source and role of the four input envelopes.}
\label{tab:prov-sources}
\footnotesize
\setlength{\tabcolsep}{4pt}
\begin{tabular}{@{}p{0.16\textwidth}p{0.37\textwidth}p{0.37\textwidth}@{}}
\toprule
Model & Source of the envelope & Cohort construction \\
\midrule
Remifentanil, $D=7$ &
  65-patient \texttt{nlme} data set; mixed-effects fit for the kinetic uncertainty; observed male covariates and a robust dose band &
  Dose pins, impairment and weight sub-bands, plus one truncated-Gaussian population variant \\
GLP-1RA, $D=9$ &
  Bosch Model D parameter estimates and relative standard errors; a synthetic Phase-2 obesity specification for the covariates; clinical dose regimens &
  Lifestyle, weight, response and activity restrictions, plus one shifted truncated-Gaussian variant \\
Ibuprofen, $D=15$ &
  $2.5$ to $97.5$ percentile ranges from a 3000-individual \texttt{httkpop} virtual population &
  Marginal sub-bands for renal, perfusion, clearance and binding contrasts; the population correlations are not sampled \\
Paracetamol, $D=23$ &
  Zurlinden--Reisfeld model through \texttt{pksensi}; the package sensitivity-analysis ranges in log units; asserted dose and adult-weight boxes &
  Therapeutic and overdose dose pins, and a constructed paediatric conjugation contrast \\
\bottomrule
\end{tabular}
\end{table}

\subsection{How a cohort is constructed}
\label{app:provenance-cohort-rule}

The reference cohort inherits the whole envelope.
A narrowed cohort replaces one or more uniform marginals with the lower or upper third of the corresponding envelope range; a sixth is used where a more severe ibuprofen contrast is needed.
For paracetamol, whose axes are already logarithmic, the fraction is taken in log units.
These fractions are design rules, not quantiles estimated from a patient population and not clinical cut-points.

A pin holds an input at one value and removes it from the sensitivity design, so its index is zero.
The remifentanil infusion rate of \SI{215.1}{\micro\gram\per\minute} and duration of \SI{10}{\minute} are the medians of the 65-patient data set.
The paracetamol dose pins of \SI{15}{\milli\gram\per\kilo\gram}, \SI{75}{\milli\gram\per\kilo\gram} and \SI{210}{\milli\gram\per\kilo\gram} span therapeutic to acute-overdose use, and \SI{70}{\kilo\gram} is the standard-adult value used in the unreported overdose variant.
The GLP-1RA lifestyle pin at $0.35$ is the midpoint of its envelope and is a mechanistic design choice rather than an observed patient value.

The two reported Gaussian variants replace the uniform envelope marginals with truncated Normals over the same support.
For remifentanil, the standard deviations are the fitted standard error for $k_{PT}$, the dispersion of the 38 male donors for weight and height, and an asserted $25\%$ coefficient of variation for the two dose axes.
For the GLP-1RA variant, the five fitted axes use their reported standard errors, the three covariates use the synthetic Phase-2 dispersions, and dose is centred on the maintenance dose with a standard deviation of one third of the titration band.
The shifted means used to represent high weight have no external source; they define the scenario being tested.

No cohort narrowing should be read as a validated clinical definition.
The results show how sensitivity changes under these transparent restrictions, while the underlying envelope sources determine the scale on which those restrictions operate.

\subsection{Remifentanil, $D = 7$}
\label{app:provenance-remi}

The remifentanil envelope is the only one fitted specifically for this study.
The source is the 65-patient \texttt{Remifentanil} data set distributed with the \texttt{nlme} package \citep{pinheirobates2000}, which records 1992 pooled plasma-concentration observations, patient covariates and infusion dosing for the population described by \citet{minto1997influence,minto1997pharmacokinetics}.
The forward model is the physiologically based model of \citet{abbiati2016definition}.

A first-order Laplace non-linear mixed-effects fit places population fixed effects on the eight log kinetic parameters, diagonal lognormal random effects on $k_{TP}$, $k_{PHP}$, $k_{HPP}$ and $k_{EL,Pl}$, and a proportional residual error on the concentrations.
The residual is $\sigma=0.167$, while the between-subject standard deviations are between $0.32$ and $0.50$.
Standard errors are asymptotic, obtained from the inverse conditional $8\times8$ fixed-effect Hessian with the variance components held at their estimates; the full Hessian is ill-conditioned.

An axis is fixed at its fitted value when its relative standard error exceeds $50\%$, when it rails against a bound, or when its conditional variance is non-positive.
Three kinetic parameters survive this rule, but only $k_{PT}$ is an envelope axis: $k_{TP}$ and $k_{PHP}$ have no cohort interpretation here.
The remaining five kinetic parameters are held at the population estimates below.

\begin{table}[htbp]
\centering
\caption{Remifentanil mixed-effects fit. The five axes marked \emph{fix} are held at the population estimate; the three marked \emph{sample} are the identifiable fixed effects.}
\label{tab:prov-remi-fit}
\small
\begin{tabular}{@{}lllll@{}}
\toprule
Parameter & Estimate & \%RSE & $90\%$ interval & Treatment \\
\midrule
$k_{TP}$             & 0.02028   & 5.3      & [0.01859, 0.02212] & sample \\
$k_{PT}$             & 0.09534   & 2.4      & [0.09171, 0.09911] & sample \\
$k_{PHP}$            & 1.8607    & 0.8      & [1.8361, 1.8855]   & sample \\
$k_{HPP}$            & 0.19474   & $\sim 0$ & near-point          & fix \\
$k_{EL,Pl}$          & 1.3582    & $\sim 0$ & flat                & fix \\
$\mathit{Eff}_{kid}$ & 1.0000    & rails    & at the bound 1.0    & fix \\
$\mathit{Eff}_{hep}$ & 0.93976   & $\sim 0$ & near-point          & fix \\
$k_{EL,Tis}$         & 6.846e-05 & 152      & non-identifiable    & fix \\
\bottomrule
\end{tabular}
\end{table}

The seven sampled axes are listed in \cref{tab:prov-remi-axes}.
Sex is fixed to male because it determines the blood-flow split, body-surface area is derived from weight and height by the Mosteller formula, and age is carried for provenance only because the forward model does not read it.

\begin{table}[htbp]
\centering
\caption{Remifentanil envelope. All marginals are uniform and independent.}
\label{tab:prov-remi-axes}
\small
\begin{tabular}{@{}>{\raggedright\arraybackslash}p{0.22\textwidth}p{0.12\textwidth}p{0.19\textwidth}>{\raggedright\arraybackslash}p{0.35\textwidth}@{}}
\toprule
Axis & Nominal & Envelope range & Source of the range \\
\midrule
Plasma-to-tissue rate  & 0.09534 & 0.048 -- 0.09917 & upper $90\%$ limit from the fit; lower poor-transfer floor asserted \\
Renal clearance eff.   & 1.00    & 0.30 -- 1.00     & unimpaired upper value; impaired lower floor asserted \\
Hepatic clearance eff. & 0.940   & 0.30 -- 1.00     & unimpaired upper value; impaired lower floor asserted \\
Body weight            & 81.4    & 63 -- 106        & observed range of the 38 male patients \\
Height                 & 178.4   & 160 -- 193       & observed range of the 38 male patients \\
Infusion rate          & 215.1   & 100 -- 470       & outward-rounded $2.5$ to $97.5$ percentile band \\
Infusion duration      & 10      & 5 -- 20          & outward-rounded $2.5$ to $97.5$ percentile band \\
\bottomrule
\end{tabular}
\end{table}

\begin{table}[htbp]
\centering
\caption{Remifentanil cohorts used in the reported results. A pinned dose is the protocol value above. The \texttt{reference\_gauss} cohort uses truncated Normals on $k_{PT}$, weight, height, infusion rate and infusion duration, while the two efficiency axes retain their uniform marginals.}
\label{tab:prov-remi-cohorts}
\small
\begin{tabular}{@{}>{\raggedright\arraybackslash}p{0.22\textwidth}>{\raggedright\arraybackslash}p{0.48\textwidth}>{\raggedright\arraybackslash}p{0.18\textwidth}@{}}
\toprule
Cohort & Restriction relative to the envelope & Type \\
\midrule
Envelope       & none & whole envelope \\
Fixed protocol & infusion rate $215.1$; duration $10$ & pins \\
Renal impair.  & dose pinned; renal clearance efficiency $0.300$ -- $0.533$ & pin and lower third \\
High weight    & dose pinned; body weight $91.7$ -- $106$ & pin and upper third \\
\texttt{reference\_gauss} & fitted/donor Gaussian marginals, truncated to the envelope & Gaussian \\
\bottomrule
\end{tabular}
\end{table}

The registry also retains hepatic impairment and two shifted Gaussian variants, but they are not part of the five-cohort summaries reported in the paper.

\subsection{GLP-1 receptor agonist, semaglutide, $D = 9$}
\label{app:provenance-glp}

The GLP-1RA model is the semaglutide implementation of \citet{bosch2024quantification}, coupled to the body-composition model of \citet{hall2011quantification}.
The point values are the reference implementation's defaults, while the relative standard errors for the fitted axes are taken from Table~3, Model~D of \citet{bosch2024quantification}.
The five mechanistic axes in the envelope are $T50$, $SSEC50$, GLP potency, lifestyle-effect decay and the lifestyle amplitude.
The three sampled covariates are baseline weight, height and baseline physical activity; sex and age are fixed.

The covariate moments come from the Phase-2 obesity-candidate specification in the companion GLP analysis, not from a named clinical data set.
It specifies $\mathrm{BMI}_0\sim N(35,4^2)$, height $\sim N(1.72,0.08^2)$ and $\mathrm{PAL}_{init}\sim\mathrm{TN}(1.55,0.20^2)$, with body weight derived from $\mathrm{BMI}_0\,\mathrm{height}^2$ and then sampled directly in the present design.
The envelope extends body weight to \SI{175}{\kilo\gram} so that the high-weight cohort remains inside the box, and it raises the activity floor to $1.45$ because the energy-balance solve becomes ill-posed below about $1.4$ under a strong drug effect.

The dose axis is the semaglutide titration-to-maintenance range, $0.25$ to \SI{2.4}{\milli\gram}, with weekly dosing fixed per compound \citep{bosch2024quantification,wilding2021onceweekly}.
The lifestyle amplitude uses the absolute support $[0,0.70]$ so that the weak and strong lifestyle scenarios are contained in the envelope.
The GLP potency upper limit is doubled for the poor-responder scenario, while the confidence interval for lifestyle decay is replaced by a $\pm40\%$ band because it is too narrow to create a useful contrast.

\begin{table}[htbp]
\centering
\caption{GLP-1RA envelope, $D=9$. All marginals are uniform and independent. The relative standard errors for the fitted axes are from Table~3, Model~D of \citet{bosch2024quantification}; the covariate and dose ranges are design specifications.}
\label{tab:prov-glp-axes}
\footnotesize
\setlength{\tabcolsep}{3pt}
\begin{tabular}{@{}>{\raggedright\arraybackslash}p{0.29\textwidth}p{0.13\textwidth}p{0.20\textwidth}>{\raggedright\arraybackslash}p{0.29\textwidth}@{}}
\toprule
Axis & Nominal & Envelope range & Source of the range \\
\midrule
Tolerance time-constant, d & 1439.3 & 485.1 -- 2393.5 & $90\%$ interval, \%RSE 40.3 \\
Tolerance amplitude SSEC50, pmol/L & 3798.4 & 1786.5 -- 5810.4 & $90\%$ interval, \%RSE 32.2 \\
GLP potency EC50, pmol/L & 48.06 & 28.14 -- 96.13 & lower confidence limit; upper limit $2\times$ estimate \\
Lifestyle decay Kred, d$^{-1}$ & 0.001953 & 0.001172 -- 0.002734 & asserted $\pm40\%$ band \\
Lifestyle amplitude & 0.2240 & 0.00 -- 0.70 & absolute support for lifestyle scenarios \\
Baseline weight, kg & 103.8 & 60 -- 175 & synthetic Phase-2 specification; upper limit raised \\
Height, m & 1.72 & 1.50 -- 2.05 & synthetic Phase-2 specification \\
Baseline activity level & 1.55 & 1.45 -- 2.10 & numerical floor; active-population ceiling \\
Dose, mg & 2.40 & 0.25 -- 2.40 & clinical titration-to-maintenance regimen \\
\bottomrule
\end{tabular}
\end{table}

\needspace{4\baselineskip}
The two activity-effect parameters have unusable entries in the published uncertainty table: the cells duplicate unrelated rows and are dimensionally inconsistent with the parameters they describe.
They are therefore held at their implementation defaults and are not envelope axes.

\begin{table}[htbp]
\centering
\caption{GLP-1RA cohorts used in the reported results. The \texttt{obese\_gauss} cohort applies the truncated-Gaussian base on all nine axes and shifts the body-weight mean to \SI{140}{\kilo\gram}; the shift is a design value.}
\label{tab:prov-glp-cohorts}
\footnotesize
\begin{tabular}{@{}>{\raggedright\arraybackslash}p{0.22\textwidth}>{\raggedright\arraybackslash}p{0.48\textwidth}>{\raggedright\arraybackslash}p{0.18\textwidth}@{}}
\toprule
Cohort & Restriction relative to the envelope & Type \\
\midrule
Envelope       & none & whole envelope \\
Weak lifestyle & lifestyle amplitude $0.000$ -- $0.233$ & lower third \\
Strong lifestyle & lifestyle amplitude $0.467$ -- $0.700$ & upper third \\
Obese          & lifestyle amplitude pinned at $0.35$; body weight $136.7$ -- $175$ & pin and upper third \\
Poor responder & lifestyle amplitude pinned at $0.35$; GLP potency $73.47$ -- $96.13$ & pin and upper third \\
Sedentary      & lifestyle amplitude pinned at $0.35$; baseline activity $1.450$ -- $1.667$ & pin and lower third \\
\texttt{obese\_gauss} & truncated-Gaussian base; body-weight mean moved to $140.0$ & Gaussian \\
\bottomrule
\end{tabular}
\end{table}

The registry retains the active variant and additional Gaussian variants for development; the reported GLP set above is the one used by the paper's numerical summaries.

\subsection{Ibuprofen, $D = 15$}
\label{app:provenance-ibuprofen}

The ibuprofen envelope is the $2.5$ to $97.5$ percentile spread of a 3000-individual virtual population generated with the physiologically based toxicokinetic model and population generator of the \texttt{httk} package \citep{pearce2017httk,ring2017identifying}.
The two ADME axes, metabolic clearance and plasma free fraction, carry the package's in-vitro and population uncertainty; the remaining axes carry physiological variability.

The virtual population is generated through a correlated covariate-to-physiology map, but the GSA uses the resulting marginal ranges independently.
Every sampled volume and flow is body-weight normalised, while raw body weight, age and chemistry-determined partition coefficients are held fixed so that independent draws do not break the model's allometry.
Gut permeability and haematocrit are excluded from the current 15-axis design for different implementation reasons: ibuprofen uses a measured bioavailability rather than the permeability value, and haematocrit is folded into the blood-to-plasma ratio before the per-row overrides reach the solver.

\begin{table}[htbp]
\centering
\caption{Ibuprofen envelope, $D=15$. All marginals are uniform and independent, and every range is the $2.5$ to $97.5$ percentile spread of the same 3000-individual \texttt{httkpop} population.}
\label{tab:prov-httk-axes}
\footnotesize
\begin{tabular}{@{}>{\raggedright\arraybackslash}p{0.34\textwidth}p{0.20\textwidth}p{0.28\textwidth}@{}}
\toprule
Axis & Nominal (p50) & Envelope range \\
\midrule
Metabolic clearance & 0.009951 & 0.000529 -- 0.08929 \\
Fraction unbound    & 0.007603 & 0.001033 -- 0.02927 \\
Fraction absorbed   & 0.9732   & 0.8890 -- 0.9891 \\
Cardiac output      & 13.60    & 7.717 -- 35.12 \\
GFR (renal)         & 0.2517   & 0.1150 -- 0.4784 \\
Gut flow fraction   & 0.1599   & 0.1408 -- 0.1794 \\
Kidney flow fraction & 0.1792  & 0.1559 -- 0.2052 \\
Liver flow fraction & 0.06506  & 0.05858 -- 0.07143 \\
Arterial volume     & 0.02056  & 0.01159 -- 0.03052 \\
Gut volume          & 0.01335  & 0.006458 -- 0.04886 \\
Kidney volume       & 0.003582 & 0.001466 -- 0.007224 \\
Liver volume        & 0.01924  & 0.007723 -- 0.03986 \\
Lung volume         & 0.006183 & 0.002234 -- 0.02187 \\
Rest-of-body volume & 0.04984  & 0.03286 -- 0.06661 \\
Venous volume       & 0.04175  & 0.02354 -- 0.06197 \\
\bottomrule
\end{tabular}
\end{table}

\begin{table}[htbp]
\centering
\caption{Ibuprofen cohorts used in the reported results. All are narrowed uniform marginals; none pins an axis.}
\label{tab:prov-httk-cohorts}
\footnotesize
\begin{tabular}{@{}>{\raggedright\arraybackslash}p{0.24\textwidth}>{\raggedright\arraybackslash}p{0.52\textwidth}>{\raggedright\arraybackslash}p{0.16\textwidth}@{}}
\toprule
Cohort & Restriction relative to the envelope & Type \\
\midrule
Envelope          & none & whole envelope \\
Elderly           & GFR lower third; cardiac output lower third & two lower thirds \\
Hepatic impair.   & metabolic clearance lower third; fraction unbound upper third & lower and upper third \\
Hypoalbuminaemia  & fraction unbound upper third & upper third \\
Enzyme induction  & metabolic clearance upper third & upper third \\
\bottomrule
\end{tabular}
\end{table}

The registry also retains severe CKD, CYP2C9 poor-metaboliser and obese variants, which are mechanistic contrasts rather than samples drawn from separate clinical populations and are not part of the reported five-cohort set.

\subsection{Paracetamol, $D = 23$}
\label{app:provenance-apap}

The paracetamol model is the physiologically based model of \citet{zurlinden2016physiologically}, imported through the \texttt{pksensi} package \citep{hsieh2020pksensi}.
No per-parameter standard errors are available for the model.
The 21 metabolic and clearance axes therefore inherit the sensitivity-analysis ranges used in the package's published workflow \citep{hsieh2018applying}, sampled uniformly in natural-log units.
Eleven ranges use the nominal plus or minus $1.96$ in log units; the other ten are the wider explicit boxes supplied by that convention, typically for capacities, clearances or cofactor synthesis.

Dose and body weight are the two population axes.
They are sampled uniformly over $10$ to \SI{250}{\milli\gram\per\kilo\gram} and $50$ to \SI{100}{\kilo\gram}, respectively, which are asserted boxes chosen to contain the dose pins.
Because the envelope is a documented sensitivity convention rather than a fitted uncertainty distribution, no Gaussian cohort is defined for this model.

{\footnotesize
\begin{longtable}{@{}>{\raggedright\arraybackslash}p{0.31\textwidth}p{0.16\textwidth}p{0.31\textwidth}@{}}
\caption{Paracetamol envelope, $D=23$. The first 21 axes are uniform in natural-log units; dose and body weight are uniform in natural units. A dagger marks a nominal outside the sampled range because the nominal is the model default rather than the centre of the envelope.}\label{tab:prov-apap-axes} \\
\toprule
Axis & Nominal & Envelope range \\
\midrule
\endfirsthead
\caption[]{\emph{continued}} \\
\toprule
Axis & Nominal & Envelope range \\
\midrule
\endhead
\bottomrule
\endfoot
Gut transit              & $-1.0$        & $-2.96$ -- $0.96$ \\
Gut peak transit         & $-3.0$        & $-4.96$ -- $-1.04$ \\
CYP Km                   & $7.6$         & $5.64$ -- $9.56$ \\
CYP capacity to NAPQI    & $8.4\dagger$  & $-2.0$ -- $5.0$ \\
SULT Km APAP             & $3.0$         & $1.04$ -- $4.96$ \\
SULT Ki                  & $20$          & $18.04$ -- $21.96$ \\
SULT Km PAPS             & $3.0$         & $1.04$ -- $4.96$ \\
Sulfation capacity       & $10$          & $0.0$ -- $10.0$ \\
UGT Km                   & $9$           & $7.04$ -- $10.96$ \\
UGT Ki                   & $23$          & $21.04$ -- $24.96$ \\
UGT Km UDPGA             & $2.3$         & $0.34$ -- $4.26$ \\
Glucuronidation capacity & $7.3$         & $0.0$ -- $10.0$ \\
AG Km                    & $9.9$         & $7.94$ -- $11.86$ \\
AG Vmax                  & $30\dagger$   & $7.0$ -- $15.0$ \\
AS Km                    & $10$          & $8.04$ -- $11.96$ \\
AS Vmax                  & $20\dagger$   & $7.0$ -- $15.0$ \\
UDPGA synthesis          & $-30\dagger$  & $0.0$ -- $13.0$ \\
PAPS synthesis            & $-30\dagger$  & $0.0$ -- $13.0$ \\
Renal CL parent          & $-2.0$        & $-6.0$ -- $1.0$ \\
Renal CL AG              & $-2.0$        & $-6.0$ -- $1.0$ \\
Renal CL AS              & $-2.0$        & $-6.0$ -- $1.0$ \\
Oral dose, mg/kg          & $1\dagger$    & $10$ -- $250$ \\
Body weight, kg           & $54.6$        & $50$ -- $100$ \\
\end{longtable}
}

All other model quantities are held at the vendored defaults, including molecular weights, tissue volumes, blood-flow fractions, partition coefficients and basal cofactor rates.
Glutathione is not an axis: the model fixes its initial state, and the UDPGA and PAPS synthesis axes are not substitutes for a glutathione input.

\begin{table}[htbp]
\centering
\caption{Paracetamol cohorts used in the reported results. The paediatric restriction is motivated by the developmental shift towards sulfation reported by \citet{miller1976acetaminophen}; the third-width sub-bands are a study design choice.}
\label{tab:prov-apap-cohorts}
\small
\begin{tabular}{@{}>{\raggedright\arraybackslash}p{0.25\textwidth}>{\raggedright\arraybackslash}p{0.51\textwidth}>{\raggedright\arraybackslash}p{0.16\textwidth}@{}}
\toprule
Cohort & Restriction relative to the envelope & Type \\
\midrule
Envelope          & none & whole envelope \\
Therapeutic       & oral dose \SI{15}{\milli\gram\per\kilo\gram} & pin \\
Overdose          & oral dose \SI{210}{\milli\gram\per\kilo\gram} & pin \\
Paediatric conj.  & sulfation capacity upper third; glucuronidation capacity lower third & two thirds \\
\bottomrule
\end{tabular}
\end{table}

The registry retains supratherapeutic and fixed-weight overdose variants, but they are not part of the four-cohort summaries reported in the paper.

\subsection{Exceptions and omitted axes}
\label{app:provenance-fallbacks}

The departures from the primary source routes are collected in \cref{tab:prov-fallbacks}.
They are part of the input definition and are stated so that a sensitivity result is not mistaken for a data-derived uncertainty statement.

\begin{table}[htbp]
\centering
\caption{Departures from the primary envelope sources.}
\label{tab:prov-fallbacks}
\footnotesize
\begin{tabular}{@{}>{\raggedright\arraybackslash}p{0.16\textwidth}>{\raggedright\arraybackslash}p{0.39\textwidth}>{\raggedright\arraybackslash}p{0.35\textwidth}@{}}
\toprule
Model & Issue & Treatment in the study \\
\midrule
Remifentanil & Five kinetic fixed effects are near-point, flat, railed or have relative standard error $152\%$; impairment has no fitted lower bound & Fix the five kinetics at their MLEs and assert $0.30$ impairment floors; assert the lower $k_{PT}$ floor \\
GLP-1RA & The Kred interval is too narrow, the poor-responder tail is not in the fitted interval, and the activity-effect rows of Table~3 are corrupt & Use a $\pm40\%$ Kred band, double the GLP potency estimate for the upper limit, use absolute LSCI support, and hold the two activity-effect parameters fixed \\
Ibuprofen & \texttt{httkpop} supplies correlated virtual individuals, while the GSA requires independent axes; two supplied axes do not reach the current design & Use only the percentile marginals, sample them independently, and omit gut permeability and haematocrit \\
Paracetamol & No fitted standard errors exist and the source ranges are a sensitivity-analysis convention & Inherit the \texttt{pksensi} log-scale ranges and assert the dose and adult-weight boxes \\
\bottomrule
\end{tabular}
\end{table}

Six paracetamol nominals lie outside their sampled ranges: CYP capacity to NAPQI, AG Vmax, AS Vmax, UDPGA synthesis, PAPS synthesis and the placeholder oral-dose nominal.
These are defaults of the vendored implementation, not centres that the envelope is required to contain.
The planned chronic-alcoholic paracetamol cohort is also absent because its parameter region caused a hard integration failure, so no converged reference could be computed for it.

\section{Supporting figures}
\label{app:figures}

The figures collected here support claims made in the main text without being needed to follow the argument.

\subsection{Why the indices are reduced over the whole horizon}
\label{app:figures-time}

\begin{figure}[p]
\centering
\makebox[\textwidth][c]{\includegraphics[width=1.35\textwidth]{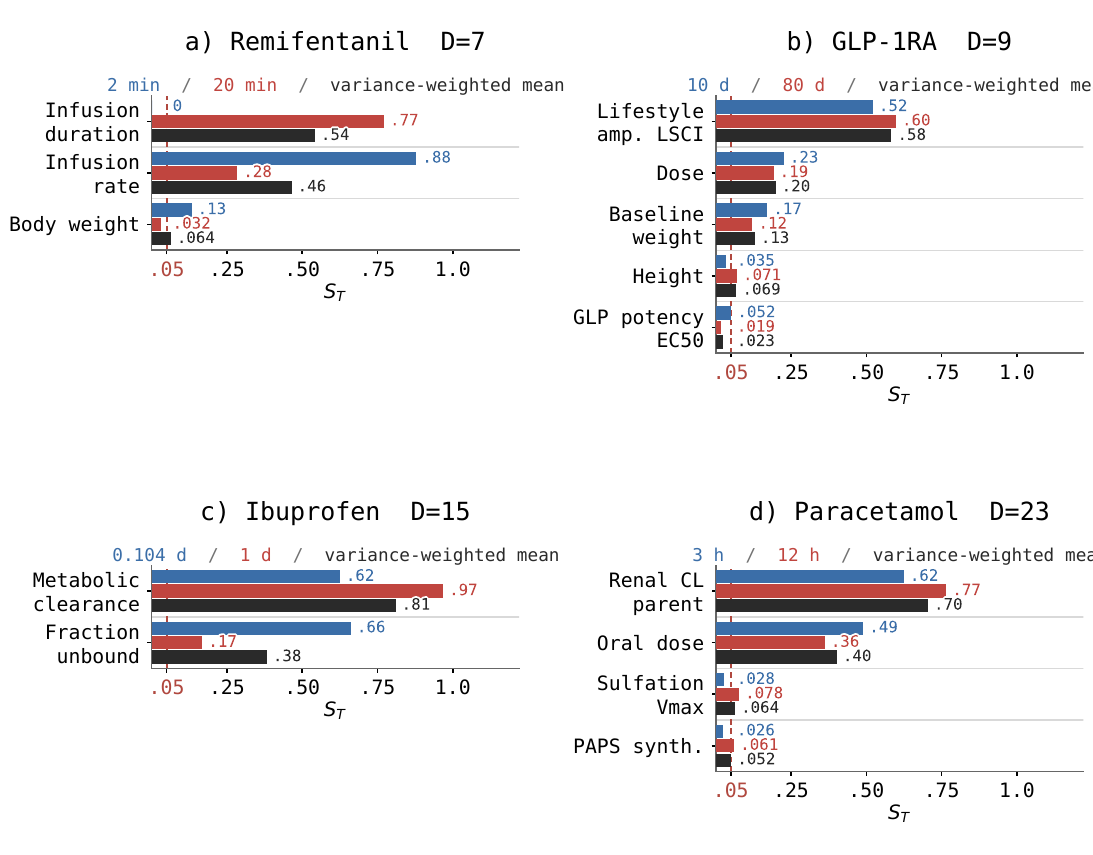}}
\caption{Converged total-effect index of the whole-envelope cohort at an early instant, a late instant
and the horizon mean, for the four models in dimension order.
The instants are named in the key above each panel, and the dashed rule is the five per cent importance threshold.
Rows are the axes reaching that threshold in at least one of the three vectors.
Values below $0.10$ carry three decimals.
}
\label{fig:st-time}
\end{figure}

\cref{sec:reduction} reduces every per-timepoint index to a mean over the horizon weighted by the output variance, rather than reading the indices at a chosen instant.
\cref{fig:st-time} is the evidence for that choice: two of the four models change their leading axis within the horizon, remifentanil between the infusion rate and the infusion duration and ibuprofen between the unbound plasma fraction and metabolic clearance, so neither instant is a defensible answer on those two.
On remifentanil the weighted reduction also returns body weight at $0.064$, above the importance threshold, where an unweighted mean over the same horizon returns $0.047$.
The GLP-1RA model is the control, its three vectors agreeing within $0.08$ on every axis.

\subsection{The gate quantities against training size}
\label{app:figures-spread}

\begin{figure}[p]
\centering
\makebox[\textwidth][c]{\includegraphics[width=1.10\textwidth]{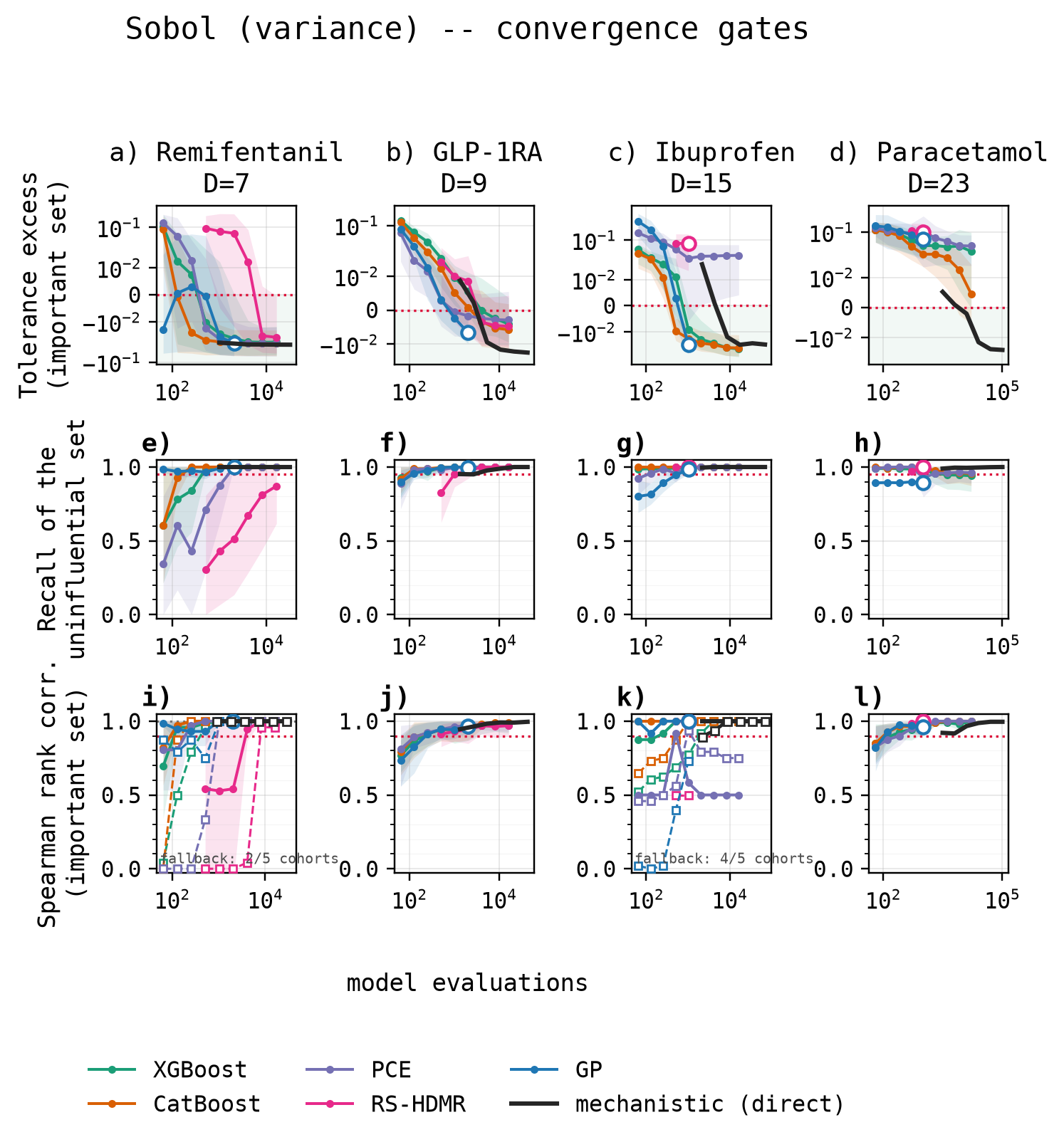}}
\vspace{-1.5em}
\caption{The three convergence-gate quantities of \cref{sec:gate} against training size for the five surrogates and the direct mechanistic sweep.
Rows are tolerance excess, unimportant-set recall and important-set Spearman correlation; columns are the four models.
Shading is the cohort spread; open squares mark the fewer-than-three-important-axes fallback.
}
\label{fig:convergence}
\end{figure}

\cref{fig:convergence} gives the three convergence-gate quantities of \cref{sec:gate} against training
size, for the five surrogates against the direct mechanistic sweep, with the band showing spread across
the cohorts of each model.

\subsection{Method against method, parameter by parameter}
\label{app:figures-agreement}

\begin{figure}[p]
\centering
\makebox[\textwidth][c]{\includegraphics[width=1.35\textwidth]{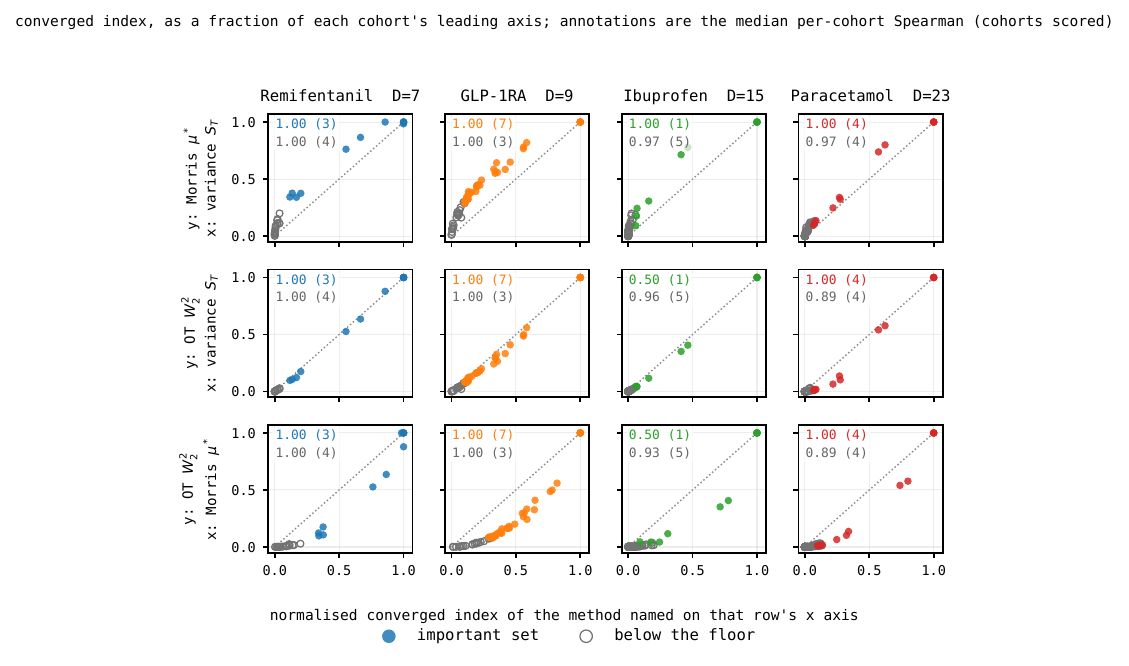}}
\caption{Method against method, parameter by parameter, for the three method pairs (rows) on the four models in dimension order (columns).
Each point is one parameter of one cohort, with every index normalised by its own cohort's largest, so a coordinate is that fraction of the cohort's leading axis; the leading axis of every cohort therefore sits at one on both axes by construction, which is an artefact of the normalisation and carries no information.
Filled points are the variance-based important set at $\ST \geq 0.05$ and hollow points its complement, both scopes defined by the variance-based threshold because it is the only one of the three on an absolute scale.
The annotation is the median over cohorts of the per-cohort Spearman correlation on the important set, with the number of contributing cohorts beneath it; cohorts carrying fewer than three axes in scope are excluded, a rank correlation over two points carrying no information.
Read that count before the correlation: the ibuprofen important-set entry rests on one cohort of three parameters, where a Spearman can take only the values $1.00$, $0.50$, $-0.50$ and $-1.00$.
}
\label{fig:agreement-scatter}
\end{figure}

\cref{fig:agreement-scatter} draws the points behind the median rank correlations and weight ratios of \cref{sec:results-agreement}, so that the shape of a disagreement is visible and a correlation carried by three axes is not read as one carried by thirty.

\end{document}